\documentclass[12pt]{article}
\usepackage[utf8]{inputenc}
\usepackage[margin=1in]{geometry}
\usepackage{setspace}

\usepackage{lscape}
\usepackage{array}
\usepackage{enumerate}
\usepackage{enumitem}
\usepackage{eufrak}
\usepackage{amsmath,amssymb,amsthm}
\usepackage{amsmath}
\usepackage{bbm}
\usepackage[demo]{graphicx}
\usepackage{caption}
\usepackage{subcaption}
\usepackage[table,xcdraw]{xcolor}
\usepackage{algpseudocode}
\usepackage{multicol}
\usepackage{flafter}
\usepackage{mathtools}
\usepackage{authblk}
\usepackage{color,hyperref}
\hypersetup{colorlinks=true,urlcolor=blue, citecolor=purple}
\usepackage{cleveref}
\usepackage{float}
\usepackage{subcaption}
\usepackage{amsmath}
\usepackage{tikz}
\usepackage{mathdots}
\usepackage{yhmath}
\usepackage{cancel}
\usepackage{bm}
\usepackage{color}
\usepackage{siunitx}
\usepackage{array}
\usepackage{multirow}
\usepackage{amssymb}
\usepackage{tabularx}
\usepackage{booktabs}
\usetikzlibrary{fadings}
\usetikzlibrary{patterns}
\usetikzlibrary{shadows.blur}
\usetikzlibrary{shapes}
\usepackage{physics}
\usepackage{amsmath}
\usepackage{tikz}
\usepackage{mathdots}
\usepackage{yhmath}
\usepackage{cancel}
\usepackage{color}
\usepackage{siunitx}
\usepackage{array}
\usepackage{multirow}
\usepackage{amssymb}
\usepackage{textcomp}
\usepackage{gensymb}
\usepackage{tabularx}
\usepackage{booktabs}
\usepackage{setspace}
\usepackage{float}
\usepackage{cleveref}
\usepackage{physics}
\usepackage{amsmath}
\usepackage{tikz}
\usepackage{mathdots}
\usepackage{yhmath}
\usepackage{cancel}
\usepackage{siunitx}
\usetikzlibrary{fadings}
\usetikzlibrary{patterns}
\usetikzlibrary{shadows.blur}
\usetikzlibrary{shapes}
\usepackage{fancyhdr}
 \usepackage{booktabs}
 \usepackage{multirow}
\usetikzlibrary{fadings}
\usetikzlibrary{patterns}
\usetikzlibrary{shadows.blur}
\usetikzlibrary{shapes}
\usepackage[ruled,vlined]{algorithm2e}
\usepackage[round]{natbib}
\usepackage{tcolorbox}
\usepackage{tabularx}
\usepackage{listings}
\usepackage{booktabs}
\usepackage{parskip}
\usepackage{amssymb}
\usepackage{multirow}

\def\R{\mathbb{R}}

\def\bY{\bm{Y}}
\def\bX{\bm{X}}
\def\bC{\bm{C}}
\def\bZ{\bm{Z}}
\def\bR{\bm{R}}

\renewcommand{\vec}[1]{\mathbf{1}}
\DeclareMathOperator*{\argmin}{arg\min}
\newcommand{\E}{\mathbb{E}}
\newcommand{\prob}{\mathbb{P}}
\newcommand{\btheta}{\bm{\theta}}
\newcommand{\boldeta}{\bm\eta}
\newcommand{\normal}{\mathcal{N}}
\numberwithin{equation}{section}

\newtheorem{theorem}{Theorem}[section]
\newtheorem{corollary}{Corollary}[theorem]

\newtheorem{lemma}[theorem]{Lemma}
\newtheorem{assumption}{Assumption}[section]

\theoremstyle{definition}
\newtheorem{definition}{Definition}[section]

\theoremstyle{remark}

\definecolor{codegreen}{rgb}{0,0.6,0}
\definecolor{codegray}{rgb}{0.5,0.5,0.5}
\definecolor{codepurple}{rgb}{0.58,0,0.82}
\definecolor{backcolour}{rgb}{0.95,0.95,0.92}

\lstdefinestyle{mystyle}{
    backgroundcolor=\color{backcolour}, 
    commentstyle=\color{codegreen},
    keywordstyle=\color{magenta},
    numberstyle=\tiny\color{codegray},
    stringstyle=\color{codepurple},
    basicstyle=\ttfamily\footnotesize,
    breakatwhitespace=false,         
    breaklines=true,                 
    captionpos=b,                    
    keepspaces=true,                 
    numbers=left,                    
    numbersep=5pt,                  
    showspaces=false,                
    showstringspaces=false,
    showtabs=false,                  
    tabsize=2
}

\usepackage{amsmath}

\title{Bayesian High-dimensional Linear Regression with Sparse Projection-posterior}
\author{Samhita Pal, Subhashis Ghoshal}
\date{}
\begin{document}

\maketitle
\allowdisplaybreaks

\abstract{We consider a novel Bayesian approach to estimation, uncertainty quantification, and variable selection for a high-dimensional linear regression model under sparsity. The number of predictors can be nearly exponentially large relative to the sample size. We put a conjugate normal prior initially disregarding sparsity, but for making an inference, instead of the original multivariate normal posterior, we use the posterior distribution induced by a map transforming the vector of regression coefficients to a sparse vector obtained by minimizing the sum of squares of deviations plus a suitably scaled $\ell_1$-penalty on the vector. We show that the resulting sparse projection-posterior distribution contracts around the true value of the parameter at the optimal rate adapted to the sparsity of the vector. We show that the true sparsity structure gets a large sparse projection-posterior probability. We further show that an appropriately recentred credible ball has the correct asymptotic frequentist coverage. Finally, we describe how the computational burden can be distributed to many machines, each dealing with only a small fraction of the whole dataset. We conduct a comprehensive simulation study under a variety of settings and found that the proposed method performs well for finite sample sizes. We also apply the method to several real datasets, including the ADNI data, and compare its performance with the state-of-the-art methods. We implemented the method in the \texttt{R} package called \texttt{sparseProj}, and all computations have been carried out using this package.}


\maketitle

\section{Introduction}
\label{sec:intro}
High-dimensional linear regression models with numerous predictors, potentially exceeding the observations in number, have garnered significant research interest. Most predictors remain inactive in such models, leading to sparsity in the regression coefficient vector. This feature enables reliable estimation of coefficients by leveraging the underlying low-dimensional structure. Introducing a penalty function to the objective function is a common approach to dealing with such problems, encouraging the minimizer to yield sparser solutions. The popular method LASSO \citep{tibshirani1996regression,zhang2008sparsity} enforces an $\ell_1$-norm constraint on the coefficient vectors, resulting in exact zeros at some coordinates. Variations of the LASSO include the Minimax Concave Penalty (MCP) \citep{zhang2010nearly}, Smoothly Clipped Absolute Deviation (SCAD) \citep{fan2001variable}, Dantzig selector \citep{candes2007dantzig}, adaptive LASSO \citep{zou2006adaptive}, non-negative garrotte \citep{breiman1995better,yuan2007non} estimators. Earlier research explored LASSO's asymptotic properties, such as consistency and limiting distribution, in fixed-dimensional settings \citep{fu2000asymptotics}. A bootstrap LASSO was introduced, but \cite{chatterjee2010asymptotic} demonstrated that it lacks consistency when one or more components of the coefficient vector are zero.

Bayesian methods for high-dimensional linear models under sparsity have been also developed. The Bayesian LASSO \citep{park2008bayesian,hans2009bayesian} employs a Laplace prior for each coefficient. While the posterior mode is the LASSO, the posterior distribution is not supported on sparse vectors and fails to concentrate near the true parameter vector. A spike-and-slab prior  \citep{mitchell1988bayesian,ishwaran2005spike} can address the issue by introducing sparsity in coefficients through a point mass at zero. It automatically selects a smaller subset of predictors as the active set in each posterior draw of the regression coefficients, resulting in multiple models in the posterior. Each model corresponds to an active predictor set; hence, computing all model posterior probabilities is very intensive. The Stochastic Search Variable Selection (SSVS) method \citep{george1993variable} uses Gibbs sampling to address the issue, but it is still a slow process. Continuous shrinkage priors replace a spike and slab mixture with a single density using a global scale parameter across all components to obtain a thick tail and a local scale parameter to create a high concentration at zero. Hence, they are also referred to as global-local priors. Some notable continuous shrinkage priors include the horseshoe \citep{pmlr-v5-carvalho09a}, normal-gamma  \citep{brown2010inference}, double-Pareto \citep{armagan2013generalized}, Dirichlet-Laplace \citep{bhattacharya2015dirichlet}, and R2-D2 \citep{zhang2022bayesian}. 


 Variational inference \citep{wainwright2008graphical} provides a computationally faster method to approximately compute posterior distributions through optimization. \cite{ray2022variational, ormerod2017variational,huang2016variational} studied the mean-field spike and slab variational Bayes approximation for high-dimensional linear regression, whereas \cite{mukherjee2022variational} studied the naive mean-field approximation to the posterior distribution arising from product priors. A general $\alpha$-variational inference technique for the high-dimensional linear model under sparsity was provided in \cite{yang2020alpha}. These variational methods also possess estimation or selection consistency. Moreover, \cite{zhang2020convergence} extensively studied convergence rates of these variational Bayes methods in the high-dimensional linear regression setting. The paper \cite{han2019statistical} conducted a non-asymptotic analysis on the approximation of the posterior distributions in linear models involving latent variables. However, to our knowledge, uncertainty quantification of these methods has not been investigated, except for \cite{yang2020variational} and \cite{bai2020nearly}. The former considered posterior distributions obtained by updating an appropriate empirically-centered Gaussian prior and showed valid frequentist coverage of their credible balls under the orthogonal design matrix. On the other hand, the latter proposed a variational approach for heavy-tailed shrinkage priors but failed to show coverage.

Some authors also studied the asymptotic properties of posterior distributions in high-dimensional linear regression models. The paper \cite{castillo2015bayesian} obtained the posterior contraction rate of a spike-and-slab prior with a Laplace slab akin to the LASSO's convergence rate and also showed that if signals are sufficiently strong, then the posterior selects the model with the correct set of preditors with high probability. They stressed the importance of the tail thickness of the slab density to prevent excessive shrinkage that may compromise the posterior contraction rate. Using an empirical Bayes approach to select the prior mean of the slab distribution instead of setting it to zero, \cite{belitser2020empirical} showed that optimal posterior contraction can be obtained using a conjugate normal slab distribution.  Posterior contraction and variable selection properties of continuous shrinkage priors were established by \cite{song2017nearly}.

In this paper, we introduce an innovative Bayesian approach to obtain a posterior distribution supported on sparse subspaces using a conjugate prior on the regression coefficient, which is easy to compute, and the posterior has optimal contraction and variable selection properties. Moreover, credible regions with asymptotically correct frequentist coverage can be obtained from the resulting posterior distribution. This approach differs from traditional Bayesian methods in that a restriction like sparsity is not imposed in the prior. Thus, conjugate priors may be used, significantly simplifying the analysis. The inference is conducted by the induced posterior distribution of a sparsity-generating map, immersing full vectors in a sparse region. The resulting  ``sparse projection-posterior'' samples are easily obtained by conjugate sampling corrected by an optimization step. The idea of using such an immersion map to transform sample draws from a conjugate posterior distribution on an unrestricted parameter space to the desirable constricted space was used previously in the contexts of monotone shape regression \cite{lin2014bayesian,chakraborty2021convergence, chakraborty2021coverage,wang2022coverage} and differential equation models \cite{bhaumik2015bayesian,bhaumik2017efficient,bhaumik2022two}.

Under the standard conditions of bounded predictor variables and compatibility conditions used for LASSO's convergence, we demonstrate that the sparse projection-posterior contracts at the same rate as the LASSO. For the LASSO, consistency of variable selection is established assuming the beta-min condition and the `irrepresentable' condition \citep{wainwright2006sharp, yuan2006model, meinshausen2009lasso}. We show that, under these conditions, the sparse projection-posterior probability of the model consisting of the true set of active predictors approaches one. A notable limitation of LASSO is its inability to provide standard errors and confidence regions for coefficients with estimated zeros. 
A debiasing technique was employed to obtain confidence intervals \citep{zhang2014confidence, van2014asymptotically, javanmard2018debiasing} at the expense of sparsity.
\cite{chatterjee2011bootstrapping} developed a modified residual bootstrap technique for constructing confidence sets with good coverage properties. Various aspects of coverage in the sparse linear regression setting were studied by \cite{nickl2013confidence,cai2017confidence}.    
While Bayesian methods hold promise for quantifying uncertainty, adjustments are often needed to match their credibility with asymptotic frequentist coverage. The article \cite{belitser2020empirical} showed that a Bayesian credible ball of the optimal size adapted to the sparsity of the regression coefficient has adequate frequentist coverage at all parameter points satisfying an {\em excessive bias restriction} condition. We show that a recentered sparse projection-posterior can construct credible intervals with valid frequentist coverage for each regression coefficient. This is equivalent to employing a debiased version of the original immersion map only for this purpose. Moreover, the sparse projection-posterior method is tailored for distributed computing, which is typically impossible for traditional Bayesian methods. Its unique approach avoids MCMC sampling, requiring only conjugate Gaussian prior draws, promising faster results.


Using an immersion map adds more flexibility to the Bayesian paradigm, similar to other extensions such as the empirical Bayes or the variational method. The variational approach finds a projection of the posterior distribution within a given class of distributions, thus transforming the original posterior distribution to a related distribution. In this sense, the sparse projection-posterior, or more generally, an ``immersion posterior" induced by an immersion map relevant to the context, is conceptually similar to the variational approach. In both contexts, the transformation is obtained by an optimization step. The main difference is that, in the variational method, the optimization is done at the distributional level, while in the immersion posterior approach, the optimization is on the arguments of the distribution. Further, in the variational approach, the original posterior distribution is not calculated, and the variational posterior is intended to approximate it. However, in the immersion posterior approach, a simple unrestricted posterior is computed typically through conjugacy, but it is never intended for inference. The optimization step in the immersion posterior approach is typically non-iterative and more straightforward than in the variational method. This results in a more explicit description of the immersion posterior that allows for studying finer properties like model selection and coverage. Finally, we note that even though the variational posterior was initially intended to be an approximation to the original posterior, such approximation results are not available, so the variational posterior is eventually considered as an alternative random measure to be used to make an inference instead of the actual posterior distribution. This also prompts the need to study the convergence properties of the variational posterior. The immersion posterior provides another typically simpler alternative with many desirable convergence properties.

The paper is organized as follows. In \Cref{Method}, we introduce the proposed novel Bayesian approach to inference for a high-dimensional linear regression problem using the sparse projection-posterior distribution. Its contraction rate, variable selection, and coverage properties are established in \Cref{Results}, followed by numerical results in \Cref{Simulation}. A discussion summarizing the aspects of the proposed method is presented in \Cref{Conclusion}. The proofs of the theorems are provided in Supplement 1 of the Appendix section. A recipe for distributed computing suitable for handling large datasets and addressing privacy concerns by distributing the computation in several local machines is presented in \Cref{distributed}, while Supplement 3 contains some additional numerical results.

\section{Methodology}\label{Method}
\subsection{Motivation}
For a high-dimensional linear regression, constructing a prior on the regression coefficients with support on a sparse domain leads to a posterior that is hard to compute and analyze theoretically because of the varying restrictions imposed by sparsity. The procedure would be much easier to compute and study if the sparsity restrictions were not imposed before sampling from the posterior, but instead, if a correction map was used to convert a posterior sample to a sparse vector. This can be done using a sparsity-inducing map on the full posterior from the entire parameter space. This is not a projection map because of the presence of a penalty term in its objective function, but the resulting induced {\em sparse projection-posterior} may be viewed as an immersion posterior in the sense of \cite{wang2022coverage}. A natural choice of a sparse transformation that maps the whole parameter space to a sparser subspace is obtained by minimizing the sum of a weighted squared $\ell_2$-norm and a standard sparsity-inducing penalty, such as the $\ell_1$-penalty. The new method can be thought of as obtained by switching the order of operation of imposing the sparsity restriction and posterior updating in a usual Bayesian method for high-dimensional regression, as schematically demonstrated below: 
\begin{figure}[htbp]
    \centering
    \includegraphics[width = 0.7\linewidth]{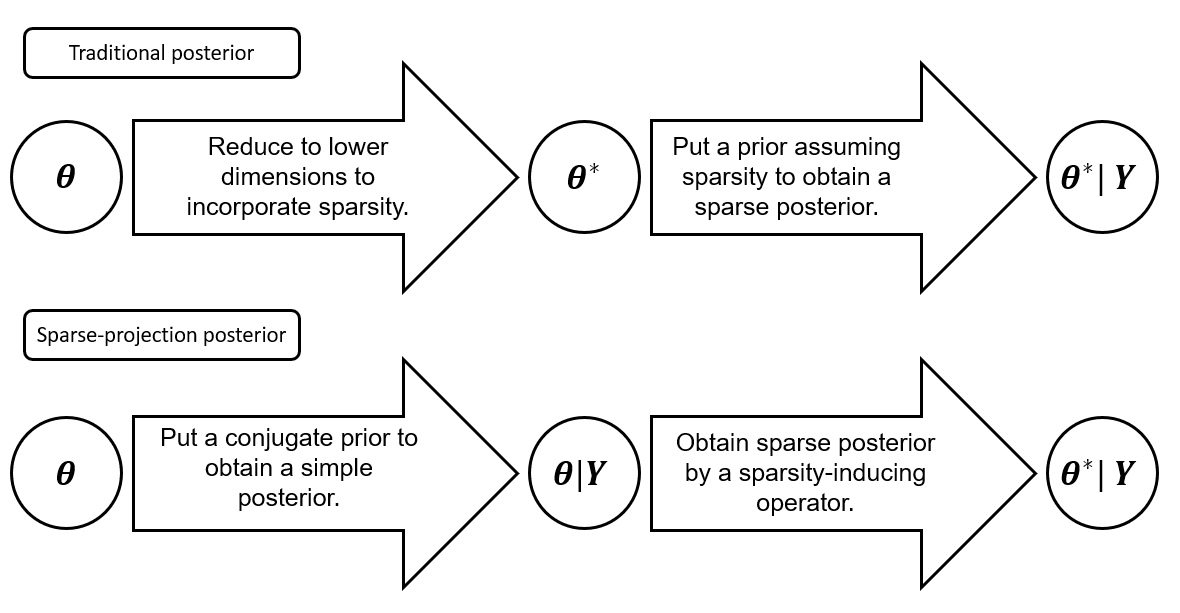}
\end{figure}

\noindent
This helps bypass the hassle of MCMC, making computations quite efficient. As a result, we do not have to worry about the convergence of the MCMC and autocorrelation of the obtained samples, and drawing much fewer samples suffices. Owing to the issue of autocorrelated posterior samples in continuous shrinkage methods such as the horseshoe, the Effective Sample Size (ESS) is very low. The ESS for the proposed method, however, will essentially be the same as the number of posterior draws owing to independence. However, in the end, both approaches lead to a data-dependent probability measure on the derived parameter space to make an inference about the parameter. Another advantage of this method is that the mapping can simultaneously be done with different tuning parameters $\lambda$ per the study's requirement or purpose. A smaller value of $\lambda$ can be used if the main goal is estimation or prediction, whereas a much larger value of the tuning parameter will be needed for consistent model selection. Optimal frequentist properties guide the choice of these maps. 

Another interpretation of the sparse projection-posterior can be given in terms of a ``super-parameter model" and a misspecified likelihood. We may view the true parameter $\btheta^0$ as a sparse projection $\btheta^0 = \iota(\bm\beta^0)$ of a ``super-parameter" $\bm\beta^0$ with no sparsity in its entries. Thus a vanilla conjugate prior can be used to make an inference on $\bm\beta$, using the misspecified likelihood obtained from the model $\bY \sim \normal_n(\bX\bm\beta, \sigma^2 \bm{I}_n)$, instead of the correct model $\bY \sim \normal_n(\bX\btheta, \sigma^2 \bm{I}_n)$. Since an inference on $\bm\beta$ is not of interest, but that on $\btheta$ is, we use the distribution induced by the map $\iota$ from the conjugate posterior for $\bm\beta$.

The soft-spike-and-slab prior and continuous shrinkage prior both use immersion maps to induce sparse posteriors on the regression parameter. The first method samples from the posterior using the stochastic search variable selection (SSVS) MCMC, and generates samples from the posterior distribution of $(\theta_1,\dots,\theta_p,\gamma_1,\ldots,\gamma_p)$, where $\gamma_1,\ldots, \gamma_p$ are variable inclusion indicators. Clearly, the sampled values of the regression parameters are dense and are made sparse by the implicit immersion map $(\theta_1,\ldots,\theta_p,\gamma_1,\ldots,\gamma_p) \mapsto (\gamma_1\theta_1,\ldots, \gamma_p\theta_p).$
Thus, in this context, $(\theta_1,\ldots,\theta_p,\gamma_1,\ldots,\gamma_p)$ are superparameters (as defined in the paper) and $\theta_j^*=\gamma_j \theta_j$, $j=1,\ldots,p$, are the parameters of the model. For continuous shrinkage priors, the thresholding map 
$\theta_j^*=\theta_j \mathbbm{1} \{|\theta_j|>\delta_n\}, j=1,\ldots,p,$ 
defines the implied immersion map with thresholding parameter $\delta_n$, which also needs to be determined and is usually chosen in an ad hoc manner. 

The main difference between our method and these is that, by using a slightly more complicated immersion map given by the sparse projection operation, we can avoid MCMC altogether and replace it with independent sampling from a multivariate normal distribution. As a result, we do not have to worry about the convergence of the MCMC and autocorrelation of the obtained samples, and drawing much fewer samples suffices. The sparse projection map is only slightly more challenging to compute, thanks to already developed algorithms using solution paths of the LASSO.

Finally, an exciting feature of the proposed sparse projection-posterior approach is that the computation can be divided among several computers, as described in Subsection \ref{distributed}. This property allows for handling very big datasets or datasets with privacy concerns without requiring the storage of the data in one machine.

\subsection{Model and Notations}

We consider the linear regression model $\bY = \bX \btheta + \bm{\varepsilon},$ where $\bX \in {\R^n \times \R^{p}}$ is the standardized design matrix, $\bY \in \R^{n}$ be the response vector, the random noise $\bm{\varepsilon}\in \R^n$ is distributed as $\normal_n(0, \sigma^2 \bm{I}_n)$ for some $\sigma>0$, and $\btheta\in \R^p$ is the vector of regression coefficients. The model dimension $p$ can grow with $n$ and be much larger than $n$. The design matrix $\bX$ is standardized so that each column has Euclidean norm $n$.

Let $\btheta^0$ denote the true value of the coefficient vector. Let $S_0=\{j: \theta^0_{j}\ne 0\}$ be the active set corresponding to $\btheta^0$ and $s_0$ be the cardinality of $S_0$, that is, the number of active predictors in the model, known as the sparsity index. Probability statements under the true distribution will be indicated by $\mathbb{P}_{\btheta^0}$. 
We write $\bC_n = n^{-1}{\bm{X}^{\mathrm{T}}\bm{X}}$ for the sample covariance matrix. For notational convenience, we assume without loss of generality that the first $s_0$ variables, denoted by $\bX_{(1)}$, are active and the true values of the coefficients corresponding to the last $p - s_0$ variables, denoted by $\bX_{(2)}$, equal zero. Then $\bC_n$ can be split as   
\begin{align}
\label{sample covariance}
 \bC_n=\begin{bmatrix}
\bC_{n(11)} & \bC_{n(12)} \\
\bC_{n(21)} & \bC_{n(22)} 
\end{bmatrix} = \frac{1}{n} \begin{bmatrix}
\bX_{(1)}^{\mathrm{T}} \bX_{(1)} & \bX_{(1)}^{\mathrm{T}} \bX_{(2)} \\
\bX_{(2)}^{\mathrm{T}} \bX_{(1)} & \bX_{(2)}^{\mathrm{T}} \bX_{(2)} 
\end{bmatrix}. 
\end{align}

 We shall use the following notations throughout the paper. The symbol 
 $\|{\bm{a}}\|_q$ stands for the $q$-norm of the vector $\bm{a} \in \R^d$ given by $(\sum_{i = 1}^{d} |a_i|^q)^{1/q}$. If $q=2$, which corresponds to the usual Euclidean norm, we shall generally drop the index 2 from the notation. For a $d_1\times d_2$-matrix $\bm{M}$, by  $\bm{M}_A$, we denote the submatrix containing the columns corresponding to a given set $A \subset \{1,2,\dots,d_2\}$.  In particular, when $A=\{j\}$ for some $j\in \{1,\ldots,p\}$, we write 
$\bm{M}^{(j)}$ for the $j$th column of $\bm{M}$ and write $\bm{M}^{(-j)}$ for the submatrix of $\bm{M}$ without the $j$th column. A random variable $X_n$ weakly converges to $X$ is denoted by $X_n \rightsquigarrow X$, and the $\gg$ or $\ll$ signs are used to describe much larger or much smaller comparisons, respectively. Let $a_n =\mathcal{O}(b_n)$ mean the sequence $a_n$ is bounded by a multiple of $b_n$, $a_n\asymp b_n$ mean both $a_n= \mathcal{O}(b_n)$ and $b_n= \mathcal{O}(a_n)$. We use $\|\bm{M}\|_F$ to denote the Frobenius norm of the matrix $\bm{M} \in \R^{m \times n}$ defined by $\sqrt{\sum_{i = 1}^m \sum_{j = 1}^n |M_{ij}|^2}$ and an $L_q$ norm of the matrix is the vector-induced norm defined by $\|\bm{M}\|_q = \sup_{\bm{x} \neq \textbf{0}} \|\bm{M} \bm{x}\|_p/\|\bm{x}\|_p$.

\subsection{Prior Specification and Posterior Distribution}

We now specify the prior distribution for the parameters. Given $\sigma>0$, let $\btheta$ be given the conjugate prior $\btheta|\sigma \sim \normal_{p}\big(\bm{0}, a_n^{-1} \sigma^2\bm{I}_{p}\big)$ for some constant $a_n$ depending on $n$. 
 Then, the posterior distribution of the parameter vector given $\sigma$ can then be easily computed as 
 \begin{align}
 \label{pos_dist}
   \btheta|(\bX,\bY,\sigma)\sim  \normal_{p} \big(\hat{\boldsymbol\theta}^\mathrm{R}, \sigma^2 \big(\bX^{\mathrm{T}} \bX + a_n\bm{I}_{p}\big)^{-1}\big),
\end{align} 
where the posterior mean is defined as $\hat{\boldsymbol\theta}^\mathrm{R} \coloneqq \big(\bX^{\mathrm{T}} \bX + a_n\bm{I}_{p}\big)^{-1} \bX^{\mathrm{T}} \bY $, owing to its resemblance with the ridge-regression estimator with penalty $a_n$. The prior choice made here does not take into account the sparsity of the coefficient vector, and, as a result, the obtained posterior does not lead to sparse samples of the parameter of interest. Sparsity is introduced in the next step, where posterior samples of dense vectors are projected to a lower-dimensional space using a sparsity-inducing map, say $f: \R^p \mapsto \R^d, \enskip d < p$, such that the unrelated components are zeroed out, giving rise to sparse posterior samples as desired. 

The marginal likelihood for $\sigma$ given by $\ell(\sigma) \propto \sigma^{-n} \exp[-\bY^{\mathrm{T}}(\bm{I}_n+ a_n \bX \bX^{\mathrm{T}})^{-1} \bY/(2\sigma^2) ]$ is obtained from the marginal distribution 
$\bY \sim \normal_n(\bm{0}, \sigma^2 (\bm{I}_n+ a_n \bX \bX^{\mathrm{T}}))$. An inverse-Gamma prior is conjugate to this likelihood. However, it will be seen that the resulting posterior is inconsistent if $p$ is not small relative to $n$. In Subsection~\ref{sigma}, we rectify it using an immersion map. 

\subsection{Sparse Projection}

To transform dense posterior samples to confine to a space with the desired sparsity, we apply a map $$f:\btheta \mapsto \btheta^*= \argmin_{u} \{ L(\btheta, \bm{u}) + \lambda_n \mathcal{P}(\bm{u})\},$$
where $L$ is a loss function, $\lambda_n$ is a tuning parameter depending on $n$ and $\mathcal{P}(\cdot)$ is a penalty function that enforces sparsity of the solution, and the extent of sparsity depends on the choice of $\lambda_n$. Our choice of the loss tries to capture the effect of the departure of $\bX \btheta^*$ from $\bX \btheta$, leading to the squared error loss $L(\btheta, \bm{u})=n^{-1}\|{\bX \btheta - \bX \bm{u}}\|^2 $, although a more general choice $(\btheta - \bm{u})^{\mathrm{T}} \bm{D} (\btheta - \bm{u})$ is also sensible. We choose $\mathcal{P}(\bm{u})=\|{ \bm{u}}\|_1$, the LASSO penalty, but other choices will also meet the intent. The posterior distribution of $\bX \btheta$ given $\sigma$ is given by 
\begin{align}
\label{posterior of X-theta}
\normal_n \big(\bX \hat{\boldsymbol\theta}^\mathrm{R} , \sigma^2 \bm{H}(a_n)\big), \; \bm{H}(a_n) = \bX \big(\bX^{\mathrm{T}} \bX + a_n \bm{I}_{p}\big)^{-1} \bX^{\mathrm{T}}.
\end{align}
Unlike the hat matrix in the formula for the least square estimator, $\bm{H}(a_n)$ is not a projection matrix. The posterior distribution of $\boldeta = \bX \btheta - \bX \theta^0$ is also  Gaussian with the same variance, but with mean $\boldsymbol \mu=\bX (\hat{\boldsymbol\theta}^\mathrm{R} - \btheta^0)$. For a given draw $\bm\theta$ from its posterior distribution given $\sigma$, the sparse projection map $\btheta\mapsto \btheta^*$ is given by 
\begin{align}
\label{sparse_proj}
    \boldsymbol \theta^* = \argmin_{\bm{u}} \{ n^{-1} \|{\bX \btheta - \bX \bm{u}}\|^2 + \lambda_n \|{\bm{u}}\|_1\}.
\end{align}

It should be noted here that the map described in \eqref{sparse_proj} is a specific choice, and the theories developed later in this paper are based on this choice. There could be several other combinations of loss functions and penalties that will guide the theoretical results. For our map, the sparse projection-posterior for $\btheta$ can be described as follows. Let $\sigma$ be a draw from the marginal posterior for $\sigma$ and let $\sigma^*=\iota(\sigma)$, the image under the immersion map $\iota$ for $\sigma$ described in Subsection~\ref{sigma}. Then $\bm\theta$ is drawn from $\normal_{p} ((\bX^{\mathrm{T}} \bX + a_n\bm{I}_{p})^{-1} \bX^{\mathrm{T}} \bY, (\sigma^*)^2 (\bX^{\mathrm{T}} \bX + a_n\bm{I}_{p})^{-1})$ and mapped to $\btheta^*$ through the sparse projection map. The resulting induced posterior is used to make inferences. The next section establishes its convergence, selection, and uncertainty quantification properties. The sampling scheme of the sparse projection-posterior method is described in Algorithm \ref{algo}.
    \begin{algorithm}[]
        \caption{Projection-posterior for $\btheta$}\label{algo}
    \begin{algorithmic}[]
            \State \textbf{Step 1}: Draw $\sigma$ from the marginal posterior for $\sigma$ and let the image under the immersion map for $\sigma$ be $\sigma^*=\iota(\sigma).$
            \State \textbf{Step 2}: Draw $\bm\theta$ from the posterior $\normal_{p} (\hat{\btheta}^\mathrm{R}, (\sigma^*)^2 (\bX^{\mathrm{T}} \bX + a_n\bm{I}_{p})^{-1}).$
            \State \textbf{Step 3}: Map to $\btheta^*$ through the map $\btheta^* = \argmin_{\bm{u}} \{ n^{-1} \|\bX \btheta - \bX \bm{u}\|^2 + \lambda_n \|\bm{u}\|_1\}.$
        \end{algorithmic}
    \end{algorithm}

\subsection{Distributed Computing}
\label{distributed}
Frequently, enhanced computational speed involves using multiple computers. Sometimes, it is essential due to enormous data sizes, making single-machine storage impossible, or privacy concerns prohibiting raw data transfer. The typical parallelization approach divides data into subsets processed concurrently on various processors. While this speeds up operations, communication costs hinder its effectiveness. Techniques have emerged to minimize communication by merging outcomes only in the final phase, resulting in simpler and faster algorithms. The challenge lies in designing nearly communication-free algorithms that match the statistical correctness of the concurrent methods on the complete dataset at a significantly reduced computational cost. In our context, when distributed over several computers, communication is only required in one step, where the necessary pieces are aggregated to obtain the required quantity. Suppose the data is divided into $m$ machines that collect (or receive) $n_1,\ldots,n_m$ data units respectively, where $\sum_{j=1}^m n_j=n$. Let $(\bX_j,\bY_j)$ denote the data units in the $j$th machine in the vector form, $j=1,\ldots,m$, while $(\bX,\bY)$ stand for the entire set of data. Clearly, $\bX^{\mathrm{T}}\bX= \sum_{j=1}^m \bX_j^{\mathrm{T}}\bX_j$ and $\bX^{\mathrm{T}}\bY= \sum_{j=1}^m \bX_j^{\mathrm{T}}\bY_j$. Since $ \bX_j^{\mathrm{T}}\bX_j$ and $\bX_j^{\mathrm{T}}\bY_j$ can be computed in the $j$th computer without communicating the entire raw data of the $j$th segment, the central computer can compute $\bX^{\mathrm{T}}\bX$ and $\bX^{\mathrm{T}}\bY$ and hence draw samples from \eqref{pos_dist} and that of the marginal posterior distribution of $\sigma$, which may be corrected by an immersion map as discussed earlier. It may be noted that the auxiliary computers perform computations only once and then communicate the summary to the central processor, where all other computations, including matrix inversion, sample draws, and optimization steps, are executed. The inversion should be done by the singular value decomposition of $\bX$. 

It is to be noted that the distributed algorithm computes the same posterior, and hence, no new results about its asymptotic properties are needed for its justification.  
The sparse projection-posterior can be computed through the distributed computation technique, making it unique among Bayesian methods or their modifications. None of the available Bayesian methods, including those based on spike-and-slab or continuous shrinkage priors, are compatible with distributed computing.

\section{Main Results}
\label{Results}

Below, we provide results on the contraction of the posterior at the true value of the parameter and consistency of variable selection. The set of assumptions required for the corresponding results is introduced immediately preceding the results. The assumptions made for the projection-posterior approach are closely related to those used to derive the convergence properties of the LASSO. It should be kept in mind that the assumptions can be changed as per the choice of the penalty for similar posterior convergence results to hold if one chooses to work with any other projection map. Moreover, we do not use the Gaussian posterior distribution for any of the proofs. We will only need sub-Gaussianity of $\bm\eta$ to address the ultrahigh-dimensional case where $\log p$ is exponential in a power of $n$. For a more modest polynomial growth in $p$, the sub-Gaussianity condition can be significantly weakened to the condition of finiteness of a sufficiently high moment, depending on the relative comparison of $p$ and $n$, provided we use a tuning parameter of a different magnitude.
\subsection{Consistency}
\begin{definition}
\label{consistency}
The posterior distribution $\Pi (\cdot|\bY)$ is said to be estimation consistent (respectively, prediction consistent) at the true value $\btheta^0$ of the parameter $\btheta$ if for all $\epsilon>0$, $\Pi (\|\btheta-\btheta^0\|>\epsilon|\bY) \to 0$ (respectively, $\Pi (\|\boldsymbol{X}\btheta-\boldsymbol{X}\btheta^0\|>\epsilon|\bY) \to 0$) in probability under $\btheta^0$. 

The posterior distribution $\Pi (\cdot|\bY)$ is said to contract at the rate $\epsilon_n$ at $\btheta^0$ if for any sequence $M_n\to \infty$, $\Pi (\|\btheta-\btheta^0\|>M_n \epsilon_n|\bY) \to 0$ in probability under $\btheta^0$. 

The posterior distribution $\Pi (\cdot|\bY)$ is said to be selection consistent if $\Pi (\{j:\boldsymbol{\theta}_j \ne 0\}= 
\{j:\boldsymbol{\theta}_{0j} \ne 0\}|\bY) \to 1$ in probability under $\btheta^0$. The posterior distribution $\Pi (\cdot|\bY)$ is said to be sign-consistent if $\Pi (\{\mathrm{sign}(\boldsymbol{\theta}) = \mathrm{sign}(\btheta^0)\}|\bY) \to 1$ in probability under $\btheta^0$, where $\mathrm{sign}(x)=1, 0, -1$ respectively for $x>0$, $x=0$ and $x<0$, and the function is applied coordinate-wise for a vector. Sign consistency is stronger than selection consistency because it correctly obtains the set of active predictors and captures the signs of the non-zero regression coefficients.
\end{definition}

\subsubsection{Estimation and Prediction Consistency}

Define the prediction risk of $\boldsymbol{\theta}^*$ to be $n^{-1} \| \bX (\boldsymbol{\theta}^* - \btheta^0)\|^2$ and measure the closeness between the sparse projection $\boldsymbol{\theta}^*$ and the true coefficients by the $\ell_1$-error of the estimated coefficients  $\| \boldsymbol \theta^* - \btheta^0\|_1 = \sum_{j = 1}^{p} |\boldsymbol \theta^*_j - \btheta^0_{j}|$. Theorem~\ref{consis_thm} below demonstrates that under a compatibility condition used for the LASSO, the posterior probability of the sparse projection $\btheta^*$ will concentrate around the true value $\btheta^0$, and bound the prediction rate under the following assumptions.

\begin{assumption}[Bounded design condition]
\label{design}
 The predictor variables are bounded, that is, there exists $M_1 > 0$ such that $|x_{ij}| \leq M_1$ for all $i = 1,\dots,n$, $j= 1,\dots,p$.
\end{assumption}

The condition is automatically met if we standardize the design points.

\begin{assumption}[Compatibility condition]
\label{compat}
For $S\subset \{1,2,\ldots,p\}$, the set of predictors $\bX$ is said to satisfy the compatibility condition at set $S$ if $$\|\bm{v}_S\|_1^2 \leq \frac{s_0}{\phi_0^2} \frac{1}{n} \|\bX \bm{v}\|^2 \mbox{ for all } \bm{v} \in \R^p, $$ $$\mbox{ such that } \|\bm{v}_{S^c}\|_1 \leq L \|\bm{v}_S\|_1 $$ 
for some constant $L>0$ and compatibility constant $\phi_0>0$ depending on $S$ and $L$. 
\end{assumption}

This assumption essentially means that the genuinely active predictors cannot be correlated too much. The condition is weaker than the restricted isometry condition \citep{candes2007dantzig}. Like in the theory for the LASSO, we apply the condition with $L = 3$. In the Bayesian context for a spike-and-slab prior, \cite{castillo2015bayesian} established the same rates under the condition with $L = 7$.

\begin{assumption}[Bounded true mean condition]
\label{mean_assum} 
The maximum of the absolute value of the expected response under the truth is bounded, that is, $\displaystyle\max_{1\le i\le n} |\E_{\btheta^0}(Y_i)| = \mathcal{O}(1)$.
\end{assumption}

It is reasonable to assume that the absolute expected value of the response under the truth is bounded for all sample points.

\begin{assumption}[Non-collinearity condition]
\label{sd_assum}
When $p>n$, $\mathrm{rank}(\bX) = n$ and the singular values $d_1, \dots, d_n$ of $\bX$ satisfy 
$\min\{ d_j^2: j=1,\ldots,n\} \gg n a_n$. 
\end{assumption}

This assumption is made to tackle the $\bm{H}(a_n)$ matrix and ensures it does not behave very differently from the actual hat matrix.  

Spike-and-slab or continuous shrinkage posterior may not need all the conditions our method needs for their convergence properties. This is analogous to the conditions needed for an $\ell_0$-penalized estimation versus its convex relaxation with the $\ell_1$-penalty giving rise to the LASSO. Spike-and-slab or its variants can be considered Bayesian analogs of $\ell_0$-penalization methods, requiring exploring all possible models. There is a trade-off between exponential computing complexity versus a few additional restrictions. The assumptions required for obtaining similar results by the continuous shrinkage prior are more or less similar to our method \citep{song2017nearly}. On the other hand, the modified bootstrap LASSO method, which is consistent and can provide uncertainty quantification for the LASSO estimator assumes fixed $p$. Moreover, the sparse projection-posterior can be justified by frequentist yardsticks, while most other Bayesian alternatives do not have similar justifications yet. 

\begin{theorem}[Estimation and prediction rate]
\label{consis_thm}
\label{pred_thm}
For $\lambda_n \asymp  \sqrt{(\log p)/n}$, under Assumptions \ref{design}, \ref{compat}, \ref{mean_assum}, \ref{sd_assum} for every sequence $M_{n} \to \infty$,  we have 
$$\Pi \big(\| \boldsymbol \theta^* - \btheta^0\|_1 \geq M_{n} s_0 \lambda_n \big| \bY \big) \to 0
, $$ $$ \Pi \big( n^{-1} \|\bX \big( \btheta^* - \btheta^0\big)\|^2 \geq M_{n} s_0 \lambda_n^2 \big| \bY \big) \to 0$$
in probability under the true distribution.
\end{theorem}

\Cref{consis_thm} shows that when the tuning parameter used in the sparsity-inducing map is of the order $\sqrt{(\log p)/n}$, the posterior probabilities that the $\ell_1$-norm of the difference between the true parameter value and the sparse projections of the posterior samples will remain bounded above by $s_0 \lambda_n$ and that the Euclidean distance between the expected response under the true distribution and the corresponding predictor using the sparse projection-posterior is bounded by $s_0 \lambda_n^2$ converge in probability to $1$ as the sample size grows indefinitely. This, in particular, implies posterior and prediction consistencies provided that $(\log p)/n\to 0$. The last condition restricts the growth of $p$ relative to $n$, but only to nearly exponential in $n$, which will be assumed throughout the paper. For the spike-and-slab type priors, \cite{castillo2015bayesian} derived comparable rates, but additionally assuming a revised compatibility condition (see Definition 2.2 and Definition 2.3 in \cite{castillo2015bayesian}) under a setup they called the `sparse lambda regime' for the choice of the prior.

\subsubsection{Variable Selection Consistency}

We show that the sparse projection-posterior is sign-consistent under a slightly stronger growth restriction on $p$ relative to $n$, that is, $p = \mathcal{O}(e^{n^{b_3}})$ for some $b_3<1$. The strong irrepresentable condition in Assumption~\ref{irrep} below helps avoid scenarios where zeros are matched, but opposite signs are used to estimate a model and allow recovery of support and signed signal magnitude. Given that the magnitude of the active predictors is bounded away from $0$ by the threshold in Assumption~\ref{betamin} below, Theorem~\ref{th:thm2} below establishes that having observed the data, the sparse projection-posterior method consistently selects the true model and additionally matches the signs of the non-zero coefficients under the following assumptions.

\begin{assumption}[Non-singularity condition]
\label{grow-p}
 The eigenvalues of the design matrix corresponding to the relevant covariates are bounded from below, that is, there exists a constant $M_2 > 0$ such that $\bm \alpha^{\mathrm{T}} (n^{-1} \bX^{\mathrm{T}}_{(1)} \bX_{(1)})\bm{\alpha} \geq M_2$  for all $\|\bm{\alpha} \| = 1$. 
 \end{assumption}
 
 The assumption controls the behavior of the inverse of $\bC_{n(11)}$. The correlation between variables can be controlled by restricting the eigenvalues of subsets of the design matrix to fall within a certain controlled interval. 

\begin{assumption}[Beta-min condition]
\label{betamin}
 There exists $0 \leq b_1 < b_2 < 1$ and $M_3 > 0$ such that $s_0 = \mathcal{O}(n^{b_1})$ and $ \enskip n^{(1 - b_2)/2} \min_{1\le j \le s_0} |\theta_j| \geq M_3.$
\end{assumption}

 This is a common assumption that bounds the number of active components, making sure that $s_0 < n$ holds (with much to spare), and additionally considers that the signal strengths of all the variables in the active set $S_0$ are larger than a specified value. 

\begin{assumption}[Strong Irrepresentable Condition]
\label{irrep}
 There exists a vector of positive constants $\boldsymbol{\nu} = (\nu_1, \dots, \nu_{p - s_0})^{\mathrm{T}}$ such that $\displaystyle \max_{1 \leq j \leq p - s_0} \big| \big( \bC_{n(21)}\bC^{-1}_{n(11)} \mathrm{sign} (\btheta^0_{(1)}) \big)_j \big| \leq 1 - \nu_j$.  
 \end{assumption}
 
  A weaker version of this assumption simply replaces the positive fractional constant $\nu$ by $1$, that is, 
 $$\displaystyle\max_{1 \leq j \leq p - s_0} \big| \big( \bC_{n(21)}\bC^{-1}_{n(11)} \text{sign} (\btheta^0_{(1)}) \big)_j \big| < 1.$$ 
 This assumption ensures that the total dependence of the noise variables on the signals cannot be larger than the irrepresentability constant (or $1$ in the weak case). It restricts the extent to which the important predictors in the model can represent the irrelevant variables to facilitate proper distinction between noise and signal.

\begin{theorem}
\label{th:thm2}
Let $\lambda_n \propto n^{(b_4-1)/2}$ and $p = \mathcal{O}(e^{n^{b_3}})$, where $0 \leq b_3 < b_4 < b_2 - b_1$. Then, under Assumptions \ref{design}, \ref{grow-p}, \ref{betamin}, \ref{irrep}, we have that  
$\Pi\big(\{\mathrm{sign}(\boldsymbol{\theta}) = \mathrm{sign}(\btheta^0)\}| \bY \big) \to 1$ in probability under the true distribution. 
\end{theorem}

 It will be seen that $\Pi \big(\{\mathrm{sign}(\boldsymbol{\theta}) = \mathrm{sign}(\btheta^0)\}| \bY \big) \geq \Pi ( A_n \cap B_n | \bY )$, where 
 \begin{align}
     A_n &= \big\{ \big| \bC^{-1}_{n(11)} \bZ_{n(1)} \big| \leq \sqrt{n} \big( |\btheta^0_{(1)}| - \big| \frac{\lambda_n}{2n} \bC^{-1}_{n(11)} \mathrm{sign}(\btheta^0_{(1)}) \big| \big) \big\} \label{define A_n},\\
     B_n & = \big\{ \big| \bC_{n(21)}\bC^{-1}_{n(11)} \bZ_{n(1)} - \bZ_{n(2)} \big| \leq \frac{\lambda_n}{2 \sqrt{n}} \boldsymbol\nu \big\}. 
      \label{define B_n}
\end{align}
 The event $A_n$ implies recovery of the signs of the true signals $\btheta^0_{(1)}$, and given $A_n$, the event $B_n$ implies shrinking the coefficients of the irrelevant covariates to $0$, that is, $\boldsymbol{\theta}^*_{(2)} = 0$. The sizes of these two events are traded off by the tuning parameter $\lambda_n$; the lesser its value, the larger is $A_n$, and the smaller is $B_n$ and vice versa. Bounding the $\Pi(A_n^c|\bY)$ and $\Pi(B_n^c|\bY)$ appropriately, the theorem can be established. A point to note here is that the penalty required to achieve this result is much stronger than the one needed for estimation or prediction consistency. 
 
\subsection{Uncertainty Quantification}
\subsubsection{Component-wise credible regions for $p \gg n$}
To compute credible bands in a high-dimensional setting, we adopt a Bayesian version of the bias-correction technique introduced by \cite{zhang2014confidence} for the LASSO. This correction compensates for the LASSO's bias, albeit at the expense of sparsity. We apply a different immersion map, which includes this Bayesian bias-correction map, resulting in a non-sparse posterior distribution.

Let the LASSO-residuals $\bR^{(j)}$ be obtained by regressing the $j$th column $\bX^{(j)}$ of the design $\bX$ on the remaining columns, denoted by $\bX^{(-j)}$, using the LASSO regularization. With $\lambda^{\bX}_j$ standing for the tuning parameter, the LASSO of $X_j$ on  $\bX^{(-j)}$ and the resulting residuals are respectively given by $\hat{\bm \gamma}_{(j)} = \argmin_{\bm{\gamma} \in \R^{(p-1)}} \| \bX^{(j)} - \bX^{(-j)} \bm\gamma\|^2_2/n + \lambda^{\bX}_j \|\bm{\gamma}\|_1, $ and $\bR^{(j)} = \bX^{(j)} - \bX^{(-j)} \hat{\bm \gamma}^{(j)}. $
The tuning parameter $\lambda^{\bX}_j$  may be chosen the same as the original LASSO tuning parameter $\lambda_n$. 
Unlike the least square residuals in low-dimension, the LASSO residuals are not orthogonal to $\bX^{(-j)}$. In the present  situation, \cite{zhang2014confidence} defined the $j$th component of the debiased estimator by $\hat{\theta}_j^{\mathrm{DB}}={\bY^{\mathrm{T}} \bR^{(j)}}/{{\bX^{(j)}}^{\mathrm{T}} \bR^{(j)}}
=\hat{\theta}_j^{\mathrm{L}}+({{{\bX^{(j)}}^{\mathrm{T}}} \bR^{(j)}})^{-1}{(\bY-\bX \hat{\bm \theta}^{\mathrm{L}})^{\mathrm{T}} \bR^{(j)}},$
where $\hat{\bm\theta}^{\mathrm{L}}$ is the LASSO estimator for $\bm\theta$. In high dimension ($p \gg n$), most subsets of $n$ predictors will likely be linearly independent, and hence  $\mathrm{rank}(\bX) = n$.
If rank$(\bX) = n$, $s_0 = o({\sqrt{n}}/{\log p})$, Assumptions~\ref{design}, \ref{compat}, \ref{mean_assum}, \ref{sd_assum} hold and the tuning parameters $\lambda_n, \lambda_n^{\bX}$ are both of the order of $\sqrt{(\log p) /n}$, then their estimator has an asymptotic normal distribution \citep{zhang2014confidence}: 
\begin{align}
\label{debiased_LASSO_CLT}
    \hat{\sigma}^{-1} \frac{|{\bX^{(j)}}^{\mathrm{T}} \bR^{(j)}|}{\|\bR^{(j)}\|} (\hat{\theta}_j^{\mathrm{DB}} - \theta^0_{j}) \rightsquigarrow \normal(0,1),
\end{align}
for an estimator $\hat{\sigma}$ of $\sigma$, and consequently, $$\lim_{n \rightarrow \infty }\prob_{\btheta^0}\left(|\hat{\theta}_j^\mathrm{DB} - \theta^0_j | < \Phi^{-1} \left(1 - \alpha/2 \right)\frac{\hat{\sigma} \|\bR^{(j)}\|}{|{\bX^{(j)}}^{\mathrm{T}} \bR^{(j)}|} \right) = 1 - \alpha$$. 

For uncertainty quantification, the form of the debiased LASSO estimator $\hat{\theta}_j^{\mathrm{DB}}$  suggests using the immersion map $\iota^{\mathrm{DB}}:\bm{\theta}\mapsto \bm{\theta}^{{**}}$ whose $j$th component $\theta^{**}_j$ is given by 
\begin{align}
\label{debiased immersion}
    \theta^*_j + \frac{{\bR^{(j)}}^{\mathrm{T}}(\bX \btheta - \bX \boldsymbol{\theta}^*)}{{\bR^{(j)}}^{\mathrm{T}} \bX^{(j)}}
    = \frac{(\bX \btheta)^{\mathrm{T}} \bR^{(j)}}{{\bX^{(j)}}^{\mathrm{T}} \bR^{(j)}} - \sum_{k \neq j} P_{jk} \theta^*_k, 
\end{align}
where $P_{jk} = {{\bX^{(k)}}^{\mathrm{T}} \bR^{(j)}}/{{\bX^{(j)}}^{\mathrm{T}} \bR^{(j)}}$, instead of the sparse-projection map used for estimation and variable selection. We shall call $\iota^{\mathrm{DB}}$ the debiased sparse projection map and the corresponding induced distribution the debiased sparse projection-posterior distribution. In the following results, we show a type of the Bernstein-von Mises theorem that the debiased sparse projection-posterior distribution of $\sqrt{n}(\theta_j^{**}-\theta_j^0)$ can be approximated by a normal distribution with the mean given by the normalized debiased LASSO estimator of \cite{zhang2014confidence} and the variance asymptotically equivalent to the asymptotic variance of the debiased LASSO estimator.  Consequently, the asymptotic frequentist coverage of a credible ball for $\theta_{j}^{**}$ will agree with the corresponding credibility for all $j \in \{1,2,\dots,p\}$. The property extends immediately to any fixed-dimensional linear function of $\bm{\theta}$.

\begin{theorem}[Bernstein-von Mises Theorem]
\label{th:thm3}
If Assumptions \ref{design}, \ref{compat}, \ref{mean_assum}, \ref{sd_assum} hold and $s_0 = o({\sqrt{n}}/{\log p})$ and $\mathrm{rank}(\bX) = n$, then 
$$\max_{1 \leq j \leq p} \sup_B \Big| \Pi \big({\sqrt{n}  (\theta^{**}_j - \theta^0_{j})}\in B |\bY \big)  - \normal_p \big(B; m_j, \Sigma_{jj} \big) \Big| \to 0,$$ 
where 
\begin{align}
    m_j &= \sqrt{n} \big( \frac{{\bR^{(j)}}^{\mathrm{T}}\bX \hat{\boldsymbol\theta}^\mathrm{R}}{{\bR^{(j)}}^{\mathrm{T}}\bX^{(j)}} - \sum_{k = 1}^{p} P_{jk}\theta^0_{k} \big), 
    \label{DBLasso lim mean} \\  
    \Sigma_{jj} & = n \sigma_0^2 \big( \frac{{\bR^{(j)}}^{\mathrm{T}} \bm{H}(a_n) \bR^{(j)}}{|{\bX^{(j)}}^{\mathrm{T}} \bR^{(j)}|^2} \big).
    \label{DBLasso lim variance}
\end{align}
\end{theorem}

Define $C_j=[\tilde{\theta}_j-q_{j,\alpha},\tilde{\theta}_j+q_{j,\alpha}] $ to be the symmetric $(1-\alpha)$-credible interval for $\theta_j$, $j=1,\ldots,p$, where $\tilde{\theta}_j$ is the median of the posterior distribution of $\theta_j^{**}$ and $q_{j,\alpha}$ is the $(1-\alpha/2)$-quantile of the posterior distribution of $|\theta_j^{**}-\tilde{\theta}_j|$. The following result shows that these can be regarded as approximate $(1-\alpha)$-confidence intervals.

\begin{corollary}
\label{coverage} (Coverage of credible interval for $\theta_j$) Under Assumptions \ref{design}, \ref{compat}, \ref{mean_assum}, \ref{sd_assum},
   $$ \prob_{\theta^0} ( \theta^0_j \in C_j) \to  1 - \alpha \text{ uniformly for all } j = 1,2,\dots,p.$$
\end{corollary}

From the proof, it will also follow that the conclusion holds for the equal-tailed $(1-\alpha)$-credible interval $C_j=[l_j,u_j]$ for $\theta_j$, $j=1,\ldots,p$, where $\Pi (\theta_j^{**}< l_j)=\alpha/2$ and $\Pi (\theta_j^{**}> u_j)=\alpha/2$.

\subsubsection{Credible ellipsoid}\label{ellipsoid}

Joint credible sets for the relevant predictors in the model automatically address any interrelationships between the coefficients of interest. When individual component-wise credible intervals are provided, these are ignored. We construct a sparse, credible ellipsoid for $\btheta$ in this subsection. 

Let the model formed by the true active variables be denoted by $F_n = \{ \btheta^*: \mathrm{sign}(\boldsymbol{\theta}^*) = \mathrm{sign}(\btheta^0) \}$. We know from Theorem~\ref{th:thm2} that the true model is selected with probability tending to one when the sample size goes to infinity, that is, $\Pi (\btheta^*\in  F_n | \bY) \to 1$ in probability under the true distribution. Thus, theoretically, we can identify the set $F_n$ from posterior sampling with large sample sizes. Conditioned on $\btheta^* \in F_n$, the probability of any event, say $A$, satisfies 
$$\prob(A|F_n) = \frac{\prob(A \cap F_n)}{\prob(F_n)} \rightarrow \prob(A).$$ Using this idea, we can now concentrate only on the coverage properties of the covariates selected by the projection method with high probability (tending to 1). Without loss of generality, the set of selected active predictors is $\{1,\ldots,s_0\}$ and so $\bX_{(1)} \in \R^{n \times s_0}$ is the matrix containing only the important predictors for all $n$ samples. 

Let $\hat S$ stand for the highest maximum sparse projection posterior probability model. 
Restricted to the selected model $\hat S$, let $\btheta_{\hat S}^{\mathrm{PS}}$ stand for the vector of components of $\btheta$ in $\hat S$ only. Then the conditional posterior distribution of $\btheta_{\hat S}^{\mathrm{PS}}$ given $\sigma$ based on only the selected predictors $\hat S$ is $\normal_{\hat s}(\hat\btheta_{\hat S}^{\mathrm{R,PS}}, {\sigma}^2(\bX_{\hat S}^\mathrm{T} \bX_{\hat S} + a_n \bm{I}_{\hat s})^{-1})$, where 
$\hat\btheta_{\hat S}^{\mathrm{R,PS}}=
(\bX_{\hat S}^\mathrm{T} \bX_{\hat S} + a_n \bm{I}_{\hat s})^{-1} \bX_{\hat S}^\mathrm{T}\bY$, the post-selection ridge regression estimator based on the selected predictors $\hat S$ and $\hat s$ is the cardinality of $\hat S$. For a given credibility level $1-\alpha$, define the post-selection credible ellipsoid $D_{n,1-\alpha}^{\mathrm{PS}}$ to be 
$$ \bigg\{\begin{pmatrix}\btheta_{\hat S} \\ \bm{0}_{\hat S^c}\end{pmatrix} : (\btheta_{\hat S} - \hat{\btheta}^\mathrm{R,PS}_{\hat S})^{\mathrm{T}}(\bX_{\hat S}^\mathrm{T} \bX_{\hat S} +  a_n \bm{I}_{\hat S}) (\btheta_{\hat S} - \hat{\btheta}^\mathrm{R}_{\hat S}) \leq r_{\alpha} {\sigma}^2\bigg\},$$ 
where 
the size $r_{\alpha}$ is chosen such that $\Pi(\btheta \in D_{n,1-\alpha}^{\mathrm{PS}}|\bY) = 1-\alpha$. We claim that $D_{n,1-\alpha}^{\mathrm{PS}}$ also serves as an asymptotic $(1-\alpha)$-confidence region. Informally, this holds because by Theorem~\ref{th:thm2}, $\hat S=S_0$ with probability tending to one, provided the conditions of the theorem are satisfied so that the setup effectively reduces to that of the Bernstein-von Mises theorem with a fixed set of parameters. More elaborately, the posterior is essentially the one based on the correct set of predictors $S_0$ with mean the ridge regression estimator based on predictors in $S_0$, and dispersion matrix approximately $\sigma_0^2 n^{-1} \bC_{(11)}$. 
By \Cref{ridge CLT}, the ridge regression estimator is asymptotically normal centered at $\btheta_{S_0}^0$ and dispersion matrix $\sigma_0^2 n^{-1} \bC_{(11)}$. The approximate Bayes-frequentist agreement leads to the correct asymptotic coverage. 
The discussion is formalized in the next theorem.

\begin{theorem}\label{thm:coverage ellipsoid}
    Let $S_0$ be fixed, and Assumptions \ref{design}, \ref{sd_assum}, \ref{grow-p}, \ref{betamin} and \ref{irrep} hold. Let 
    $r_\alpha$ stand for the posterior $(1-\alpha)$-quantile of $ (\btheta_{\hat S}^{\mathrm{PS}} - \hat{\btheta}^\mathrm{R,PS}_{\hat S})^{\mathrm{T}} (\bX_{\hat S}^{\mathrm{T}} \bX_{\hat S}+a_n \bm{I}_{\hat s}) (\btheta_{\hat S}^{\mathrm{PS}} - \hat{\btheta}^\mathrm{R,PS}_{\hat S})$. Then 
 the coverage 
    $\prob_{\btheta^0}(\btheta^0 \in D_n)\to 1-\alpha$, 
    where $D_{n}$ is 
    $$\{\btheta_{\hat S}: (\btheta_{\hat S} - \hat{\btheta}^\mathrm{R,PS}_{\hat S})^{\mathrm{T}} (\bX_{\hat S}^{\mathrm{T}} \bX_{\hat S}+a_n \bm{I}_{\hat s}) (\btheta_{\hat S} - \hat{\btheta}^\mathrm{R,PS}_{\hat S}) \leq r_{\alpha}\}\times \{\bm{0}_{\hat S^c}\}.$$  
\end{theorem}


Oftentimes, $\Pi(F_n|\bY)$ may not be close to $1$ unless the sample size is impractically large since the number of possible models is exceptionally high. However, the only requirement is that the correct model $F_n$ has the highest posterior probability. This happens more commonly --- $\Pi(F_n|\bY)>1/2$ gives a sufficient condition.

\subsection{Tuning}

As seen above, recovery and prediction accuracy both require the tuning parameter $\lambda_n \asymp \sqrt{{(\log{p})}/{n}}$, whereas correct model selection is guaranteed when $\lambda_n \propto n^{{(b_4-1)}/{2}}$ with some $b_3 \leq b_4 < b_2 - b_1$. As discussed earlier, different $\lambda_n$ can be chosen for different purposes, giving different immersion posterior for different inferential problems. Typically, slightly larger values of $\lambda_n$ will give accurate model selection. The asymptotic theory guides the choice, but a data-driven choice of the tuning parameter is desirable from practical considerations. One way to regularize the model for optimal estimation and prediction could be to fit the model using the lasso with a wide range of values of $\lambda_n$ and choose that which gives the lowest prediction error. This is justified because the requirement on the tuning parameter $\lambda_n$ for prediction consistency is the same as the LASSO. Since accurate model selection requires a larger penalty, depending on how much larger the prediction error could be allowed, we may keep increasing the value of the tuning parameter until an anticipated level of sparsity is achieved. This should induce greater sparsity in the posterior distribution while keeping the prediction error within a tolerable limit.  

\subsection{Error variance}
\label{sigma}

As the sparse projection-posterior approach uses conditional draws given $\sigma^2$, the posterior distribution of $\sigma$ must be consistent to ensure the asymptotic frequentist properties of the sparse projection-posterior. However, unless $p$ is of a smaller order of $n$, the marginal vanilla posterior for $\sigma^2$ obtained from the regression model using a standard prior is inconsistent. 
Putting the non-informative prior Gamma(0,0) on the precision parameter $\tau = \sigma^{-2}$ so that the prior density is  $p(\tau) = 1/\tau$, we obtain the Gamma$({n}/{2}, {n \hat{\sigma}_n^2}/{2})$ posterior for $\tau$, where $\hat{\sigma}_n^2 = n^{-1} \bY^{\mathrm{T}} ( \bm{I}_n - \bm{H}(a_n) ) \bY$. Note that the posterior mean $\E(\tau|\bY) = {n}/({\bY^{\mathrm{T}} ( \bm{I}_n - \bm{H}(a_n) ) \bY})$ blows up as $n$ grows (cf., Lemma~\ref{variance estimate}). 
This is analogous to the inconsistency of standard estimators like the maximum likelihood estimator or the unbiased variance estimator. The technique of immersion posterior can be used to rectify the problem. This can be corrected by applying a data-dependent immersion map to $\tau$. For a data-dependent factor $\kappa$, consider the map $ \tau \mapsto \kappa \tau$. Its induced posterior distribution is thus given by $(\kappa \tau | \bY) \sim \text{Gamma}({n}/{2}, \kappa^{-1} {n \hat{\sigma}_n^2}/{2})$, which has expectation $\kappa/\hat{\sigma}_n^2$. If we choose $\kappa=\hat{\sigma}_n^2/\tilde{\sigma}^2$, where $\tilde{\sigma}^2$ is a consistent estimator of $\sigma^2$ such as in \cite{sun2012scaled}, then the posterior mean of the immersion posterior corresponding to the immersion map $\tau\mapsto \kappa \tau$ is a consistent point estimator for $\tau=\sigma^{-2}$. The posterior variance of $\kappa \tau$ is given by ${2\kappa^2}/({n\hat{\sigma}_n^4}) = {2}/({n \tilde{\sigma}^4})$, which goes to zero as $n \to \infty$ since $\tilde{\sigma}^2$ converges to the finite positive constant $\sigma_0^2$ by the consistency of $\tilde{\sigma}^2$ for $\sigma^2$. Thus, the immersion posterior for $\tau$ using the data-dependent immersion map $\tau\mapsto \kappa \tau$ is consistent by Chebyshev's inequality. 
As the correspondence between $\tau$ and $\sigma$ is smooth and one-to-one, the induced immersion posterior for $\sigma$ is also consistent. In other words, we use the immersion map $\sigma\mapsto \sigma^*=(\tilde{\sigma}/\hat{\sigma}_n)\sigma$ for $\sigma$, and the joint immersion map for $(\btheta,\sigma)$ is 
 given by $(\btheta, \sigma ) \mapsto ( \btheta^*, ({\tilde{\sigma}}/{\hat{\sigma}_n}) \sigma )$ in place of their original posterior distributions. To sample from the joint immersion posterior, we sample $\tau$ from $\mathrm{Gamma}(n/2, n \tilde{\sigma}^2/2)$, compute $\sigma=\tau^{-1/2}$, set $\sigma^*=(\tilde{\sigma}/\hat{\sigma}_n)\sigma$, draw $\btheta$ from $\normal_n \big(\hat{\boldsymbol\theta}^\mathrm{R} , {\sigma^*}^2 (\bX^\mathrm{T}\bX + a_n \bm{I}_p)^{-1}\big)$ and compute $\btheta^*$ from \eqref{sparse_proj} for the appropriate choice of $\lambda_n$ depending on the purpose such as estimation, variable selection or uncertainty quantification. Consequently, we can find a shrinking neighbourhood $\mathcal{U}_n$ of $\sigma_0$ such that $\Pi(\sigma^* \in \mathcal{U}_n|\bY) \to 1$ in probability as $n \to \infty$.

\section{Numerical Results}
\label{Simulation}

We consider the linear regression model with response $\bY \in \R^p$ and fixed design $\bX \in \R^{n \times p}$, and its rows are sampled from $\normal_p(\bm{0}, \boldsymbol\Sigma)$. For the Bayesian method, we use the median probability model (MPM) \citep{barbieri2004optimal},  which chooses those variables in the final model whose marginal posterior probability of being included is at least $0.5$ to select predictors.  We use an estimator with zeroes at all components not selected in more than 50\% MCMC iterations. The estimator takes the average of all the $R$ samples for the remaining components, where $R$ is the number of MCMC draws. A comparison between Bayesian and frequentist methods included in the study of estimation accuracy is based on the  
$\mathrm{MSE} = (\hat{\bm{\theta}} - \bm\theta^0 )^{\mathrm{T}} \bm{\Sigma} (\hat{\bm{\theta}} - \bm\theta^0).$ 
For a variable selection procedure, we define True Positives (TP) as the signals correctly identified to be non-null, False Positives (FP) as the noise variables falsely selected as active variables, True Negatives (TN) as the null variables correctly not selected by the model and finally False Negatives (FN) as the signals mistakenly left out by the model. To compare different variable selection methods, we report three quantities, namely, True Positive Rate (TPR) which is the ratio of correctly selected predictors to the total number of true signals ($s_0$), \text{TPR} = {\text{TP}}/({\text{TP + FN}}), False Discovery Proportion (FDP) which is the ratio of falsely selected predictors to the number of selected signals, \text{FDP} = {\text{FP}}/({\text{FP+TP}}) and Matthew's correlation coefficient (MCC):
\begin{align*} 
\text{MCC} = \frac{\text{TN}\times \text{TP} - \text{FN}\times \text{FP}}{\sqrt{(\text{TP+FP})(\text{TP+FN})(\text{TN+FP})(\text{TN+FN})}}.
\end{align*}  
The closer the MCC value is to $1$, the better its selection ability. For uncertainty quantification, we report the mean coverage and length of 95\% confidence/credible intervals for the $s_0$ active covariate coefficients and again for the $p - s_0$ noise, averaged over the number of replications ($M$), which is taken to be $100$ unless mentioned otherwise. For Bayesian methods, credible regions are constructed using sample quantile values from posterior draws. The estimation accuracy, model selection ability, and coverage 
performance of the projection-posterior are compared with those of the Bayesian LASSO (BLASSO), spike-and-slab LASSO (SSLASSO), posterior from the horseshoe prior, minimax-concave penalty (MCP) estimator, LASSO, debiased LASSO, and bootstrap LASSO. The last two are the only frequentist methods included in the study of confidence intervals. The competing Bayesian methods are computed based on $R = 10000$ MCMC runs to provide results comparable to frequentist methods. A burn-in of $2000$ samples is administered before collecting posterior draws. We introduce and use the \texttt{R} package named \texttt{sparseProj} \citep{sparseProj2024} to implement the sparse projection-posterior method.

\subsection{Simulation Study 1}\label{simu 1}
\subsubsection{Simulation Setup}

We consider three cases for the simulation study. In the first case, we compare the model selection and estimation performances of different Bayesian and frequentist procedures in a sparse linear regression setup, where we assume $p$ grows like $n^{2/3}$, e.g., $n = 1000$,  $p = 100$, and the number of active predictors $s_0 = 10$. The second case explores the scenario when $n$ and $p$  are similar, e.g., $n = p = 300$ and $s_0 = 10$. Lastly, in the third case, we allow $p$ to grow exponentially with $n$ like $e^{n^c}$ for some $0<c<1$. We choose $n = 100$ and $c = 0.45$ so that $p = 2000$ and set $s_0$ to $10$. Here, two scenarios are explored: one where the design $\bX$ consists of all independent covariates ($\boldsymbol\Sigma = \sigma^2 \bm{I}$, $\sigma^2 = 1$ where $\bm{I}$ is the $p \times p$ identity matrix), and the other considers an autoregressive correlation structure which assumes that the correlation between variables gradually decreases as the distance between their relative positions increases ($\bm{\Sigma}=(\!( \sigma_{ij})\!)$, $\sigma_{ij} = \rho^{|j-i|}$, where $\rho = 0.7$). The coefficient vector $\boldsymbol\theta$ is set up such that the first $s_0 = 10$ components correspond to moderately strong signals, that is, $\theta_j = 2 $ for  $j = 1,\dots,s_0$, and the remaining $p - s_0$ components are zeroes. The Bayesian LASSO has been skipped in the last two cases due to its exceptionally long run times.

\subsubsection{Tuning parameters}

The tuning parameters in LASSO and MCP are all computed by a 10-fold cross-validation. The cross-validated LASSO tuning value is used as an empirical Bayes estimate of the common parameter $\lambda$ in the prior distributions of the local shrinkage parameters. The target acceptance rate in the SSVS method is set to $0.345$. All other parameters in any model used are set to their defaults. For our method, the posterior draws of $\btheta$ from the multivariate Gaussian have been shrunk to $\btheta^*$ using the projection map with the regularization parameter same as the one chosen by CV in LASSO as the two methods share all theoretical properties.

\subsubsection{Simulation Results}

We investigated the time taken for one run of each method in this study for both small-$p$ and large-$p$ cases. In \Cref{tab:my-timetable} reported are the mean times (in seconds) to finish $R = 10000$ MCMC runs of each of the Bayesian methods along with one run of the debiased LASSO and the bootstrap LASSO with $5000$ replicates. The projection method and the bootstrap LASSO have similar computing costs. Computation time under $p = 2000$ for BLASSO is excessive (in hours) and have not been reported along with the other faster computing methods. Since we reuse the value obtained for the LASSO using cross-validation, and since most of the computational cost is used in tuning, the additional computation cost of our method over the LASSO is not substantial. For instance, running LASSO once with cross-validation for the tuning parameter for a regression problem with $p = 2000$ variables, $s_0 = 10$ active variables and $n = 100$ sample size is about 0.26 seconds. On the other hand, if we run LASSO repeatedly to project the posterior samples, we take about $\sim3.6$ seconds for 1000 posterior samples, $\sim5.2$ seconds for 2000 posterior samples, $\sim8$ seconds for 3000 posterior samples, $\sim10.5$ seconds for 4000 posterior samples and so on. It should be remembered that this time is in addition to the time for cross-validation as well as the independent sampling from the conjugate posterior.

\begin{table*}[t]
\centering
\resizebox{\textwidth}{!}{%
\begin{tabular}{cccccccc}
\hline
Model & Debiased LASSO & Bootstrap LASSO & Horseshoe & SSLASSO & SSVS & Bayesian LASSO & Projection \\ \hline
{$n = 1000, p = 100$} & \begin{tabular}[c]{@{}c@{}}20.5 (2.39)\end{tabular} & \begin{tabular}[c]{@{}c@{}}17.2 (2.73)\end{tabular} & \begin{tabular}[c]{@{}c@{}}25.7 (4.01)\end{tabular} & \begin{tabular}[c]{@{}c@{}}\textless 1 (0.00)\end{tabular} & \begin{tabular}[c]{@{}c@{}}20.8 (1.61)\end{tabular} & \begin{tabular}[c]{@{}c@{}}227.4 (15.38)\end{tabular} & \begin{tabular}[c]{@{}c@{}}17.8 (2.14)\end{tabular} \\ \hline
{$n = 100, p = 2000$} & \begin{tabular}[c]{@{}c@{}}339.533 (22.87)\end{tabular} & \begin{tabular}[c]{@{}c@{}}116.1 (11.03)\end{tabular} & \begin{tabular}[c]{@{}c@{}}234.7 (16.21)\end{tabular} & \begin{tabular}[c]{@{}c@{}}18.468 (3.59)\end{tabular} & 209.1 (16.90) & - & \begin{tabular}[c]{@{}c@{}}141.0 (12.09)\end{tabular} \\ \hline
\end{tabular}%
}
\caption{The mean (sd) time (in seconds) to complete one run of the competing methods, averaged over $M$ replications.}
\label{tab:my-timetable}
\end{table*}

\begin{figure}
    \centering
    \includegraphics[width=0.63\linewidth]{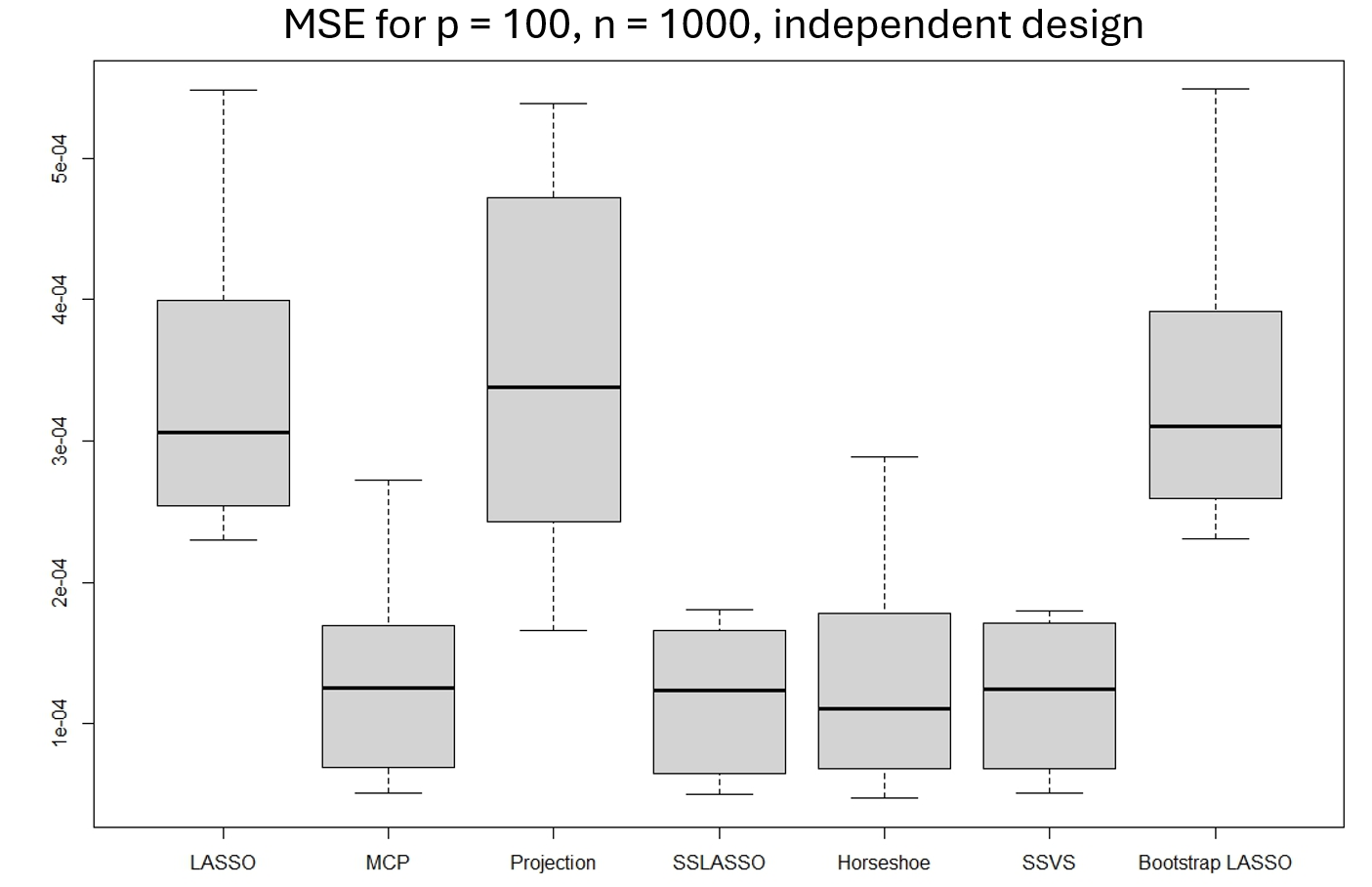}
    \includegraphics[width=0.63\linewidth]{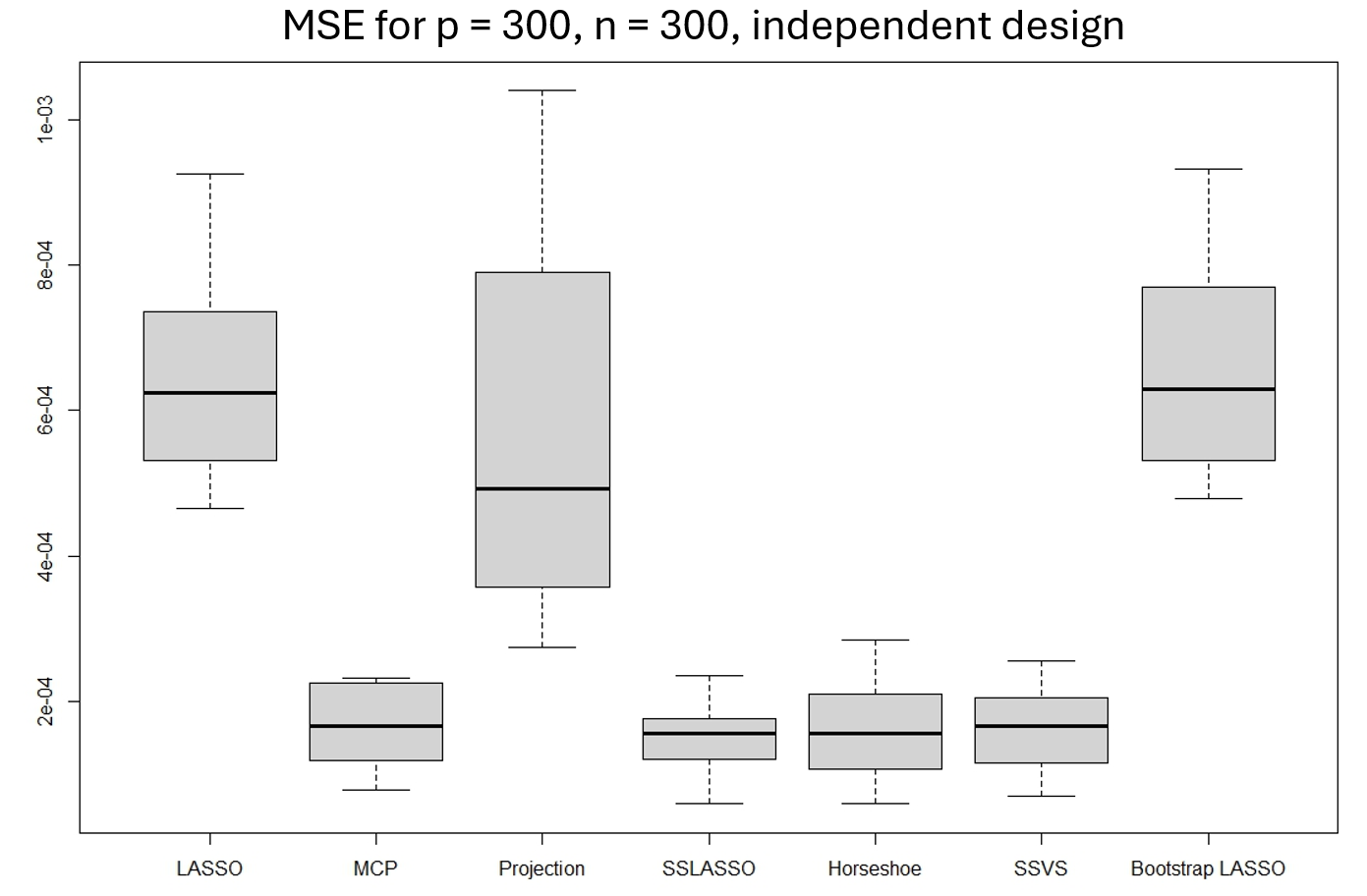}
    \includegraphics[width=0.63\linewidth]{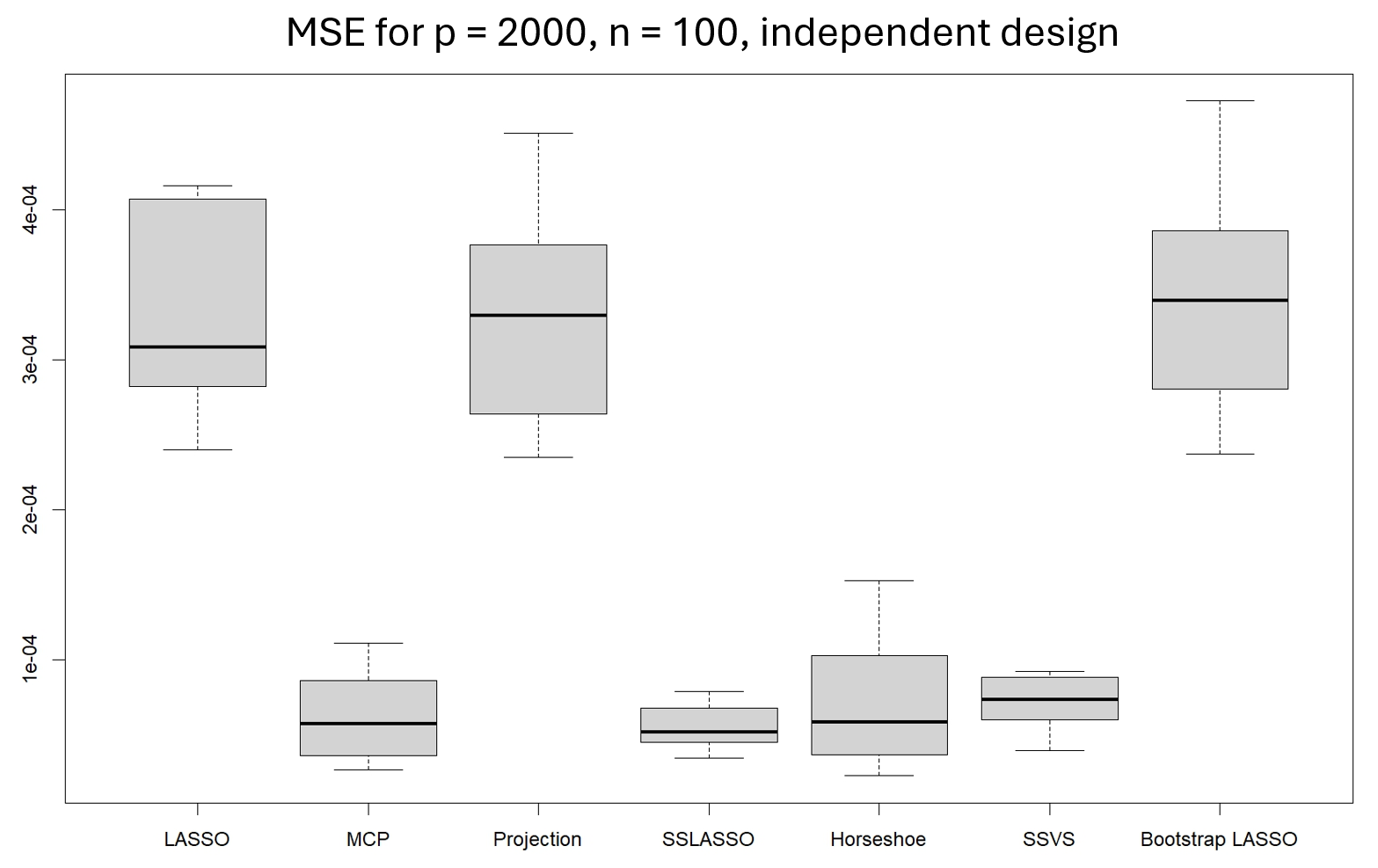}
    \caption{Boxplots of the MSEs for different methods averaged over $M = 100$ replications corresponding to the independent design case.}
    \label{fig:MSE_indep}
\end{figure}
\begin{figure}
    \centering
    \includegraphics[width=0.63\linewidth]{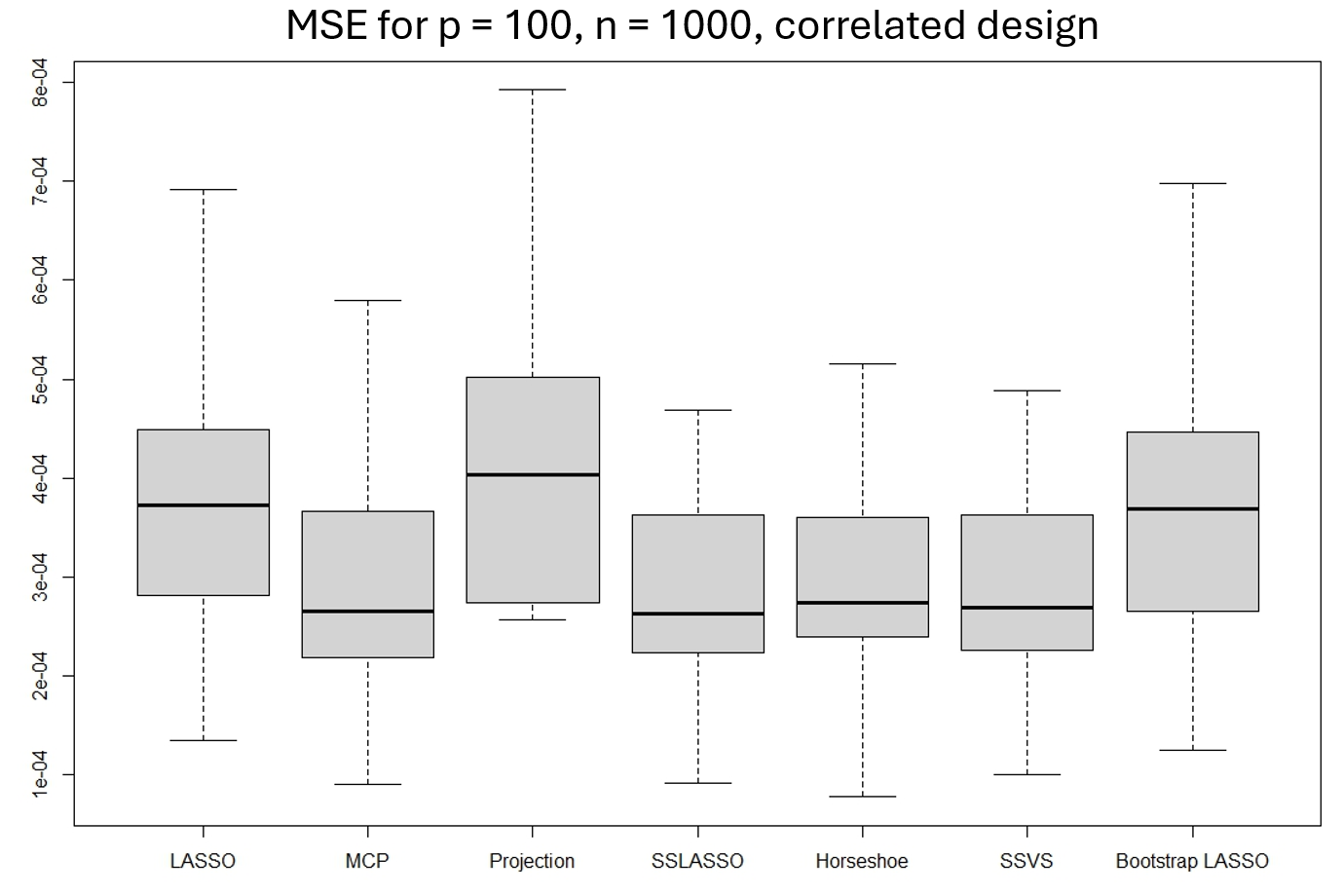}
    \includegraphics[width=0.63\linewidth]{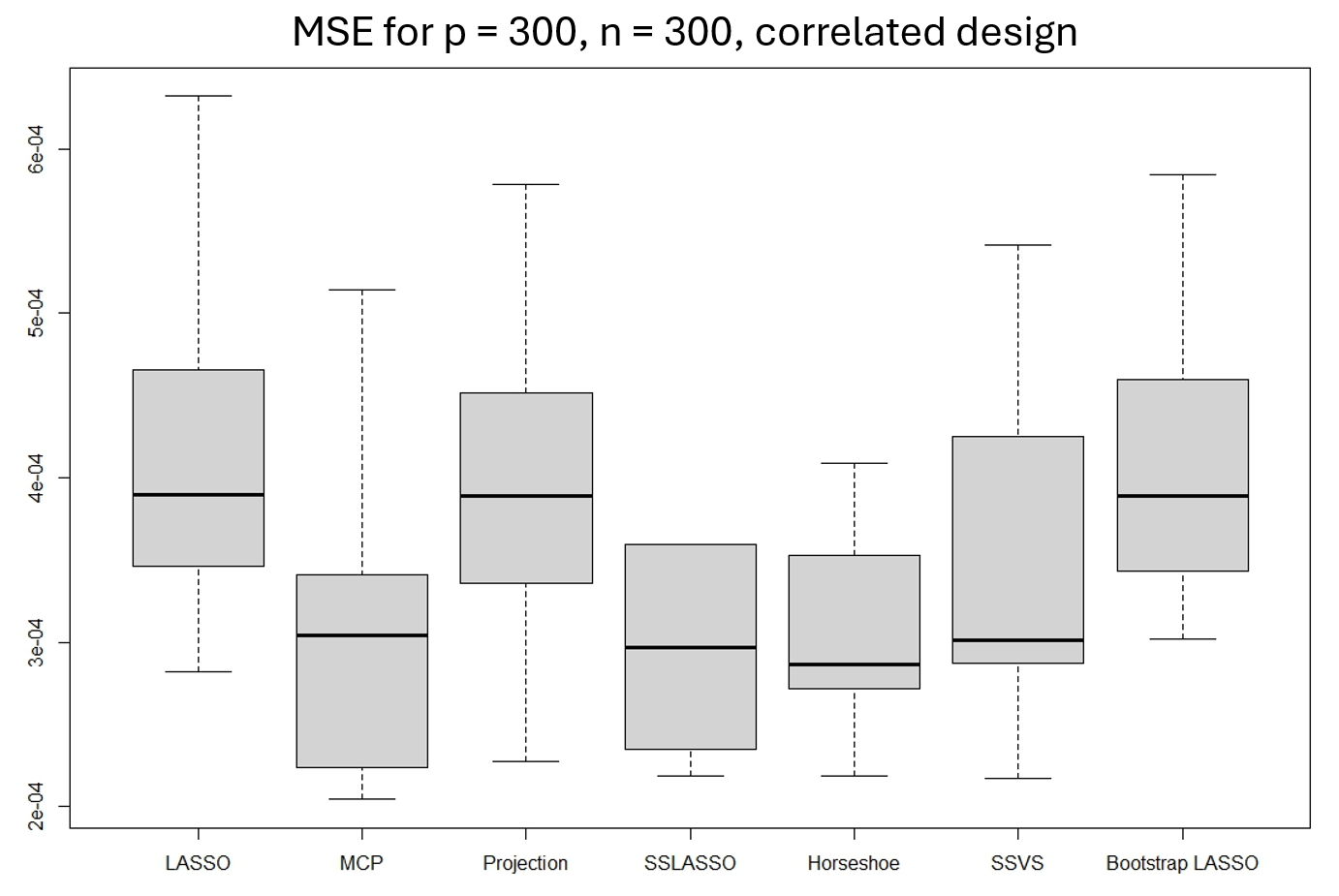}
    \includegraphics[width=0.63\linewidth]{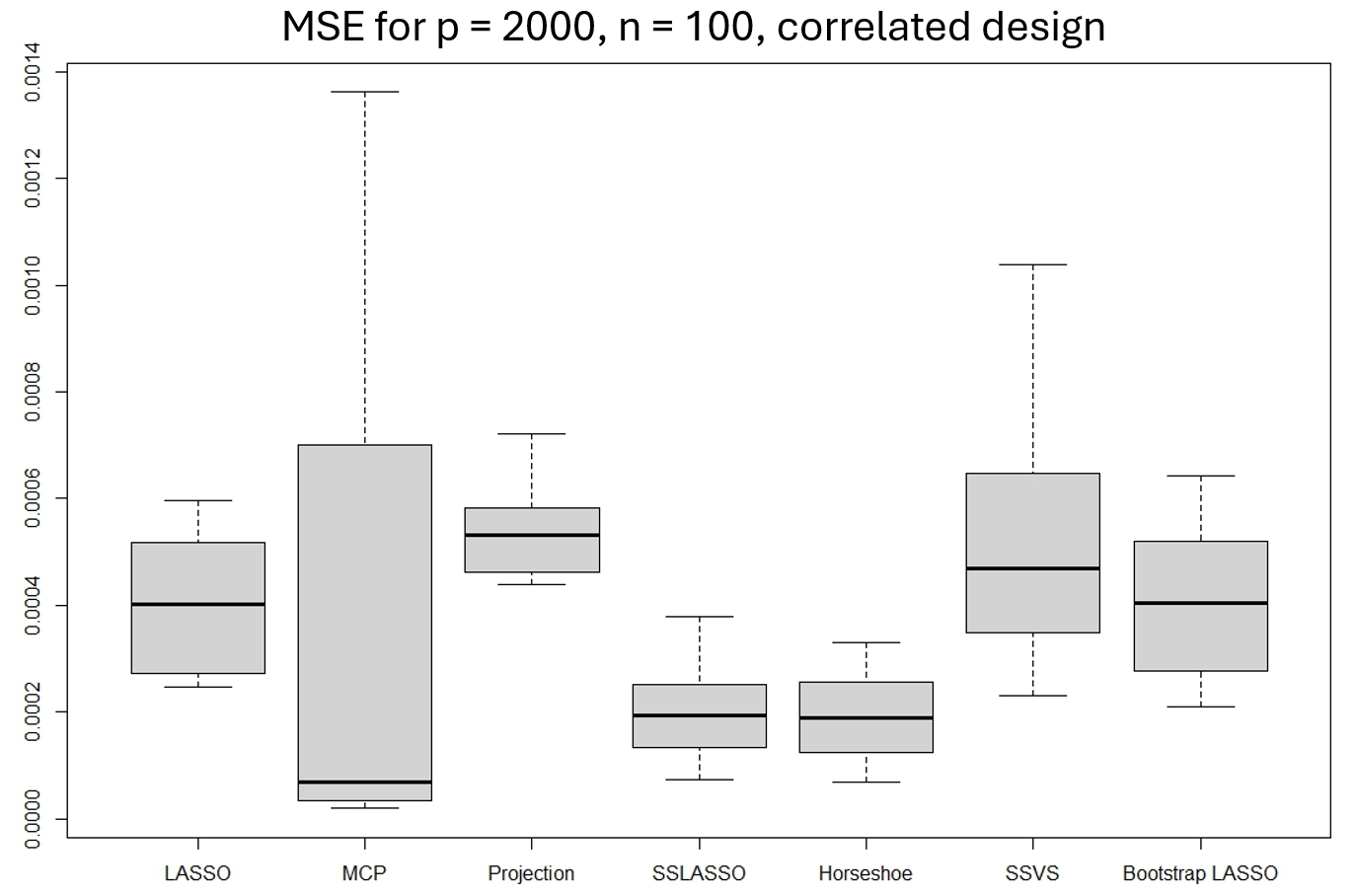}
    \caption{Boxplots of the MSEs for different methods averaged over $M = 100$ replications corresponding to the correlated design case.}
    \label{fig:MSE_corr}
\end{figure}
\begin{figure}
    \centering
    \includegraphics[width=0.63\linewidth]{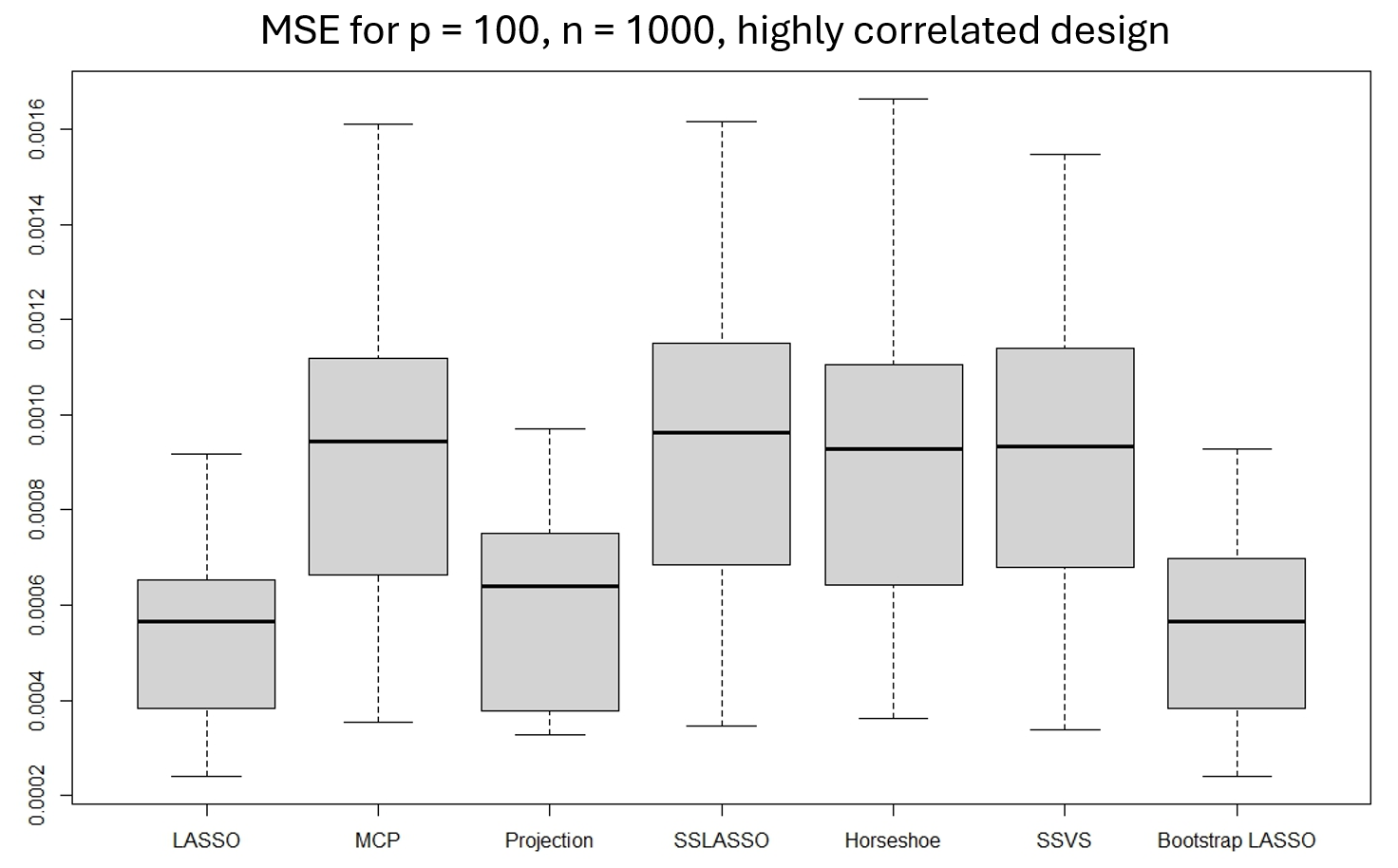}
    \includegraphics[width=0.63\linewidth]{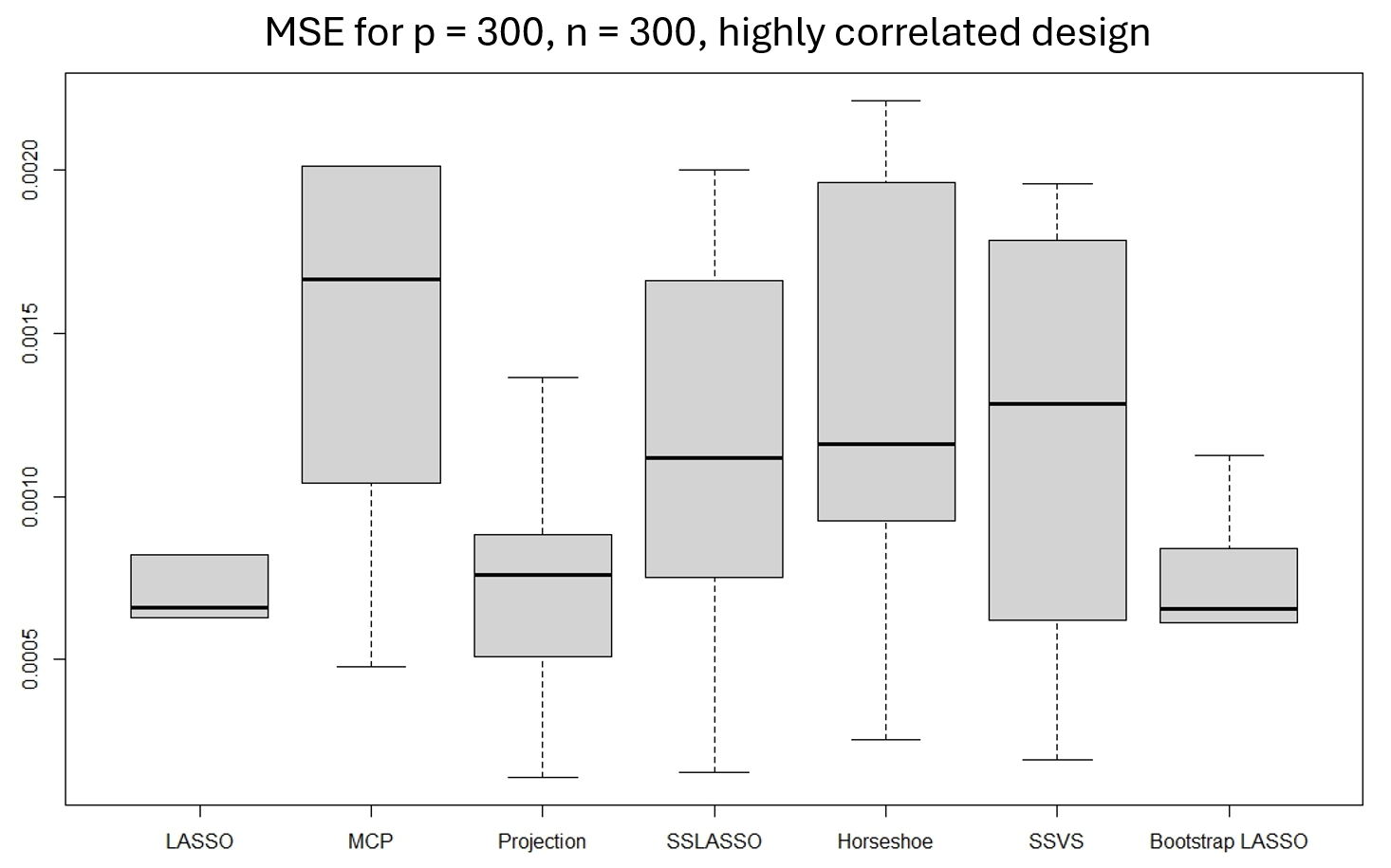}
    \includegraphics[width=0.63\linewidth]{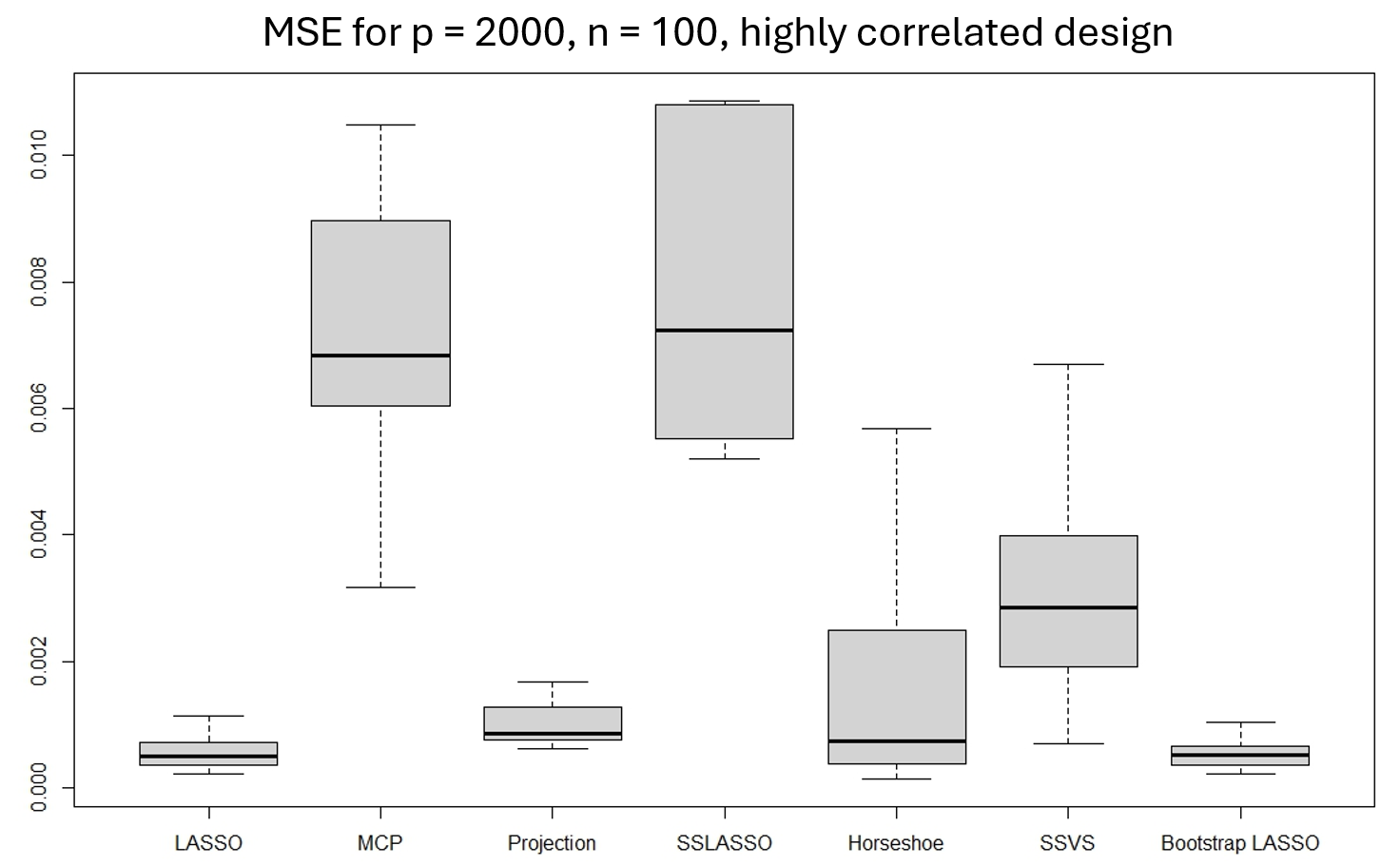}
    \caption{Boxplots of the MSEs for different methods averaged over $M = 100$ replications corresponding to the highly correlated design case.}
    \label{fig:MSE_high_corr}
\end{figure}

The estimation errors for each of the models measured in terms of MSE are plotted in \Cref{fig:MSE_indep} corresponding to the independent design case and in \Cref{fig:MSE_corr} corresponding to the correlated design case. Although the MSE of the sparse projection-posterior method is higher in the independent design regime, it does not exceed the LASSO or the bootstrap LASSO. This is expected since the sparse projection posterior heavily relies on the choice of penalization. However, one should also note that in the correlated design case, the errors for the other models increased, thereby diminishing the gap in MSE among all competing methods. This showcases that our method performs robustly even when the covariates are correlated, which is actually more realistic. To see if this trend is maintained when correlation among variables is further increased, we conducted the simulation with the AR correlation structure for the predictors, but now with an enhanced correlation parameter $\rho = 0.9$. Under such high correlation, the MSE of our method actually becomes lower than most of the competing methods. The boxplots capturing this phenomenon are given in \Cref{fig:MSE_high_corr}. Moreover, the proposed method performs robustly in terms of MSE when the true data generating process is non-normal (\Cref{tab:my-table_non-normal}), whereas methods such as horseshoe, SSVS or SSLASSO incur very high MSEs. 


\begin{table*}[t]
\centering
\resizebox{0.8\textwidth}{!}{%
\begin{tabular}{@{}ccccccccc@{}}
\toprule
\multirow{2}{*}{($n,p,s_0$)} &
  \multirow{2}{*}{Methods} &
  \multicolumn{3}{c}{Uncorrelated Design} &
   &
  \multicolumn{3}{c}{Correlated Design} \\ \cmidrule(lr){3-5} \cmidrule(l){7-9} 
 &           & TPR           & FDP            & MCC            &  & TPR           & FDP            & MCC            \\ \midrule
\multirow{5}{*}{\begin{tabular}[c]{@{}c@{}}$n = 100$\\ $p = 2000$\\ $s_0 = 10$\end{tabular}} &
  LASSO &
  1 &
  0.154 (0.183) &
  0.943 (0.114) &
   &
  1 &
  0.145 (0.017) &
  0.918 (0.105) \\
 & MCP       & 1             & 0.056 (0.125)  & 0.958 (0.072) &  & 1             & 0.056 (0.015) & 0.969 (0.065) \\
 &
  Horseshoe & 1 &
  0.013 (0.047) &
  0.990 (0.025) &
   &
  1 &
  0.015 (0.047) &
  0.992 (0.025) \\
 & SSVS      & 1  & 0              & 1 &  & 1 & 0              & 1 \\
 & SSLASSO   & 1 & 0              & 1              &  & 0.995 (0.001)             & 0.005 (0.050)             & 0.995 (0.046)              \\
 & Projection   & 1             & 0.084 (0.063)              & 0.964 (0.108)              &  & 1             & 0.018 (0.037)              & 0.991 (0.019)              \\ \midrule
 \multirow{6}{*}{\begin{tabular}[c]{@{}c@{}}$n =  300$\\ $p = 300$\\ $s_0 = 10$\end{tabular}} &
  LASSO &
  1 &
  0.090 (0.016) &
  0.961 (0.104) &
   &
  1 &
  0.074 (0.088) &
  0.959 (0.048) \\
 & MCP       & 1             & 0.086 (0.015)              & 0.973 (0.093)              &  & 1             & 0.049 (0.085) & 0.973 (0.047) \\
 & Horseshoe & 1             & 0.010 (0.011) & 0.986 (0.014) &  & 1             & 0.008 (0.042) & 0.984 (0.011) \\
 & SSVS      & 1             & 0              & 1              &  & 1             & 0              & 1              \\
 & SSLASSO   & 1             & 0              & 1              &  & 1             & 0.001 (0.000)              & 0.999 (0.005)             \\ & Projection   & 1             & 0.023 (0.014)              & 0.978 (0.105)              &  & 1             & 0.051 (0.086)              & 0.972 (0.050)             \\ \midrule
\multirow{6}{*}{\begin{tabular}[c]{@{}c@{}}$n =  1000$\\ $p = 100$\\ $s_0 = 10$\end{tabular}} &
  LASSO &
  1 &
  0.091 (0.013) &
  0.965 (0.081) &
   &
  1 &
  0.129 (0.012) &
  0.979 (0.019) \\
 & MCP       & 1             & 0.052 (0.013)              & 0.974 (0.017)              &  & 1             & 0.108 (0.011) & 0.996 (0.084) \\
 & Horseshoe & 1             & 0.006 (0.023) & 0.992 (0.013) &  & 1             & 0.004 (0.034) & 0.991 (0.007) \\
 & BLASSO    & 1             & 0.292 (0.104) & 0.819 (0.070) &  & 1             & 0.269 (0.101) & 0.833 (0.089) \\
 & SSVS      & 1             & 0              & 1              &  & 1             & 0              & 1              \\
 & SSLASSO   & 1             & 0              & 1              &  & 1             & 0.001 (0.000)              & 0.999 (0.005)             \\ & Projection   & 1             & 0.015 (0.038)              & 0.991 (0.022)              &  & 1             & 0.052 (0.042)              & 0.984 (0.030)             \\ 
 \bottomrule
\end{tabular}%
}
\caption{Variable selection performances of the different methods corresponding to the small-$p$ and large-$p$ cases. The mean (sd) of $M = 100$ replications are reported.}
\label{tab:my-table_for_var_select}
\end{table*}

In \Cref{tab:my-table_for_var_select}, we report the measures for variable selection corresponding to the small-$p$ ($n=1000,p=100$), same-$p$ ($n=300=p$) and large-$p$ ($n=100,p=2000$) regimes and both design types. The SSVS maintained its best selection performance in low- and high-dimensional cases, followed by the SSLASSO and the projection method. The model selection performance of the sparse projection posterior is robust under the correlated design setup. When $p<n$, the Bayesian LASSO selects many noise variables, causing its FDP to be the highest and, consequently, its MCC to be the lowest. The MCP does a good job of identifying the correct signals. On average, the sparse projection-posterior-based method is comparable to the horseshoe method. Overall, from this limited simulation study, it can be concluded that the proposed method has good selection properties. In addition, our method has good coverage and this makes our method preferable in comparison to the existing Bayesian methods. High-dimensional linear regression under sparsity is a very crowded domain and several ground-breaking research has been conducted in this area. Therefore, it is challenging to outperform these current approaches in all aspects; yet, our approach has an edge in terms of guaranteeing precise coverage.

\begin{figure}[htb!]
    \centering
    \includegraphics[width = 0.43\textwidth, height = 5.25cm]{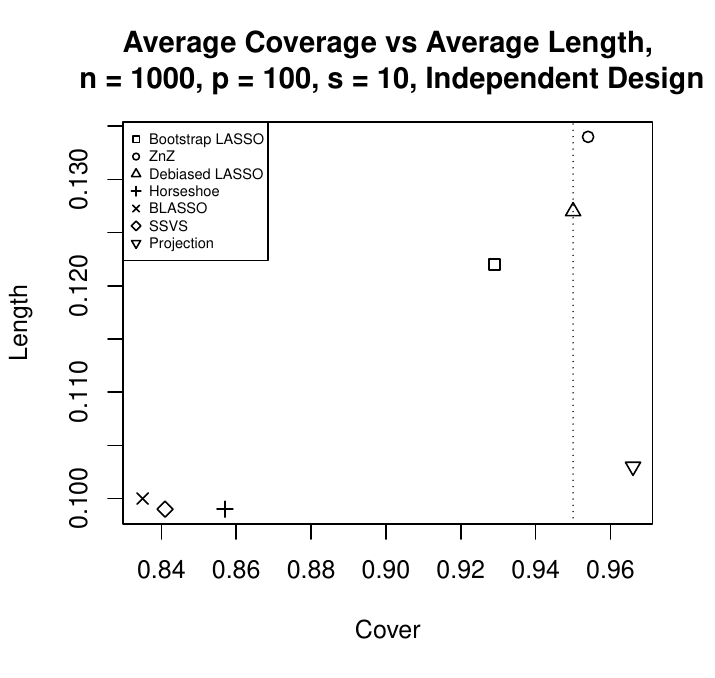}
    \includegraphics[width = 0.43\textwidth, height = 5.25cm]{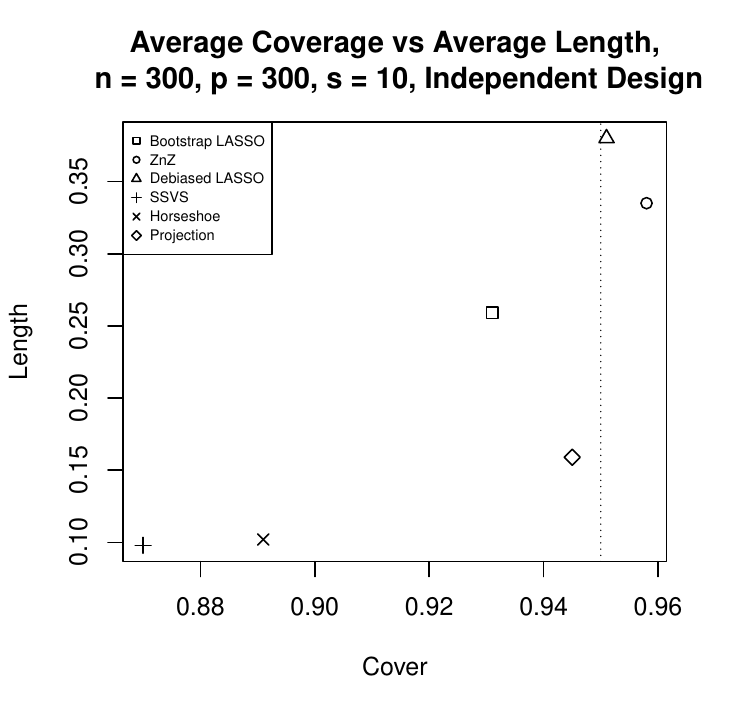}
    \includegraphics[width = 0.41\textwidth, height = 5cm]{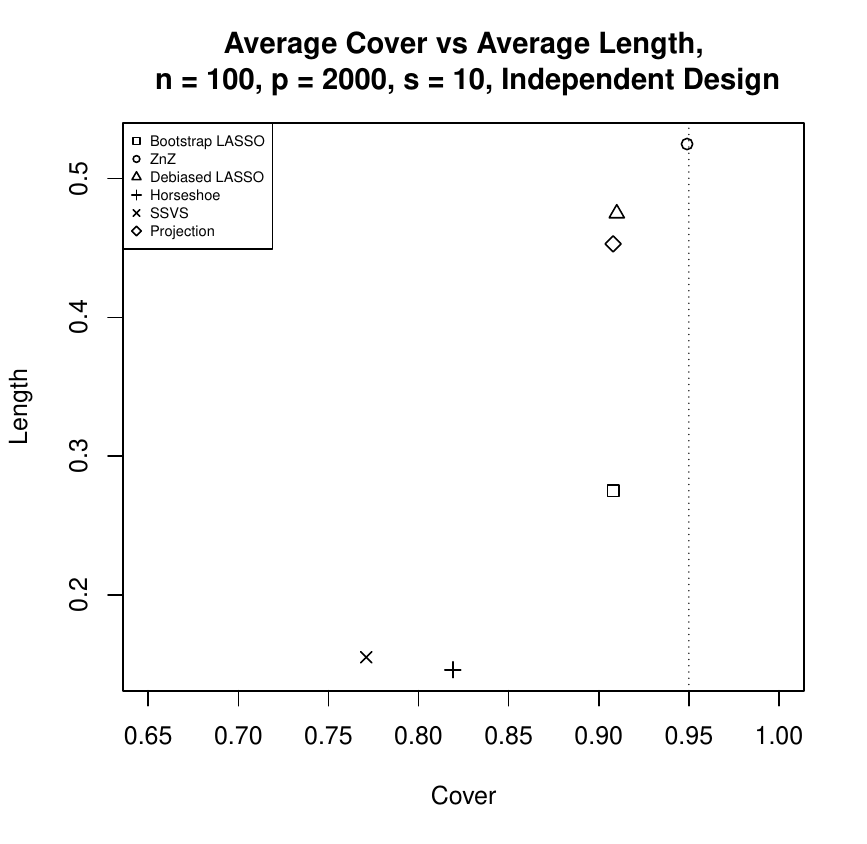}
    \includegraphics[width = 0.46\textwidth, height = 5.5cm]{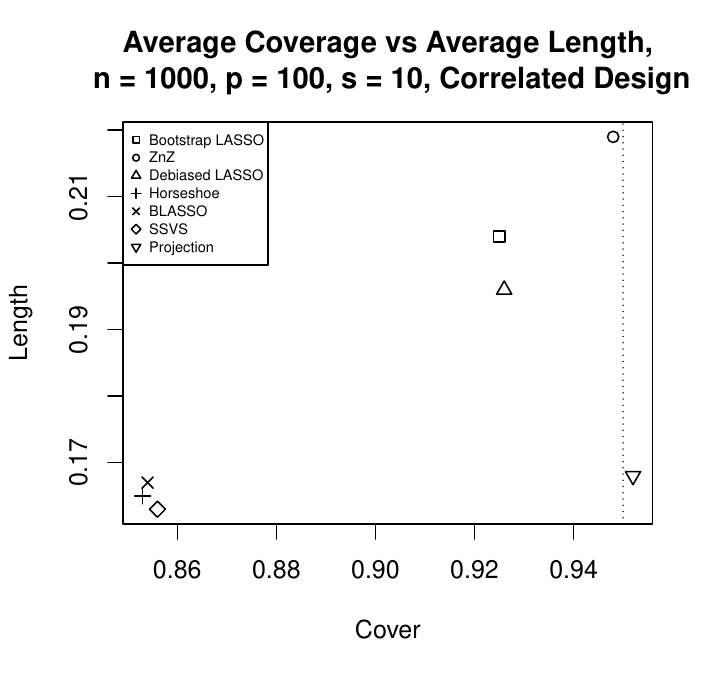}
    \includegraphics[width = 0.43\textwidth, height = 5cm]{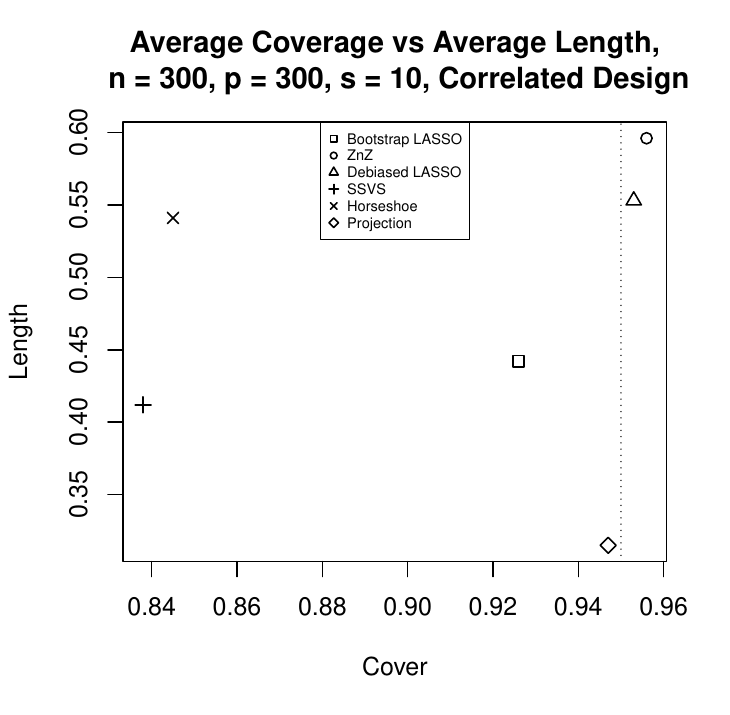}
    \includegraphics[width = 0.43\textwidth, height = 4.75cm]{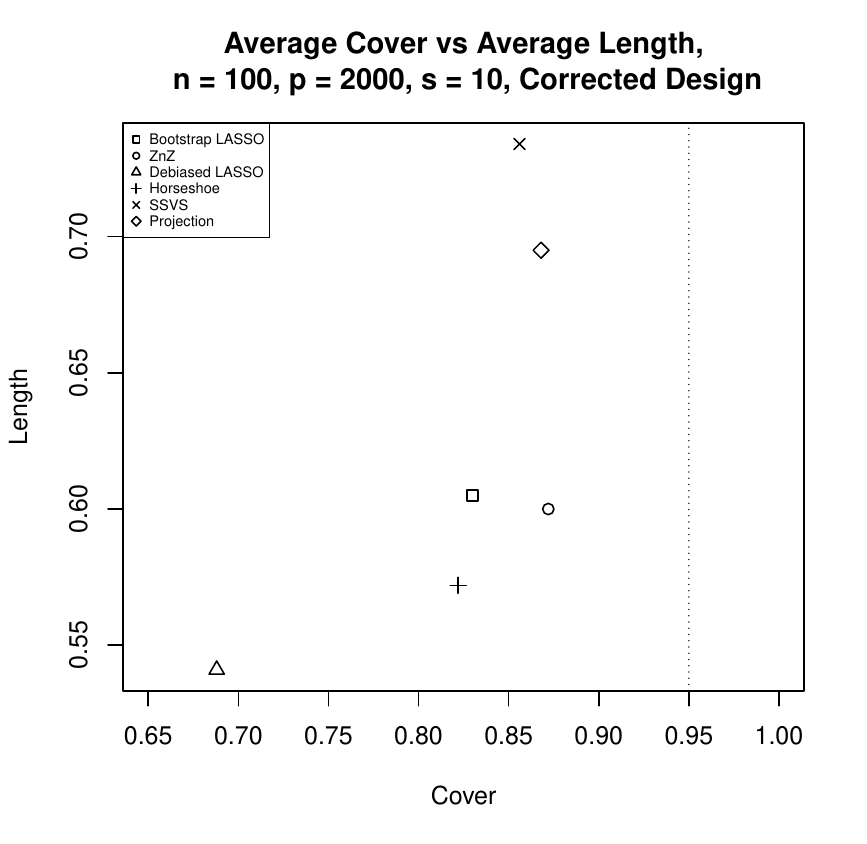}
    \caption{The average coverage probabilities for the 10 signals are plotted against the average lengths of the confidence/credible intervals for the competing methods.} 
    \label{fig:cover_vs_length}
\end{figure}

We compare the coverage of 95\% confidence or credible regions from the different methods included in the simulation. In \Cref{tab:my-table_for_CI_ind,tab:my-table_for_CI_cor}, we report the mean coverage averaged over the $M$ replicates corresponding to the uncorrelated and correlated designs respectively, but separately for active and inactive variables. In general, coverage and length-wise, all methods do better when the variables are independent. On the Bayesian side, only the proposed sparse projection-posterior method reaches close to the desired coverage. The average lengths of all the intervals increase significantly to provide the same extent of coverage when the dimension $p$ grows with the sample size $n$. The existing Bayesian methods provide coverage if the parameter is zero but fail to attain 95\% coverage for non-zero parameter values corresponding to active predictors. Also, the average coverage of signals deteriorates in the correlated design setting. However, when $p \gg n$, the only Bayesian method that provides good coverage of the signals is the sparse projection-posterior method. The plots in \Cref{fig:cover_vs_length} show the coverage and length of credible/confidence regions averaged over all the signal variables for all competing methods. For small-$p$ and $n = p$ cases, the sparse projection posterior method attains desirable coverage with a very small length, making it the best candidate. Its performance degrades when the dimension exceeds $n$ but is still better than most methods. We also show that the post-selection credible ellipsoid attains exact coverage under both independent and correlated designs. The scope to jointly quantify for the signal uncertainty helps avoid the debiasing map. Since debiasing is computationally heavy, bypassing it gives more computational edge. It should also be noted that the sparse projection posterior offers good to moderate coverage in very high dimensions even without debiasing. 

\begin{table*}[t]
\centering
\resizebox{0.8\textwidth}{!}{%
\begin{tabular}{cccccc}
\hline
\multirow{2}{*}{($n,p,s_0$)} & \multirow{2}{*}{Methods} & \multicolumn{4}{c}{Uncorrelated Design} \\ \cline{3-6} 
 &  & Signal Coverage & \begin{tabular}[c]{@{}c@{}}Length of Intervals\\ for Signals\end{tabular} & Noise Coverage & \begin{tabular}[c]{@{}c@{}}Length of Intervals\\ for Noise\end{tabular} \\ \hline
\multirow{6}{*}{\begin{tabular}[c]{@{}c@{}}$n = 100$\\ $p = 2000$\\ $s_0 = 10$\end{tabular}} & \begin{tabular}[c]{@{}c@{}}Bootstrap LASSO\end{tabular} & 0.908 (0.025) & 0.275 (0.099) & 0.984 (0.005) & 0.574 (0.002) \\
 & ZnZ & 0.949 (0.019) & 0.525 (0.101) & 0.947 (0.005) & 0.527 (0.010) \\
 & \begin{tabular}[c]{@{}c@{}}Debiased LASSO\end{tabular} & 0.910 (0.091) & 0.475 (0.095) & 0.943 (0.005) & 0.463 (0.009) \\
 & Horseshoe & 0.859 (0.026) & 0.146 (0.031) & 0.979 (0.003) & 0.442 (0.002) \\
 & SSVS & 0.771 (0.101) & 0.155 (0.081) & 0.992 (0.001) & 0.096 (0.002) \\
 & Debiased Projection & 0.908 (0.094) & 0.453 (0.023) & 0.967 (0.004) & 0.427 (0.003) \\
 & Projection & 0.899 (0.013) & 0.314 (0.034) & 0.988 (0.006) & 0.172 (0.002) \\
 & Credible Ellipsoid & 0.853 (0.112) & 0.344 (0.045) & - & - 
 \\ \hline
 \multirow{7}{*}{\begin{tabular}[c]{@{}c@{}} $n = p = 300$\\ $s_0 = 10$\end{tabular}} & Bootstrap LASSO & 0.931 (0.082) & 0.259 (0.022) & 0.972 (0.019) & 0.372 (0.011) \\
 & ZnZ & 0.958 (0.071) & 0.335 (0.240) & 0.952 (0.011) & 0.355 (0.091) \\
 & Debiased LASSO & 0.951 (0.073) & 0.380 (0.375) & 0.949 (0.016) & 0.360 (0.108) \\
 & SSVS & 0.870 (0.131) & 0.098 (0.014) & 0.999 (0.000) & 0.163 (0.032) \\
 & Horseshoe & 0.941 (0.099) & 0.102 (0.015) & 0.999 (0.000) & 0.205 (0.005) \\
 & Debiased Projection & 0.945 (0.011) & 0.159 (0.006) & 0.977 (0.001) & 0.267 (0.015) \\
 & Projection & 0.940 (0.048) & 0.301 (0.030) & 0.997 (0.001) & 0.129 (0.005) \\
 & Credible Ellipsoid & 0.927 (0.008) & 0.268 (0.025) & - & - \\\hline
\multirow{7}{*}{\begin{tabular}[c]{@{}c@{}}$n =  1000$\\ $p = 100$\\ $s_0 = 10$\end{tabular}} & \begin{tabular}[c]{@{}c@{}}Bootstrap LASSO\end{tabular} & 0.929 (0.080) & 0.122 (0.003) & 0.945 (0.024) & 0.046 (0.009) \\
 & ZnZ & 0.954 (0.076) & 0.134 (0.003) & 0.948 (0.022) & 0.133 (0.003) \\
 & \begin{tabular}[c]{@{}c@{}}Debiased LASSO\end{tabular} & 0.950 (0.082) & 0.127 (0.003) & 0.945 (0.023) & 0.126 (0.003) \\
 & Horseshoe & 0.957 (0.108) & 0.099 (0.002) & 0.991 (0.012) & 0.052 (0.003) \\
 & BLASSO & 0.835 (0.115) & 0.100 (0.002) & 0.999 (0.003) & 0.018 (0.004) \\
 & SSVS & 0.841 (0.112) & 0.099 (0.002) & 0.998 (0.000) & 0.038 (0.001) \\
 & Debiased Projection  & 0.966 (0.111) & 0.103 (0.003) & 0.949 (0.030) & 0.088 (0.002) \\ 
 & Projection & 0.942 (0.019) & 0.333 (0.026) & 0.998 (0.000) & 0.126 (0.003) \\
 & Credible Ellipsoid & 0.956 (0.016) & 0.225 (0.010) & - & - 
 \\ \hline
\end{tabular}%
}
\caption{The average (standard error) coverage and average (standard error) length of the signals  {(columns 1 and 2)} and noise variables  {(columns 3 and 4)} corresponding to $p > n$  {(Upper)}, $p = n$ {(Middle)} and $p < n$  {(Lower)} for the independent design respectively.}
\label{tab:my-table_for_CI_ind}
\end{table*}

\begin{table*}[t]
\centering
\resizebox{0.8\textwidth}{!}{%
\begin{tabular}{cccccc}
\hline
\multirow{2}{*}{($n,p,s_0$)} & \multirow{2}{*}{Methods} & \multicolumn{4}{c}{Correlated  Design} \\ \cline{3-6} 
 &  & Signal Coverage & \begin{tabular}[c]{@{}c@{}}Length of Intervals\\ for Signals\end{tabular} & Noise Coverage & \begin{tabular}[c]{@{}c@{}}Length of Intervals\\ for Noise\end{tabular} \\ \hline
\multirow{6}{*}{\begin{tabular}[c]{@{}c@{}}$n = 100$\\ $p = 2000$\\ $s_0 = 10$\end{tabular}} & \begin{tabular}[c]{@{}c@{}}Bootstrap LASSO\end{tabular} & 0.830 (0.180) & 0.605 (0.139) & 0.992 (0.004) & 0.523 (0.002) \\
 & ZnZ & 0.872 (0.146) & 0.600 (0.081) & 0.951 (0.006) & 0.586 (0.007) \\
 & \begin{tabular}[c]{@{}c@{}}Debiased LASSO\end{tabular} & 0.688 (0.147) & 0.541 (0.063) & 0.949 (0.004) & 0.564 (0.005) \\
 & Horseshoe & 0.822 (0.155) & 0.572 (0.067) & 0.966 (0.002) & 0.359 (0.007) \\
 & SSVS & 0.856 (0.122) & 0.734 (0.226) & 0.972 (0.001) & 0.312 (0.015) \\
 & Debiased Projection & 0.868 (0.131) & 0.695 (0.011) & 0.982 (0.002) & 0.548 (0.007)\\
 & Projection & 0.847 (0.119) & 0.311 (0.135) & 0.992 (0.003) & 0.151 (0.006) \\
 & Credible Ellipsoid & 0.906 (0.052) & 0.410 (0.002) & - & - \\ \hline
 \multirow{7}{*}{\begin{tabular}[c]{@{}c@{}} $n = p = 300$\\ $s_0$ = 10\end{tabular}} & Bootstrap LASSO & 0.926 (0.090) & 0.442 (0.031) & 0.966 (0.022) & 0.288 (0.009) \\
 & ZnZ & 0.956 (0.067) & 0.596 (0.212) & 0.947 (0.010) & 0.352 (0.077) \\
 & Debiased LASSO & 0.953 (0.067) & 0.553 (0.478) & 0.948 (0.014) & 0.320 (0.091) \\
 & SSVS & 0.838 (0.119) & 0.412 (0.103) & 0.998 (0.001) & 0.120 (0.150) \\
 & Horseshoe & 0.945 (0.107) & 0.541 (0.029) & 0.996 (0.001) & 0.361 (0.007) \\
 & Debiased Projection & 0.947 (0.013) & 0.315 (0.010) & 0.968 (0.007) & 0.464 (0.007) \\
 & Projection & 0.941 (0.031) & 0.361 (0.056) & 0.975 (0.002) & 0.147 (0.004) \\
 & Credible Ellipsoid & 0.941 (0.007) & 0.338 (0.013) & - & - \\ \hline
\multirow{7}{*}{\begin{tabular}[c]{@{}c@{}}$n =  1000$\\ $p = 100$\\ $s_0 = 10$\end{tabular}} & \begin{tabular}[c]{@{}c@{}}Bootstrap LASSO\end{tabular} & 0.925 (0.072) & 0.204 (0.005) & 0.953 (0.029) & 0.119 (0.015) \\
 & ZnZ & 0.948 (0.070) & 0.219 (0.005) & 0.953 (0.025) & 0.221 (0.004) \\
 & \begin{tabular}[c]{@{}c@{}}Debiased LASSO\end{tabular} & 0.926 (0.088) & 0.196 (0.005) & 0.952 (0.023) & 0.198 (0.004) \\
 & Horseshoe & 0.953 (0.102) & 0.165 (0.003) & 0.979 (0.006) & 0.061 (0.005) \\
 & BLASSO & 0.854 (0.106) & 0.167 (0.004) & 0.982 (0.002) & 0.022 (0.007) \\
 & SSVS & 0.856 (0.099) & 0.163 (0.003) & 0.991 (0.002) & 0.045 (0.003) \\
 & Debiased Projection & 0.952 (0.101) & 0.168 (0.002) & 0.969 (0.019) & 0.092 (0.008) \\
 & Projection & 0.949 (0.042) & 0.377 (0.013) & 0.996 (0.001) & 0.129 (0.002) \\
 & Credible Ellipsoid & 0.940 (0.011) & 0.211 (0.004) & - & - \\ \hline
\end{tabular}%
}
\caption{The average (standard error) coverage and average (standard error) length of the signals  {(columns 1 and 2)} and noise variables  {(columns 3 and 4)} corresponding to $p > n$  {(Upper)}, $p = n$ {(Middle)} and $p < n$  {(Lower)} for the correlated design respectively.}
\label{tab:my-table_for_CI_cor}
\end{table*}

\begin{figure}[htbp]
    \centering
    \includegraphics[width = 0.49\textwidth, height = 5.5cm]{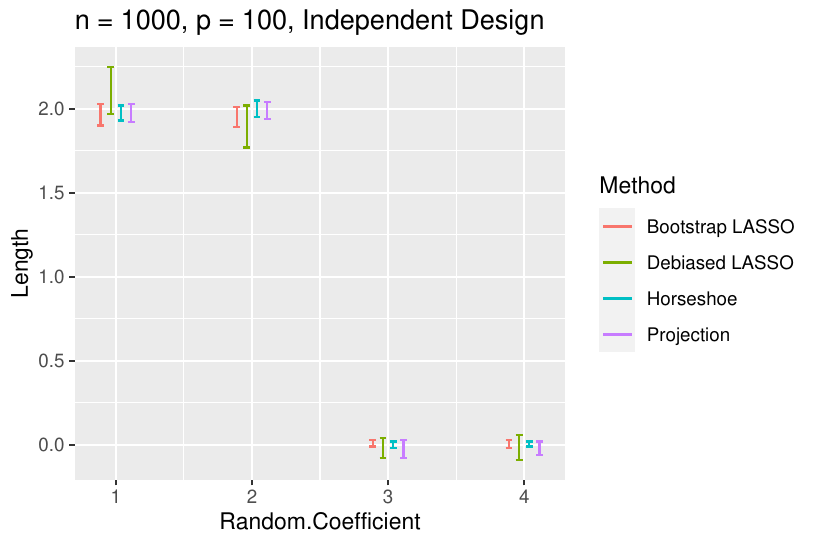}
    \includegraphics[width = 0.49\textwidth, height = 5.5cm]{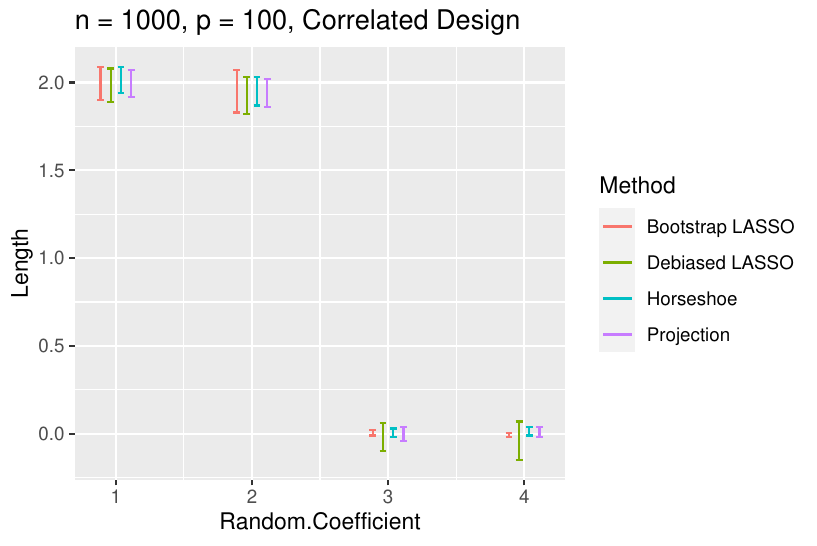}
    \includegraphics[width = 0.49\textwidth, height = 5.5cm]{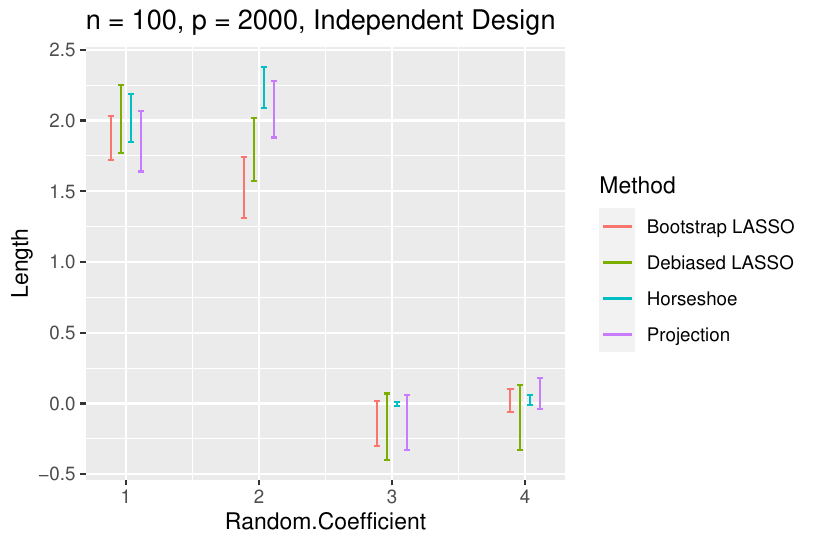}
    \includegraphics[width = 0.49\textwidth, height = 5.5cm]{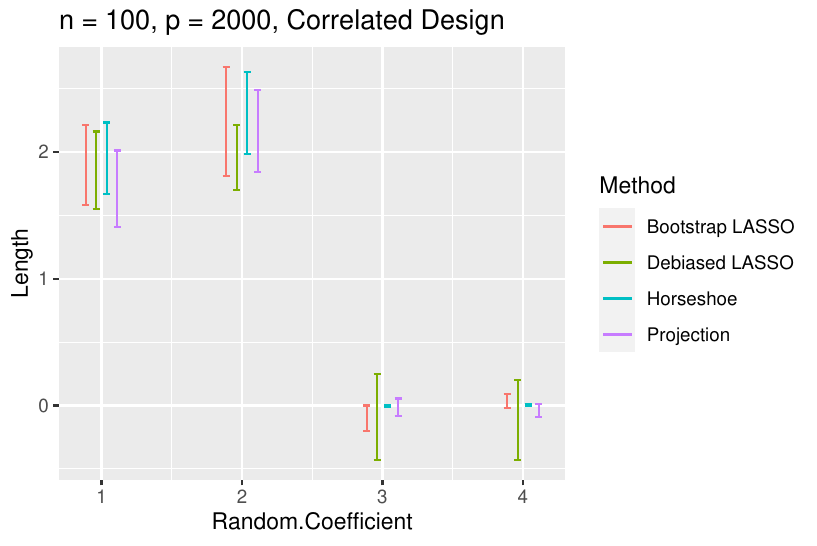}
    \caption{Plots of the randomly selected signal and noise coefficients for different methods corresponding to the independent design ($p > n$:  {lower left panel)} and the correlated design ($p > n$: {lower right panel)} and similarly, the independent design ($p < n$: {upper right panel)} and the correlated design ($p < n$: {upper left panel)}.} 
    \label{fig:random_cover}
\end{figure}

In \Cref{fig:random_cover}, we present individual coverage and interval lengths for two randomly selected signals and two randomly selected noise components in all four settings. These settings compare intervals produced by the horseshoe, sparse projection-posterior, debiased LASSO (ZnZ), and bootstrap LASSO for active (1 and 6) and noise (50 and 100) variables. Both signals have a true coefficient of 2. The plots show frequentist methods, and the Bayesian sparse posterior-projection method offers good coverage and comparable interval lengths. The horseshoe prior, in particular, exhibits slightly shorter lengths, especially for noise variables, consistent with results in Table 3. Importantly, our proposed method consistently covers the true signal. Finally in \Cref{tab:my-table_non-normal}, we show how our method maintains good coverage under different non-normal error distributions.


\begin{table*}[t]
\centering
\resizebox{\textwidth}{!}{
\begin{tabular}{ccccccccc}
\toprule
\begin{tabular}[c]{@{}c@{}}Performance\\ Metric\end{tabular} & \begin{tabular}[c]{@{}c@{}}Error\\ Model\end{tabular} & LASSO & MCP & \begin{tabular}[c]{@{}c@{}}Bootstrap \\ LASSO\end{tabular} & Horseshoe & SSLASSO & SSVS & Projection \\ \toprule
\multirow{3}{*}{MSE} & Uniform & 1.94 (0.262) & 1.84 (0.151) & 9.99 (2.85) & 4.07 (2.15) & 0.50 (1.89) & 4.12 (2.88) & 2.06 (0.376) \\
 & Laplace & 5.43 (2.946) & 2.49 (1.464) & 5.17 (2.600) & 8.20 (7.583) & 2.94 (2.244) & 4.27 (2.101) & 5.66 (2.039) \\
 & Chi-square & 4.36 (0.722) & 18.71 (1.783) & 6.26 (0.602) & 16.62 (1.744) & 12.58 (0.910) & 5.10 (1.918) & 3.52 (0.189) \\ \hline
\multirow{3}{*}{MCC} & Uniform & 0.926 (0.035) & 0.996 (0.001) & - & 0.991 (0.001) & 0.998 (0.000) & 1 (0.000) & 0.931 (0.017) \\
 & Laplace & 0.924 (0.087) & 0.977 (0.058) & - & 0.977 (0.037) & 0.986 (0.023) & 1 (0.000) & 0.967 (0.0490 \\
 & Chi-square & 0.932 (0.066) & 0.747 (0.113) & - & 0.874 (0.109) & 0.606 (0.017) & 0.706 (0.066) & 0.934 (0.059) \\ \hline
\multirow{3}{*}{\begin{tabular}[c]{@{}c@{}}Signal\\ Coverage\end{tabular}} & Uniform & - & - & 0.88 (0.06) & 0.91 (0.03) & - & 0.90 (0.01) & 0.94 (0.02) \\
 & Laplace & - & - & 0.86 (0.12) & 0.90 (0.14) & - & 0.91 (0.07) & 0.93 (0.13) \\
 & Chi-square & - & - & 0.80 (0.05) & 0.83 (0.15) & - & 0.70 (0.07) & 0.78 (0.05) \\ \hline
\multirow{3}{*}{\begin{tabular}[c]{@{}c@{}}Signal \\ Length\end{tabular}} & Uniform & - & - & 0.692 (0.028) & 0.771 (0.028) & - & 0.373 (0.021) & 0.752 (0.011) \\
 & Laplace & - & - & 0.832 (0.147) & 0.982 (0.133) & - & 0.410 (0.013) & 1.018 (0.010) \\
 & Chi-square & - & - & 2.253 (0.164) & 3.430 (0.295) & - & 2.880 (0.100) & 1.781 (0.044) \\ \bottomrule
\end{tabular}}
\caption{MSE, MCC, average signal coverage and average length of signal intervals reported for different methods with $M = 100$ replications. MSEs are reported in the order of $10^{-4}$ with standard error reported in the order of $10^{-6}$.}
\label{tab:my-table_non-normal}
\end{table*}

\subsection{Simulation Study 2: Distributed Computing}
\label{simu 2}

This section shows the usefulness of the sparse projection-posterior method when the sample size is too huge. We will conduct the following simulation to verify this claim. Here, we generate the design matrix from the normal distribution with $n = 100000$ rows and $p = 500$ columns. We set the sparsity to $s_0 = 20$ and generate the signals from the Uniform(1,3) distribution. We randomly divide the data into 100 splits to contain 1000 observations. Then, the $m$-th split of the data, containing the sub-matrix $\bX_m \in \R^{1000 \times 500}$ and the sub-vector $\bY_m \in \R^{1000}$ is sent to the $m$-th GPU to compute $\bX_m^\mathrm{T}\bX_m$ and $\bX_m^\mathrm{T}\bY_m$, where $m = 1,2,\dots,100$. Finally, these pieces are combined to get $\bX^\mathrm{T}\bX = \sum_{m = 1}^{100} \bX_m^\mathrm{T}\bX_m$ and $\bX^\mathrm{T}\bY = \sum_{m = 1}^{100} \bX_m^\mathrm{T}\bY_m$ following \Cref{distributed}. Over 100 iterations, the average time taken for parallelly computing $\bX_m^\mathrm{T}\bX_m$ and $\bX_m^\mathrm{T}\bY_m$ along with the cost of communication is only $0.17$ seconds. If, on the other hand, we used the entire data in a single machine to compute $\bX^\mathrm{T}\bX$ and $\bX^\mathrm{T}\bX$ directly, the average time taken over 100 replications is $18.32$ seconds. Although the rest of the method, including sampling from the vanilla posterior (10000 posterior samples) followed by the sparsity map, is carried out on a single machine, this initial distributed computing gives quite a computational edge to our method. Moreover, the \texttt{horseshoe} package in \texttt{R} that implements the horseshoe method cannot allocate the huge data size and hence is rendered incapable in such scenarios. We have not tried Bayesian LASSO or SSVS in this section as we already mentioned their high computation complexity in \Cref{simu 1}. The average estimation error (MSE) for the LASSO, sparse-projection, and sparse-projection distributed are respectively 0.0017, 0.00084, and 0.00097. The average MCC in the same order is 0.923, 0.998, and 0.981. 

\subsection{Simulation Study 3: Polygenic Risk Score Analysis}\label{simu 3}

We present a preliminary PRS study, an important real-life application of high-dimensional linear regression methods. LDpred (\cite{vilhjalmsson2015modeling}) that uses a spike-and-slab prior, PRS-CS (\cite{ge2019polygenic}) that uses the Strawderman-Berger GL prior (\cite{berger1996choice}) and lassosum (\cite{mak2017polygenic}) that uses a penalized regression technique are few of the commonly used methods. In addition, we implement the proposed sparse projection-posterior technique to obtain the polygenic risk scores of the phenotype of interest.

\begin{figure}[t]
    \centering
    \includegraphics[width = 0.3\textwidth, height = 5cm]{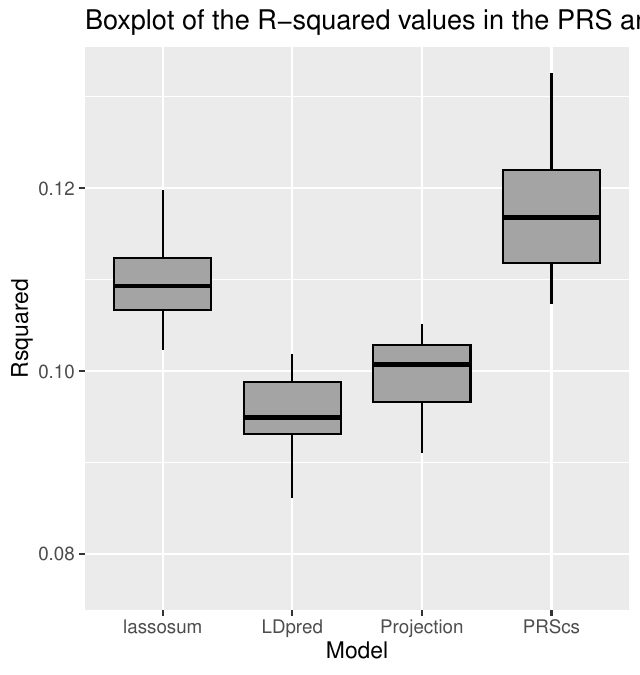}
    \caption{Boxplots of $R^2$ corresponding to the four competing methods.}
    \label{fig:my_label_PRS}
\end{figure}

Mathematically, letting $\hat{\theta}_i$ be the effect size estimate of the $i$th  variant obtained in the three different ways described above and $G_i$ be the coded marker at the $i$th locus, we have $\text{PRS} = \sum_{i = 1}^m \hat{\theta}_i G_i$. PRS analysis requires two sets of data: the target data containing individual-level genotype and phenotype data of a group of people (sample size $n$) and the base data comprising GWAS effect size estimates and corresponding p-values for the same phenotype and similar ancestry. The goal is to capture the phenotypic variability through SNP variability, known as SNP-heritability. The predictive power of PRS tends to increase with $n$. In this context, the linkage disequilibrium (LD), simply the location-based association between SNPs on the genome, should be considered for better predictions. We perform PRS-CS in \texttt{python} using the \texttt{PRScs} package (\cite{ge2019polygenic}) and use \texttt{R} for both LDpred and lassosum that require packages \texttt{bigsnpr} and \texttt{bigstatsr} (\cite{choi2020tutorial,prive2020efficient}). The posterior mean effect size can be expressed using the LD matrix and the GWAS summary statistics within the base data in both cases. The sparse effect sizes are then used to predict the phenotype in the target data. We use quality-controlled simulated target data based on the 1000 Genomes Project European samples and a quality-controlled base GWAS data on simulated height (phenotype of interest) provided in \cite{choi2020tutorial}. The target data contains relevant information for 483 individuals. We divide the dataset into 21 parts, each containing 23 samples, and use 20 as training and 1 as testing. Thus, we get 21 sets of $R^2$ and prediction errors and plot their distributions in \Cref{fig:my_label_PRS}. The prediction accuracies of the three state-of-the-art methods seem comparable to our proposed method, with the projection method doing slightly better than the LDpred. An interesting study could check if the prediction accuracies increase when the target data sample size increases.

\section{Real Data Analysis 1}

\begin{figure}[htbp]
    \centering
    \includegraphics[width = 0.23\textwidth, height = 4cm]{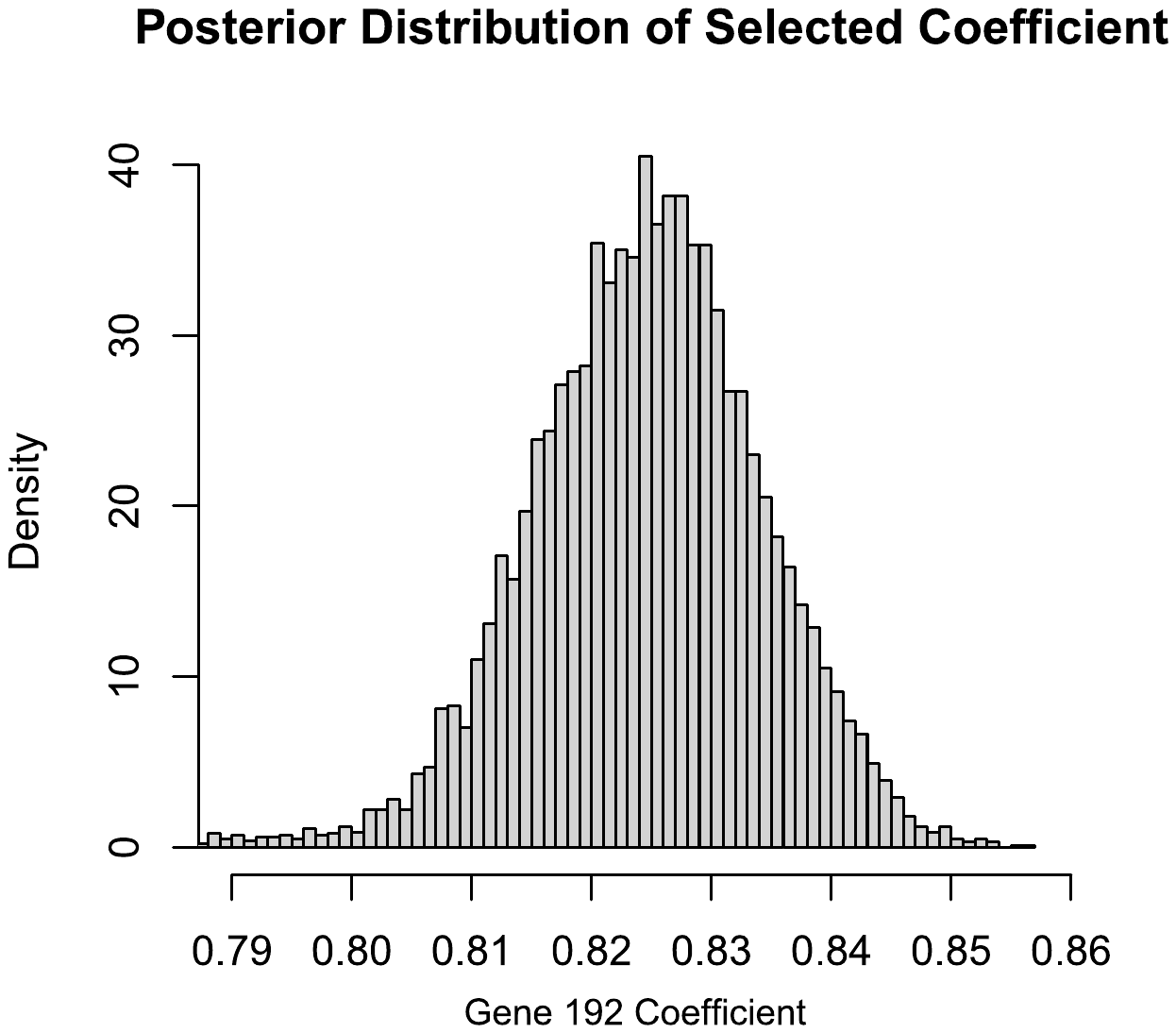}
    \includegraphics[width = 0.23\textwidth, height = 4cm]{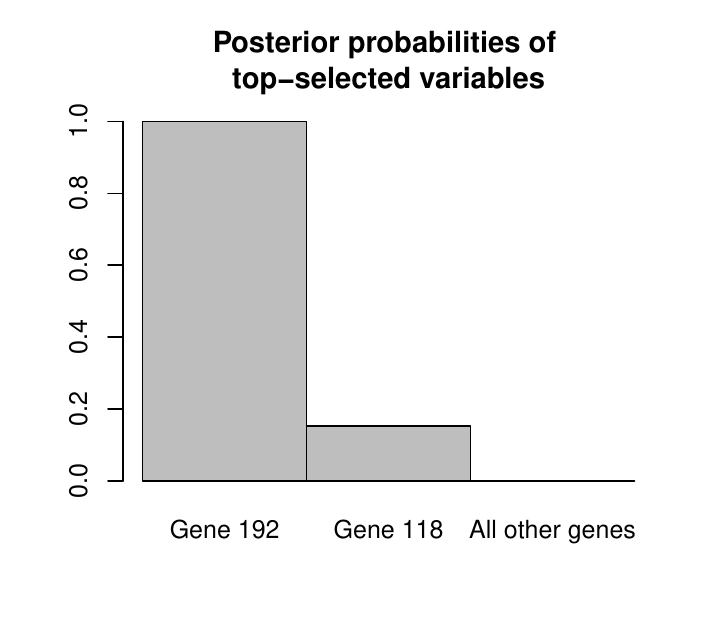}
    \includegraphics[width = 0.27\textwidth, height = 5cm]{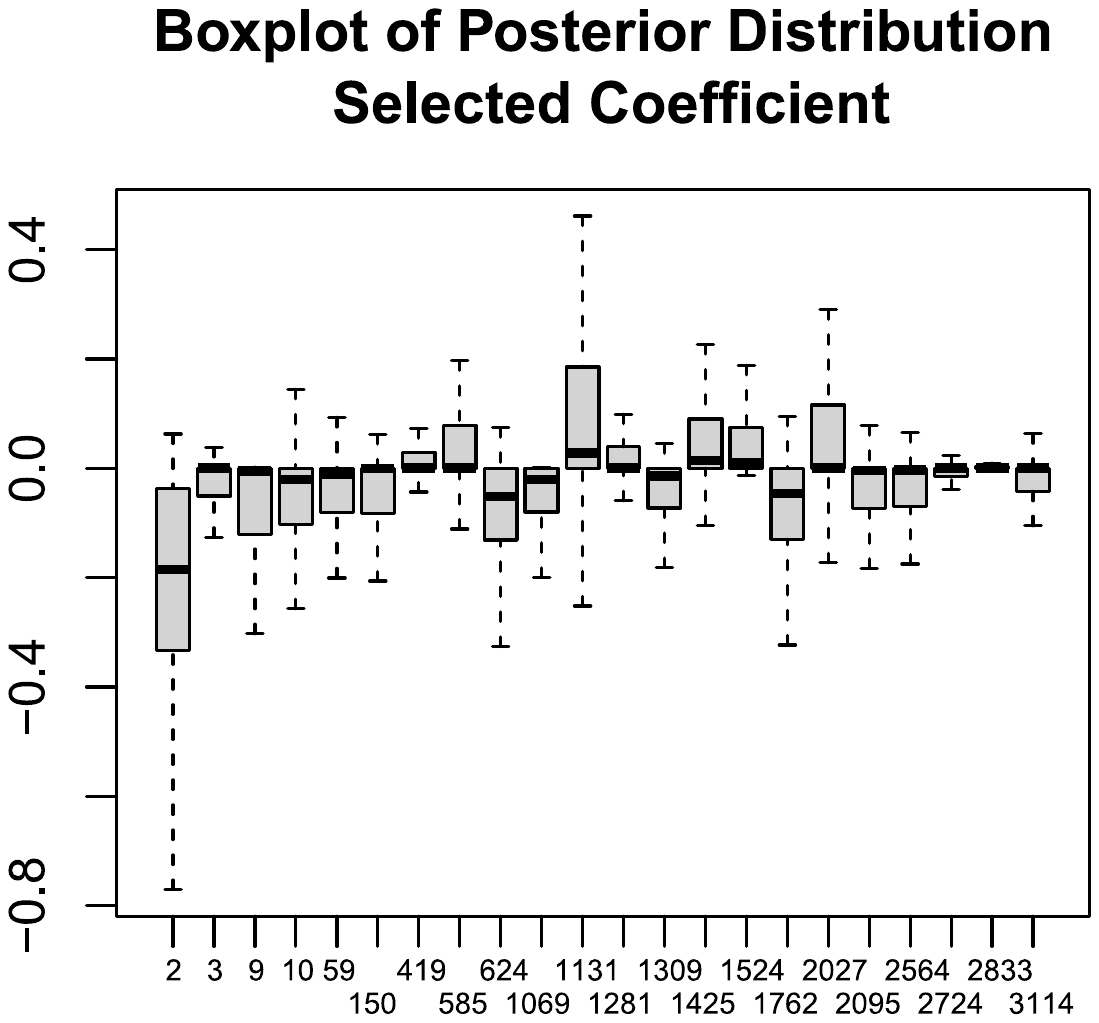}
    \includegraphics[width = 0.20\textwidth, height = 5cm]{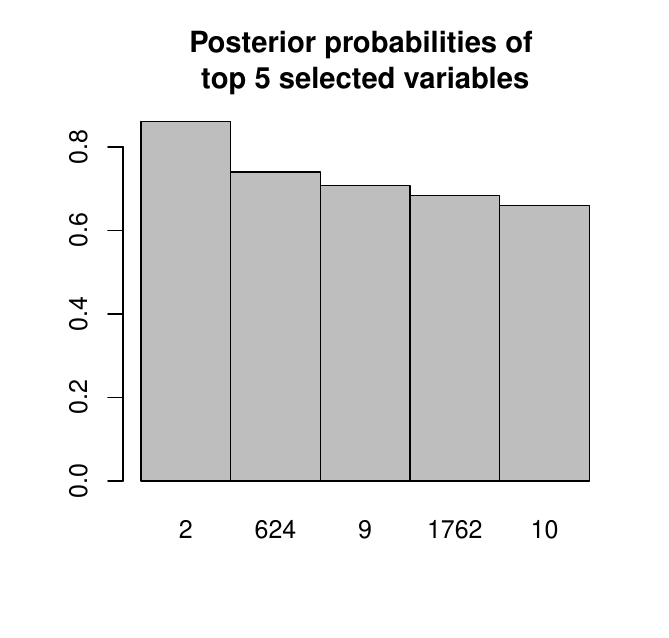}
    \caption{Posterior characteristics of the sparse projection-posterior method applied to the \texttt{flare }gene expression and the \texttt{riboflavin} data.} 
    \label{fig:real_data}
\end{figure}

First, we use the Bardet-Biedl syndrome gene expression data, an open-access dataset in the \texttt{flare} package in \texttt{R}. The expression data on 200 genes was collected from the eye tissue samples of 120 rats. The response is the expression data from the TRIM32 gene. The goal here is to detect the genes, if any that regulate the function of the discovered causal gene TRIM32. On applying the sparse projection-posterior method using the LASSO cross-validated penalty parameter, the MPM selects only gene 192, the only variable the LASSO selected. However, unlike the LASSO, our method is not limited to selection and estimation; instead, it quantifies the associated uncertainties. The first two plots in \Cref{fig:real_data} show the entire induced posterior distribution of the estimated selected gene and the top-selected genes. Also, a 95\% credible interval for the selected variable is obtained as $(0.7553066, 0.8421185)$. The computation time is only $\sim$ 13 seconds for 10000 induced posterior draws.


We consider the \texttt{riboflavin} data available in \cite{buhlmann2014high} for the second real data analysis. It deals with the problem of modeling the logarithm of the riboflavin production rate using the logarithm of the expression levels of $p = 4,088$ genes, observed for only $n = 71$ subjects. The only Bayesian method comparable to our proposed method is the horseshoe; the Bayesian LASSO and SSVS are too slow for high dimensions. The MPM of the horseshoe is the full model, whereas the sparse projection-posterior selects only 22 variables. Although the estimation by the horseshoe is more accurate, its model selection property lags. The LASSO selects 32 variables, and taking a queue from the simulation, we suspect the selected model could be an over-selection. To verify if the sparser model produced by the sparse projection-posterior method adequately explains the response, we studied the prediction errors in 10 random splits of the data, with 60 data points in the train set and the remaining 11 observations in the test set. The average prediction error for LASSO was 97.43, and that for the sparse projection-posterior was 90.93. The third and fourth plots in \Cref{fig:real_data} respectively show the full posterior distributions of the selected variables and the proportion of posterior samples containing the top 5 most selected models. We also provide the debiased-projection credible sets for the top 5 selected variables due to lack of space, noting that these can be extended to all variables. The right endpoint of these intervals is almost always 0 since their estimates are negative; hence, these intervals will likely cover the actual values. 

\begin{table*}[t]
\centering
\resizebox{\textwidth}{!}{%
\begin{tabular}{cccccc}
\hline
Top Selected Variables & 2 & 624 & 9 & 1762 & 10 \\ \hline
Debiased-projection credible interval & (-0.32578,0) & (-0.08465,0) & (-0.26747,0) & (-0.21748,0) & (-0.22362,0) \\ \hline
\end{tabular}%
}
\caption{Credible Intervals using the debiased-projection posterior for the top 5 most selected variables by the sparse-projection posterior method.}
\label{tab:my-table_dlasso_cred_real_data}
\end{table*}

\section{Real Data Analysis 2}
We are interested to know which genes if at all any, are most significant in regulating the total Alzheimer's Disease Assessment Scale (ADAS) cognitive scores of 729 individuals from the ADNI 2 and ADNI GO phases collected for the Alzheimer’s Disease Neuroimaging Initiative\footnote[1]{Data used in preparation of this article were obtained from the Alzheimer’s Disease Neuroimaging Initiative
(ADNI) database (\href{https://adni.loni.usc.edu/}{https://adni.loni.usc.edu/}). As such, the investigators within the ADNI contributed to the design
and implementation of ADNI and/or provided data but did not participate in the analysis or writing of this report.}. The primary goal of ADNI has been to
test whether serial magnetic resonance imaging (MRI), positron emission tomography (PET), other
biological markers, and clinical and neuropsychological assessment can be combined to measure the
progression of mild cognitive impairment (MCI) and early Alzheimer’s disease (AD). In our study, blood-based microarray expression data on 49386 genes are present. We filter the baseline ADAS data corresponding to the subjects available in the gene expression data. Previous studies have implications regarding the relation of blood-biomarker gene expression data with the stages of AD \citep{almansoori2024predicting, lee2020prediction, li2018systematic} using logistic LASSO or other machine learning algorithms. We identify genes based on their high association with the ADAS score. We work with the expression data using the genes retained after performing a Sure Independent Screening (SIS) pre-processing \citep{sisRpackage}, that removed variables having a very low correlation with the response based on some threshold. Moreover, we take this opportunity to discuss the benefit of the distributed computing property of our method. The proposed sparse projection-posterior is the only Bayesian linear regression method under sparsity to accommodate distributed computation. In real-world scenarios, data collection in different phases is quite common. We treat data from the ADNI 2 and ADNI GO phases as separate datasets. Our method only requires the summary measures of the data as explained in \Cref{distributed} instead of loading the entire data from all phases on a single machine. This not only reduces the burden of handling huge datasets but also respects the privacy concerns that may be associated with sharing the raw data. To see how the distributed sparse-projection method compares to existing methods and the proposed sparse-projection posterior method using the entire dataset, we report the data analysis results from the LASSO, the horseshoe, the sparse-projection, and the sparse-projection distribution.

\begin{figure}[htbp]
    \centering
    \includegraphics[width = 0.5\linewidth]{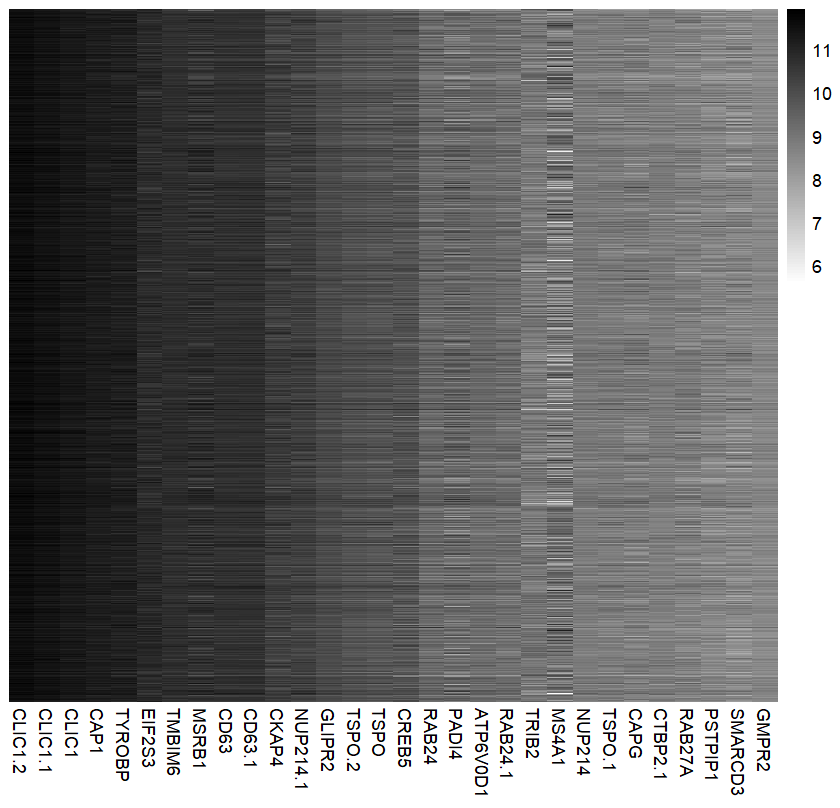}
    \caption{Heatmap of top 30 highest expressed genes.}
    \label{fig:heatmap}
\end{figure}

We divide the merged ADNI GO and ADNI 2 data into test data containing 20\% of the entire site and training data. The training dataset is split into two groups according to the two phases of ADNI. We capture the mean prediction errors of LASSO, the sparse-projection method, its distributed version, and the horseshoe method over 10 such random splits. The average prediction errors and their standard errors are as follows: LASSO 108.823 (28.065), sparse-projection 102.186 (26.151), sparse-projection distributed 102.309 (26.160) and finally the horseshoe 110.829 (22.123). In \Cref{fig:heatmap}, we plot the expression data corresponding to the top 30 genes with the highest average expressions across all subjects in the study in a decreasing fashion, with the left-most gene having the highest average expression. The sparse-projection and the sparse-projection distributed select the same gene as the LASSO. The selected gene is the CLIC1 gene, which is not only the highest average expressed gene across the subjects in ADNI 2 and ADNI GO as suggested by the heat plot in \Cref{fig:heatmap}, it has also been widely studied in association with Alzheimer's disease by \cite{carlini2020clic1,zhang2013integrated,milton2008clic1,novarino2004involvement}.
On the other hand, the estimator of the regression coefficient obtained from the horseshoe method fails to invoke sparsity and consequently selects the full model. The 95\% credible interval for the selected gene using the debiased-projection is (0.89,7.29), and the same using just the sparse-projection posterior samples is (1.30,1.34). The ADNI data analysis thus not only reiterates the relevance and effectiveness of the proposed sparse projection-posterior method and demonstrates how well it works with distributed computing.

\section{Discussion}
\label{Conclusion}

We show that the sparse projection-posterior concentrates around the true regression coefficients at an optimal rate. Our Bayesian method exhibits favorable sign-consistency, accurately selecting active predictors and determining their signs with high probability. Additionally, we establish the correct asymptotic frequentist coverage for the sparse projection-posterior credible ball through proper re-centering. The proposed method is scalable to distributed computing, efficiently handling large datasets. Simulation studies across various sample sizes and predictor dimensions support our theoretical results, revealing comparable MSE to the LASSO, correct selection of active predictors, and accurate coverage matching credibility levels in Bayesian credible sets. The extensive simulation studies show that our proposed method performs well on all these fronts. While our proposed method incurs losses in certain areas, such as a higher MSE, it compensates with strengths in others, like being comparable to LASSO and demonstrating superior MCC relative to most methods, except SSVS and SSLASSO. It excels in the coverage of credible intervals, outperforming SSVS, SSLASSO, and all other Bayesian methods, making it the best method overall. Not only in the high-dimensional penalized linear regression, this Bayesian approach shows tremendous potential in many other statistical models, where the final goal is seemingly unachievable through a traditional Bayesian method but may be easily circumvented with the help of an aptly designed projection map.

\section{Acknowledgement}
Data collection and sharing for the Alzheimer's Disease Neuroimaging Initiative (ADNI) is funded by the National
Institute on Aging (National Institutes of Health Grant U19 AG024904). The grantee organization is the Northern
California Institute for Research and Education.

\section{Funding}
This research is partially supported by ARO grant number 76643MA 2020-0945.

\section{Appendix}
\begin{lemma}
\label{lemma3}
Under Assumption \ref{sd_assum},
$$\displaystyle \max_{j=1,\ldots,n} \big |1-\frac{d_j^2}{d_j^2 + a_n} \big|
=\max_{j=1,\ldots,n} \big |\frac{a_n}{d_j^2 + a_n} \big|
=o(n^{-1}),$$  $$ \big| 1 - n^{-1} \sum_{j=1}^{n} \frac{d_j^2}{d_j^2 + a_n} \big| = o (n^{-1}),$$ 
where $d_1,\ldots,d_n$ are the singular values of $\bX$. 
\end{lemma}

\begin{proof}
 The second statement is implied by the first. 
By Assumption~\ref{sd_assum}, 
$$|1-d_j^2/(d_j^2+a_n)|\le a_n/\min\{ d_j^2: j=1,\ldots,n\}=o(1/n),$$ 
uniformly for all $j=1,\ldots,n$. 
\end{proof}

\begin{lemma}
\label{variance estimate}
    Under Condition \ref{sd_assum}, $n\hat\sigma_n^{2}\to 0$ in probability under the true distribution, where $\hat{\sigma}_n^2$ is the usual Bayesian variance estimator $n^{-1}\bY^{\mathrm{T}} ( \bm{I}_n - \bm{H}(a_n) ) \bY$. 
\end{lemma}

\begin{proof}
It suffices to show that the mean and the variance of the quadratic form $ \bY^{\mathrm{T}} ( \bm{I}_n - \bm{H}(a_n) ) \bY$ both tend to zero in probability. Recalling that $\bY=\bX \btheta^0+\bm{\varepsilon}$, the mean is given by $(\bX \btheta^0)^{\mathrm{T}} (\bm{I}_n-\bm{H}(a_n)) (\bX \btheta^0)$. By Lemma~\ref{lemma3}, all eigenvalues of $\bm{I}_n-\bm{H}(a_n) $ are uniformly $o(n^{-1})$, while by Assumption~\ref{mean_assum}, $\|\bX \btheta^0\|\le \sqrt{n} \|\bX \btheta^0\|_\infty =o(n^{1/2})$. Thus $\E\big( \bY^{\mathrm{T}} ( \bm{I}_n - \bm{H}(a_n) ) \bY\big) \to 0$. Now, $\mathrm{var}\big( \bY^{\mathrm{T}} ( \bm{I}_n - \bm{H}(a_n) ) \bY \big)$ is  
\begin{align*}
    \lefteqn{\text{tr} \big(\bm{I}_n- 2 \bX (\bX^{\mathrm{T}}\bX+a_n \bm{I}_p)^{-1} \bm{X}^{\mathrm{T}}} \\ & \quad + \bX (\bX^{\mathrm{T}}\bX+a_n \bm{I}_p)^{-1} \bm{X}^{\mathrm{T}}\bX (\bX^{\mathrm{T}}\bX+a_n \bm{I}_p)^{-1} \bm{X}^{\mathrm{T}}\big) \\
    &=\sum_{j=1}^n \big( 1-2\frac{d_j^2}{d_j^2+a_n}+\frac{d_j^4}{(d_j^2+a_n)^2} \big) \\
    &=\sum_{j=1}^n \frac{a_n^2}{(d_j^2+a_n)^2}, 
\end{align*}
which is $o(n^{-1}) $ by Lemma~\ref{lemma3}. 
\end{proof}

\begin{lemma}
\label{lemma0}
Under Assumptions \ref{design}, \ref{mean_assum}, 
\begin{eqnarray*}
\Pi \big( \{ \max_{1 \leq j \leq p} n^{-1} \big| \boldeta^{\mathrm{T}} \bX^{(j)} \big| 
 \leq \lambda_n/2 \} \big| \bY \big) 
  \geq 1 - 2e^{-((C_0/4 - 1)\log p)/2}
\end{eqnarray*}
in probability for $\lambda_n = \sigma_0 \sqrt{C_0 \log p/{n}}$ and $C_0 > 4$.
\end{lemma}

\begin{proof}
First, we note that $$\big( \boldeta - \boldsymbol{\mu} \big)^{\mathrm{T}} \bX^{(j)} | (\bY,\sigma^*) \sim \normal \big( \bm{0}, {\sigma^*}^2 {\bX^{(j)}}^{\mathrm{T}} \bm{H}(a_n) \bX^{(j)}\big),$$ where ${\bX^{(j)}}^{\mathrm{T}} \bm{H}(a_n) \bX^{(j)}$ is the $j$th diagonal element of $\bX^{\mathrm{T}} \bX \big( \bX^{\mathrm{T}} \bX + a_n \bm{I}_p\big)^{-1} \bX^{\mathrm{T}} \bX$. Moreover, using the singular value decomposition of $\bX = \bm{U} \bm{D} \bm{V}^{\mathrm{T}}$, we get 
$$\big( \bm{I}_n - \bm{H}(a_n) \big) = \bm{U} \big( \bm{I}_p - \bm{D} \big( \bm{D}^2 + a_n \bm{I}_p \big)^{-1} \bm{D}\big) \bm{U}^{\mathrm{T}}$$ 
and hence $\big(\bm{I}_n - \bm{H}(a_n) \big)$ is non-negative definite, which in turn implies $\bX^{\mathrm{T}} \bX \geq \bX^{\mathrm{T}} \bm{H}(a_n) \bX$. Consequently, the magnitude of the $j$th diagonal element of $\bX^{\mathrm{T}} \bX$ will be greater than the posterior variance of $\big( \boldeta - \boldsymbol{\mu} \big)^{\mathrm{T}} \bX^{(j)}$. Since $\bX$ is standardized to make all columns have Euclidean norm $n$, the diagonals of $(\bX^{\mathrm{T}} \bX)/n$ are all $1$. 

Since $\Pi(\sigma^* \notin \mathcal{U}_n|\bY) \to 0$, it is enough to bound 
    $\Pi \big( \{ \max_{1 \leq j \leq p} n^{-1} \big| \boldeta^{\mathrm{T}} \bX^{(j)} \big|
    \leq \lambda_n/2 \} \big| \bY, \sigma^* \big)$ by $ 2e^{-((C_0/4 - 1)\log p)/2}$ uniformly in $\sigma^* \in \mathcal{U}_n$.

Now for any $\sigma^* \in \mathcal{U}_n$, 
\begin{align}
\label{var}
  & \Pi \big( \big\{ \max_{1 \leq j \leq p} \frac{|( \boldeta - \boldsymbol{\mu} )^{\mathrm{T}} \bX^{(j)}|}{\sigma^* \sqrt{{\bX^{(j)}}^{\mathrm{T}} \bm{H}(a_n) \bX^{(j)}}} > \lambda_n/2 \big\} \big| \bY, \sigma^* \big) \nonumber\\
  & \geq \Pi \big( \big\{ \max_{1 \leq j \leq p} \frac{|( \boldeta - \boldsymbol{\mu} \big)^{\mathrm{T}} \bX^{(j)} \big|}{\sigma^* \sqrt{n}} > \lambda_n/2 \big\} \big| \bY, \sigma^* \big).
\end{align}
Putting $\lambda_n=\sqrt{C_0\log p/n}$, uniformly in $\sigma^* \in \mathcal{U}_n$, we can bound 
\begin{align*}
    \lefteqn{\Pi \bigg( \bigg\{\max_{1 \leq j \leq p} \frac{ \big| \big( \boldeta - \boldsymbol{\mu} \big)^{\mathrm{T}} \bX^{(j)} \big|}{n} \leq \sigma_0 \sqrt{\frac{C_0\log p}{4n}} \bigg\} \bigg| \bY, \sigma^* \bigg)} \\
    &\geq  1 -  \Pi \bigg( \bigg\{ \max_{1 \leq j \leq p} \frac{\big| \big( \boldeta - \boldsymbol{\mu} \big)^{\mathrm{T}} \bX^{(j)} \big|}{\sigma^* \sqrt{n{\bX^{(j)}}^{\mathrm{T}} \bm{H}(a_n) \bX^{(j)}}}\\
    & \qquad \qquad \qquad > \frac{\sigma_0}{\sigma^*} \sqrt{\frac{C_0\log p}{4n}} \bigg\} \big| \bY, \sigma^* \bigg)\\ 
    &\geq  1 - 2p  \Pi \bigg( \frac{\big( \boldeta^{\mathrm{T}} \bX^{(j)} - \boldsymbol \mu^{\mathrm{T}} \bX^{(j)} \big)}{\sigma^* \sqrt{{\bX^{(j)}}^{\mathrm{T}} \bm{H}(a_n) \bX^{(j)}}} \\
    & \qquad \qquad \qquad > \frac{\sigma_0}{\sigma^* \sqrt{n}} \sqrt{\left(\frac{C_0}{4}-1\right) \log p  + \log p} \bigg| \bY, \sigma^* \bigg)\\
    &\geq  1 - 2\exp\{ -((C_0/4 - 1)\log p)/2\}.
\end{align*}

Using Assumptions \ref{design} and \ref{mean_assum} and the fact that the maximum of $n$ i.i.d. Gaussian random variables grows at the $\mathcal{O} (\sqrt{\log n})$ rate, we can write 
\begin{align*}
    \lefteqn{\max_{1 \leq j \leq p} \frac{1}{n} \big| \boldsymbol \mu^{\mathrm{T}} \bX^{(j)} \big|} \\ 
     &= \max_{1 \leq j \leq p} \big| {\bX^{(j)}}^{\mathrm{T}} \bm{H}(a_n) \big( \bX\btheta^0 + \bm \varepsilon \big) - {\bX^{(j)}}^{\mathrm{T}} \bX \btheta^0 \big|\\
     &= \max_j \big|{\bX^{(j)}}^{\mathrm{T}} \bm{H}(a_n) \bm \varepsilon - {\bX^{(j)}}^{\mathrm{T}} \big( \bm{I}_n - \bm{H}(a_n)\big) \bX \btheta^0 \big|\\
    &\leq  \max_j \big| {\bX^{(j)}}^{\mathrm{T}} \bm{H}(a_n) \boldsymbol \varepsilon \big| + \max_j \big| {\bX^{(j)}}^{\mathrm{T}} \big( \bm{I}_n - \bm{H}(a_n)\big) \bX \btheta^0 \big|\\
    &\leq  \max_j \big[ \max_{1 \leq i \leq n} |\varepsilon_i| \|{\bX^{(j)}}^{\mathrm{T}} \bm{H}(a_n)\|_1 \\
    & \quad + \max_{1 \leq i \leq n} |\E(Y_i)| \| {\bX^{(j)}}^{\mathrm{T}} \big( \bm{I}_n - \bm{H}(a_n) \big)\|_1\big],
\end{align*}
which is bounded above by $\mathcal{O} \big( \sqrt{\log n}\big) \| {\bX^{(j)}}^{\mathrm{T}} \bm{I}_n\|_1 = \mathcal{O} \big( \sqrt{ \log n}\big)$.
\end{proof}

\begin{lemma} 
\label{lemma5}
The debiased LASSO satisfies 
$$n^{-1/2}\frac{ |{\bX^{(j)}}^{\mathrm{T}} \bR^{(j)}|}{\|\bR^{(j)}\|_2} \big|\hat{\theta}_j^{\mathrm{DB}} - (\theta^0_{j}+m_j/\sqrt{n}) \big| \leq A_1 + A_2,$$ 
where  
$A_1 = n^{-1/2} \|\bY - \bX \hat{\boldsymbol \theta}^\mathrm{R} \|_1$ and  $$A_2 = {{\sqrt{n}} {\lambda^{\bX_j}\|\btheta^0 - \hat{\boldsymbol \theta}^L\|_1}}/{({2\|\bR^{(j)}\|_2})} .$$
\end{lemma}

\begin{proof}
We observe that 
    \begin{align*}
    & n^{-1/2} \Bigg|\frac{ {\bX^{(j)}}^{\mathrm{T}} \bR^{(j)}}{\|\bR^{(j)}\|_2} \big[\hat{\theta}_j^{\mathrm{DB}} - (\theta^0_{j}+m_j/\sqrt{n}) \big] \Bigg|\\
    & = n^{-1/2} \Bigg|\frac{ {\bX^{(j)}}^{\mathrm{T}} \bR^{(j)}}{\|\bR^{(j)}\|_2} \big[ \frac{{\bR^{(j)}}^{\mathrm{T}} \bY}{{\bR^{(j)}}^{\mathrm{T}}\bX^{(j)}} - \sum_{k \neq j} P_{jk} \hat{\theta}_k^L \\ 
    & \qquad \qquad \qquad - \theta^0_{j} - \frac{{\bR^{(j)}}^{\mathrm{T}}\bX \hat{\boldsymbol\theta}^\mathrm{R}}{{\bR^{(j)}}^{\mathrm{T}}\bX^{(j)}} + \sum_{k = 1}^{p} P_{jk}\theta^0_{k} \big] \Bigg|\\
    & = n^{-1/2} \Bigg| \big[\frac{{\bR^{(j)}}^{\mathrm{T}} (\bY - \bX \hat{\boldsymbol \theta}^\mathrm{R})}{\|\bR^{(j)}\|_2} + \sum_{k \neq j} \frac{ {\bX^{(k)}}^{\mathrm{T}} \bR^{(j)}}{\|\bR^{(j)}\|_2} \big(\theta^0_{k} - \hat{\theta}_k^L \big) \big] \Bigg|\\
    & \leq n^{-1/2} \big[ \max_{1 \leq i \leq n} \frac{|\bR_i^{(j)}/n|}{\|\bR^{(j)}\|_2/n} \|\bY - \bX \hat{\boldsymbol \theta}^\mathrm{R} \|_2 \\
    & \qquad \qquad \qquad + \max_{k \neq j} \frac{|{\bR^{(j)}}^{\mathrm{T}} \bX^{(k)}/n|}{\|\bR^{(j)}\|_2/n}  \|\btheta^0 - \hat{\boldsymbol \theta}^L\|_1 \big]\\
    & \leq n^{-1/2} \big[\|\bY - \bX \hat{\boldsymbol \theta}^\mathrm{R} \|_2 + \frac{\lambda^{\bX_j}/2}{\|\bR^{(j)}\|_2/n} \|\btheta^0 - \hat{\boldsymbol \theta}^L\|_1 \big],
    \end{align*}
    establishing the claim. 
\end{proof}
    
\begin{lemma}
\label{lemma4} 
Under \Cref{sd_assum}, $n^{-1/2} \|\bY - \bX \hat{\boldsymbol \theta}^\mathrm{R} \|_1 = o_P(1).$
\end{lemma}
\begin{proof}

Note that
\begin{align}
\label{split}
   n^{-1/2} \|\bY - \bX \hat{\boldsymbol \theta}^\mathrm{R} \|_2 
    = & n^{-1/2} \| \big( \bm{I}_n - \bm{H}(a_n) \big) \bX \btheta^0 \|_2 \\
    &+ n^{-1/2} \| \big( \bm{I}_n - \bm{H}(a_n) \big) \bm{\varepsilon}\|_2. \nonumber
\end{align}
For the second summand, $E \big( \bm{\varepsilon}^{\mathrm{T}} \big( \bm{I}_n - \bm{H}(a_n) \big)^2 \bm{\varepsilon} \big) = \sigma^2 \text{tr} \big( \big( \bm{I}_n - \bm{H}(a_n) \big)^2 \big)$ and $\text{var}(\bm{\varepsilon}^{\mathrm{T}} \big( \bm{I}_n - \bm{H}(a_n) \big)^2 \bm{\varepsilon}) = \sigma^2 \text{tr} \big( \big( \bm{I}_n - \bm{H}(a_n) \big)^4 \big)$. Since $\big(\bm{I}_n - \bm{H}(a_n) \big)$ is non-negative definite,  by \Cref{lemma3}, $\mathrm{tr}\big(\big( \bm{I}_n - \bm{H}(a_n) \big)^2 \big) \leq  \mathrm{tr}\big( \bm{I}_n - \bm{H}(a_n) \big) = o(1)$, and similarly, $\text{tr} \big( \big( \bm{I}_n - \bm{H}(a_n) \big)^4 \big) = o(1).$ Consequently, $n^{-1/2} \| \big( \bm{I}_n - \bm{H}(a_n) \big) \bm{\varepsilon}\|_2 = o(n^{-1/2})$.

Referring to $(\bm{A})_{i,j}$ as the element in the $i{\textnormal{th}}$ row and $j{\textnormal{th}}$ column of a matrix $\bm{A} \in \R^{n \times n}$, we can bound the first term $n^{-1/2} \| \big( \bm{I}_n - \bm{H}(a_n) \big) \bX \btheta^0 \|_2$
in \eqref{split} by   
\begin{align*}
    n^{-1/2} \| \bX \btheta^0 \|_2 \bigg(\sum_{i = 1}^{n} \sum_{j = 1}^{n}  \big( \bm{I}_n - \bm{H}(a_n) \big)^2_{i,j} \bigg)^{1/2}
\end{align*}
using the Cauchy-Schwartz inequality. Finally, this upper bound is $\Big\{\textnormal{tr}\big( \bm{I}_n - \bm{H}(a_n) \big)^2\Big\}^{{1}/{2}}=o(1)$.
\end{proof}

Let $\hat\btheta^{\mathrm{R},\mathrm{PS}}_{S_0}=(\bX_{(1)}^\mathrm{T}\bX_{(1)} + a_n \bm{I}_{s_0})^{-1} \bX_{(1)}^\mathrm{T}\bY$ stand for the ridge regression estimator with predictors restricted to $S_0$ post selection of model $S_0$. 

\begin{lemma}\label{ridge CLT}
    If the $S_0$ is fixed and  $\bC_{n(11)}\to \bC_{(11)}$ as $n \to \infty$, then under Assumptions \ref{design}, $\sqrt{n}(\hat{\btheta}^\mathrm{R,PS}_{S_0} - \btheta^0_{S_0}) \rightsquigarrow \normal_{s_0}(\textbf{0}_{s_0},\sigma_0^2\bC^{-1}_{(11)} )$.
\end{lemma}


\begin{proof}

Without loss of generality, let the first $s_0$ components correspond to signal variables. 
Since $ n^{-1}(\bX_{(1)}^\mathrm{T}\bX_{(1)} 
+ a_n \bm{I}_{s_0})\to \bC_{(11)}$, it suffices to prove that 
$n^{-1/2}\bX_{(1)}^\mathrm{T}\bm{\varepsilon} \rightsquigarrow \normal_{s_0}(0, \sigma_0^2 \bC_{(11)})$.

Let $\bm{0}\ne \bm{b} \in \R^{s_0}$ be fixed. Because of the Cram\'er-Wold device, it is enough to note that 
the variance term 
\begin{align*}
     \lefteqn{\textnormal{cov}(n^{-1/2} \sum_{j=1}^{s_0} \sum_{i = 1}^n b_j x_{ij}\varepsilon_i, n^{-1/2} \sum_{j=1}^{s_0} \sum_{i = 1}^n x_{ik}\varepsilon_i)}\\ 
     & = \sigma_0^2 n^{-1} \sum_{j=1}^{s_0} \sum_{k=1}^{s_0} \sum_{i = 1}^n x_{ij}x_{ik} b_j b_k 
      \to \sigma_0^2 \bm{b}^{\mathrm{T}} \bC \bm{b}, 
 \end{align*}
and verify the conditions of Lindeberg's central limit theorem (CLT) for $n^{-1/2} \bm{b}^{\mathrm{T}} \bX_{(1)}^\mathrm{T}\bm{\varepsilon}$. 
Since $\bm{\varepsilon}$ has i.i.d. components, Lindeberg's condition reduces to  
$$\frac{n^{-1}\max_i (\sum_{j = 1}^{s_0} b_j x_{ij})^2}{n^{-1}\sum_{i = 1}^n (\sum_{j = 1}^{s_0} b_j x_{ij})^2} \to 0.$$ 
This holds since 
\begin{align*}
\frac{n^{-1} \max_i (\sum_{j = 1}^{s_0} b_j x_{ij})^2}{n^{-1} \sum_{i = 1}^n (\sum_{j = 1}^{s_0} b_j x_{ij})^2} & \leq \frac{\|\bm{b}\|^2 n^{-1}\max_i \sum_{j=1}^{s_0} x_{ij}^2 }{n^{-1} \|\bm{b}^\mathrm{T}\bX_{(1)}\|^2}\\ 
&\le  \frac{\|\bm{b}\|^2 n^{-1} s_0 M_1^2}{ \bm{b}^\mathrm{T} \bC_{n(11)}\bm{b}} \to 0,
\end{align*}
as $ \bm{b}^\mathrm{T} \bC_{n(11)}\bm{b}\to \bm{b}^\mathrm{T} \bC_{(11)}\bm{b}>0$ for any $\bm{b}\ne \bm{0}$ by the positive definiteness of $\bC_{(11)}$.  
\end{proof}

\begin{proof}[Proof of \Cref{consis_thm}]
From \eqref{sparse_proj}, since $\btheta^*$ is the minimizer, we can write 
\begin{align}
\label{thm1eq1}
  \lefteqn{  \frac{1}{2n}\|\bX \btheta - \bX \btheta^0\|^2_2 + \lambda_n \|\btheta^0\|_1 }\notag\\
  & \geq \frac{1}{2n}\|\bX \btheta - \bX \boldsymbol \theta^*\|^2_2 + \lambda_n \|\boldsymbol \theta^*\|_1 \nonumber\\
    & = \frac{1}{2n}\| \bX \btheta^0 + \boldeta - \bX \boldsymbol \theta^* \|^2_2 + \lambda_n \| \boldsymbol \theta^* \|_1 \nonumber\\
    & = \frac{1}{2n} \| \bX \btheta^0 - \bX \boldsymbol \theta^* \|_2^2 + \frac{1}{2n} \| \boldeta \|_2^2 \nonumber \\ 
    &\quad - \frac{1}{n} \boldeta^{\mathrm{T}} \bX(\btheta^0 - \boldsymbol \theta^*) + \lambda_n \| \boldsymbol \theta^* \|_1. 
\end{align}
Given the $j$th column of $\bX$, define a set $E_n = \{ \max_{1 \leq j \leq p} n^{-1} \big| \boldeta^{\mathrm{T}} \bX^{(j)} \big| \leq \lambda_0 \}$ for some constant $\lambda_0$. Then, following \Cref{lemma0} on this set, \eqref{thm1eq1} implies
\begin{align*}
    \lefteqn{ \frac{1}{2n} \| \bX \btheta^0 - \bX \boldsymbol \theta^* \|_2^2 + \lambda_n \|\boldsymbol \theta^*\|_1} \\
    &\leq  n^{-1} \boldeta^{\mathrm{T}} \bX(\btheta^0 - \boldsymbol \theta^*) + \lambda_n \|\btheta^0\|_1 \\
    &\leq  \max_{1 \leq j \leq p} \frac{|\boldeta^{\mathrm{T}} \bX^{(j)}|}{n} \| \btheta^0 - \boldsymbol \theta^* \|_1 + \lambda_n \|\btheta^0\|_1.
\end{align*}
which is bounded by $\lambda_0\| \btheta^0 - \boldsymbol \theta^* \|_1 + \lambda_n \|\btheta^0\|_1$. After decomposing the vectors into components $\btheta_{S_0}$ and $\btheta_{S_0^c}$ corresponding to the oracle set $S_0$ and its complement, we have by the triangle inequality that $\|\btheta^*_{S_0} - \btheta^0_{S_0}\|_1 \geq \|\btheta^0_{S_0}\|_1 - \|\btheta^*_{S_0}\|_1$.
Since $\btheta^0_{S_0^c} = \bm{0}_{p-s_0}$, by choosing $\lambda_0 \leq {\lambda_n}/{2}$, we can rewrite the above inequality as
\begin{align*}
    \frac{1}{2n} \| \bX \btheta^0 - \bX \boldsymbol \theta^* \|_2^2 \leq & \frac{\lambda_n}{2} \big( \|\btheta^0_{S_0} - \btheta^*_{S_0}\|_1 + \|\btheta^0_{S_0^c}\|_1 \big) \nonumber\\
    & \quad + \lambda_n \big(\|\btheta^0_{S_0}\|_1 - \|\btheta^*_{S_0}\|_1 - \|\btheta^*_{S_0^c}\|_1 \big),
\end{align*} 
which implies that 
$$\frac{1}{2n} \| \bX \btheta^0 - \bX \boldsymbol \theta^* \|_2^2 + \frac{\lambda_n}{2} \| (\boldsymbol \theta^* - \btheta^0)_{S_0^c} \|_1 \leq \frac{3 \lambda_n}{2} \| (\boldsymbol \theta^* - \btheta^0)_{S_0} \|_1
.$$
Then, making use of the compatibility condition, we arrive at the relation
\begin{align*}
    \frac{1}{2n} \| \bX \btheta^0 - \bX \boldsymbol \theta^* \|_2^2 + \frac{\lambda_n}{2} \| \boldsymbol \theta^* - \btheta^0 \|_1 & \leq 2 \lambda_n \|(\boldsymbol \theta^* - \btheta^0)_{S_0} \|_1 \\ & \leq \frac{2 \lambda_n\sqrt{s_0}}{\sqrt{n} \phi_0} \|\bX \btheta^0 - \bX \boldsymbol \theta^*\|_2 .
\end{align*}  Finally, using $2ab \leq a^2 + b^2$ with $a = 1/(2\sqrt{n})\|\bX\btheta^0 - \bX\btheta^*\|_2$ and $b = \lambda_n \sqrt{s_0}/\phi_0$, we can bound the above as
$$ \frac{1}{4n} \| \bX \btheta^0 - \bX \boldsymbol \theta^* \|_2^2 + \frac{\lambda_n}{2} \| \boldsymbol \theta^* - \btheta^0 \|_1 \leq \frac{4 \lambda_n^2 s_0}{\phi_0^2},$$ 
which implies that $\| \boldsymbol \theta^* - \btheta^0 \|_1 = \mathcal{O}_P(s_0 \lambda_n).$
\end{proof}

\begin{proof}[Proof of \Cref{th:thm2}]
We can write $\btheta^*$ as,
\begin{align}
\label{argmin}
    &\argmin_{\bm{u}} \{ \|{\bX \btheta - \bX \bm{u}}\|_2^2 + n\lambda_n \|{\bm{u}}\|_1 \} \nonumber \\
    = &\argmin_{\bm{u}} \{ \|{\bX \btheta - \bX \btheta^0 - \bX \big(\bm{u} - \btheta^0 \big)}\|_2^2 + n\lambda_n \|{\bm{u} - \btheta^0 + \btheta^0}\|_1\} \nonumber.
\end{align}
Hence $\bm{u}_1^*=\btheta^*-\bm{\theta}^0$ is given by  
\begin{align}
    \lefteqn{ \argmin_{\bm{u}_1} \{ \|{\boldeta - \bX \bm{u}_1}\|_2^2 + n\lambda_n \|{\bm{u}_1 + \btheta^0}\|_1\} } \nonumber\\
    &= \argmin_{\bm{u}_1} \{ \bm{u}_1 ^{\mathrm{T}} \bX ^{\mathrm{T}} \bX \bm{u}_1 - 2 \bm{u}_1^{\mathrm{T}} \bX^{\mathrm{T}} \boldeta + n\lambda_n \sum_{j=1}^{p} |\theta^0_{j} + u_{1j}|\}.
\end{align}
Differentiating $\bm{u}_1 ^{\mathrm{T}} \bX ^{\mathrm{T}} \bX \bm{u}_1 - \bm{u}_1^{\mathrm{T}} \bX^{\mathrm{T}} \boldeta$ with respect to $\bm{u}_1$, we get, \begin{equation}\label{kkt}
    \big[2 \bX^{\mathrm{T}} \bX \bm{u}_1 - 2 \bX^{\mathrm{T}} \boldeta \big] = 2 \sqrt{n} \{\bC_n(\sqrt{n}\bm{u}_1) - \bZ_n \}, 
\end{equation}
where 
$$\bZ_n \coloneqq n^{-1/2} \bX^{\mathrm{T}} \boldeta = n^{-1/2} \begin{bmatrix}
    \bX^{\mathrm{T}}_{n(1)} \boldeta\\
    \bX^{\mathrm{T}}_{n(2)} \boldeta
\end{bmatrix} = \begin{bmatrix}
    \bZ_{n(1)} \\
    \bZ_{n(2)} 
\end{bmatrix}.$$ 
To find a solution to \eqref{argmin}, the Karush-Kuhn-Tucker (KKT) conditions are employed. These represent a set of initial derivative tests, often referred to as first-order necessary conditions, to determine the optimality of a solution in a constrained optimization problem, assuming that specific regularity conditions are met. The stationarity condition that the gradient of the generalized Lagrangian given in \eqref{kkt} is zero at the minimizer $\bm{u}_1^*$ is characterized by 
 $$ 2 \sqrt{n} \{\bC_n(\sqrt{n}\bm{u}_1) - \bZ_n \} = - n \lambda_n \boldsymbol\gamma,$$ 
 where $\bm{\gamma} = \big(\gamma_1, \dots, \gamma_{p}\big)^{\mathrm{T}}$ is the subdifferential given by  
 $$\gamma_j \in \begin{cases}
        \textnormal{sign}(\theta^*_j),& \text{if } \theta^*_j \neq 0 \text{ for } j \leq s_0,\\
        [-1,1], & \text{if } \theta^*_j = 0 \text{ for } j >s_0.
    \end{cases}$$
    
Let $\boldsymbol\theta^* = \begin{bmatrix}
\boldsymbol\theta^*_{(1*)} \\
\boldsymbol\theta^*_{(2*)} 
\end{bmatrix} $ be the partition of the sparse projection where $\boldsymbol\theta^*_{(1*)}$ consists of all non-zero elements of $\boldsymbol\theta^*$ and $\boldsymbol\theta^*_{(2*)}$ is the collection of the zero components. This partition is different from the original partition of the data-generating truth $\btheta^0 = \begin{bmatrix}
\btheta^0_{(1)} \\
\btheta^0_{(2)} 
\end{bmatrix}$. Then, we can partition the associated matrices and vectors of \eqref{kkt} as $$2 \sqrt{n} \Bigg\{\begin{bmatrix}
\bC_{n(11*)} & \bC_{n(12*)} \\
\bC_{n(21*)} & \bC_{n(22*)} 
\end{bmatrix} \sqrt{n}  \begin{bmatrix}
\bm{u}^*_{1(1*)} \\
\bm{u}^*_{1(2*)} 
\end{bmatrix} -  \begin{bmatrix}
\bZ_{n(1*)} \\
\bZ_{n(2*)} 
\end{bmatrix} \Bigg\}.$$

Following the KKT conditions, we have 
\begin{enumerate}
    \item [(a)] $\sum_{k = 1}^2\bC_{n(1k*)}[\sqrt{n} \bm{u}^*_{1(k*)}] - \bZ_{n(1*)} = - \frac{\sqrt{n}\lambda_n}{2} \text{sign}(\boldsymbol \theta^*_{(1)})$,
    
    \item [(b)] $|\bC_{n(21*)}[\sqrt{n} \bm{u}^*_{1(1*)}] + \bC_{n(22*)}[\sqrt{n} \bm{u}^*_{1(2*)}] - \bZ_{n(2*)}| \leq \frac{\sqrt{n}\lambda_n}{2}\vec{1}.$
\end{enumerate}

For the event $\{\text{sign}(\btheta^*) = \text{sign}(\btheta^0)\}$ to happen, the partition of $\boldsymbol\theta^*$ and $\btheta^0$ should match. That is, if the sparse projection selects the variables correctly, then $\bm{u}^*_{1(2*)} = \boldsymbol\theta^*_{(2*)} - \btheta^0_{(2)} = \boldsymbol\theta^*_{(2)} - \btheta^0_{(2)} = \bm{0}$ and $\text{sign}(\btheta^0_{(1)}) = \text{sign}(\boldsymbol \theta^*_{(1)})$. Then, the revised KKT conditions, along with an added condition listed below, imply sign consistency of the sparse projection. That is, if
\begin{enumerate}
    \item [(i)] $\bC_{n(11)}[\sqrt{n} \bm{u}^*_{1(1)}] - \bZ_{n(1)} = - \frac{\sqrt{n}\lambda_n}{2} \text{sign}(\btheta^0_{(1)})$,
    
    \item [(ii)] $|\bC_{n(21)}[\sqrt{n} \bm{u}^*_{1(1)}] - \bZ_{n(2)}| \leq \frac{\sqrt{n}\lambda_n}{2}\vec{1}$,
    
    \item [(iii)] $\{|\bm{u}^*_{1(1)}| < |\btheta^0_{(1)}|\} \subseteq \{\text{sign}(\boldsymbol\theta^*_{(1)}) = \text{sign}(\btheta^0_{(1)})\}$,
\end{enumerate}
then, the events in (i), (ii), (iii) together imply $\{\text{sign}(\boldsymbol\theta^*_{(1)}) = \text{sign}(\btheta^0_{(1)})\} = \{\text{sign}(\btheta^*) = \text{sign}(\btheta^0)\} $ and 
\begin{align*}
    \lefteqn{\Pi( \{\text{sign}(\btheta^*) = \text{sign}(\btheta^0)\} | \bY) }\\
    &\geq \Pi \big( \big\{ \bC_{n(11)}[\sqrt{n} \bm{u}^*_{1(1)}] - \bZ_{n(1)} = - \frac{ \sqrt{n}\lambda_n}{2} \text{sign}(\btheta^0_{(1)}) \big\}  \\
    & \quad \bigcap \{ |\bm{u}^*_{1(1*)}| < |\btheta^0_{(1)}| \} \\
    & \quad \bigcap \big\{ \big|\bC_{n(21)}[\sqrt{n} \bm{u}^*_{1(1)}] - \bZ_{n(2)} \big| \leq \frac{ \sqrt{n}\lambda_n}{2}\vec{1} \big\}  \big| \bY  \big) \\
     & =\Pi \big( \big\{ \big| \bC^{-1}_{n(11)} \big(\bZ_{n(1)} - \frac{ \sqrt{n}\lambda_n}{2} \text{sign}(\btheta^0_{(1)}) \big) \big| \leq \sqrt{n} |\btheta^0_{(1)}| \big\} \\
    & \quad \bigcap \big\{ \big| \bC_{n(21)}\bC^{-1}_{n(11)} \bZ_{n(1)} - \bZ_{n(2)} \big| \\
    & \quad \leq \frac{ \sqrt{n}\lambda_n}{2} \big(\vec{1} - \big| \bC_{n(21)}\bC^{-1}_{n(11)}\text{sign}(\btheta^0_{(1)}) \big| \big)\big\} \big| \bY \big) \\
     & = \Pi \big( A_n \cap \Tilde{B}_n | \bY \big), 
\end{align*}
where, by the irrepresentable condition in \Cref{irrep}, 
\begin{align*}
    & \Tilde{B}_n \coloneqq \big\{ \big| \bC_{n(21)}\bC^{-1}_{n(11)} \bZ_{n(1)} - \bZ_{n(2)} \big| \\ & \qquad \leq \frac{ \sqrt{n}\lambda_n}{2} \big(\vec{1} - \big| \bC_{n(21)}\bC^{-1}_{n(11)}\text{sign}(\btheta^0_{(1)}) \big| \big)\big\} \supseteq B_n.
\end{align*}  
On the event $B_n$, every component of the vector on the left-hand side of the inequality is less than the respective component of the vector on the right-hand side. Then, we conclude that 
\begin{align*}
    \Pi(  \{\text{sign}(\btheta^*) = \text{sign}(\btheta^0)\} | \bY) & \geq \Pi \big( A_n \cap B_n | \bY \big)\\ & = 1 - \big[ \Pi\big(A_n^c | \bY \big) + \Pi\big(B_n^c | \bY \big) \big].
\end{align*}
Define the following quantities 
\begin{align*}
    & \bm{W}_{1n} = n^{-1/2}\bC^{-1}_{n(11)} \bX^{\mathrm{T}}_{n(1)},\\
    & \bm{V}_1 = n^{-1}\sigma^2 \bC_{n(11)}^{-1} \bX_{n(1)}^{\mathrm{T}} \bm{H}(a_n) \bX_{n(1)} \bC_{n(11)}^{-1},\\
    & \bm{W}_{2n} = n^{-1/2} \big( \bC_{n(21)} \bC_{n(11)}^{-1} \bX_{n(1)}^{\mathrm{T}} - \bX^{\mathrm{T}}_{n(2)} \big),\\
    & \bm{V}_2 = \bC_{n(21)} \bm{V}_1 \bC_{n(21)} - n^{-1}\sigma^2 \bX_{n(2)}^{\mathrm{T}} \bm{H}(a_n) \bX_{n(2)},
\end{align*}    
where $\bm{W}_{2n}\in \R^{(p - s_0) \times n}.$ Given the data, we have 
\begin{align*}
\bC^{-1}_{n(11)} \bZ_{n(1)} \big| \bY \sim \normal_{s_0}(\bm{W}_{1n}\boldsymbol{\mu}, \bm{V}_1), \\ 
\bC_{n(21)} \bC_{n(11)}^{-1} \bZ_{n(1)} - \bZ_{n(2)} \big| \bY \sim \normal_{p - s_0} \big( \bm{W}_{2n} \boldsymbol\mu, \bm{V}_2 \big).  
\end{align*}
Let $\bm{W}_{2n} = \big( \bm{w}_{2n}^{(1)}, \dots, \bm{w}_{2n}^{(p - s_0)}\big)^{\mathrm{T}}$ such that $\bm{w}_{2n}^{(j)} \in \R^n \enskip \text{ for all } j = 1, \dots, p - s_0$. Thus, the individual components of both vectors defined above have finite variances. Then, defining $a_{nj}$ as the $j$th component of the $s_0$-dimensional vector $\bC^{-1}_{n(11)} \bZ_{n(1)} - \bm{W}_{1n} \boldsymbol{\mu}$ and $b_{nj}$ as the $j\textnormal{th}$ component of $\bC_{n(21)} \bC_{n(11)}^{-1} \bZ_{n(1)} - \bZ_{n(2)} - \bm{W}_{2n} \boldsymbol\mu$, we have, $\sup_{\sigma^* \in \mathcal{U}_n} \E(a_{nj}|\bY, \sigma^*)^2 < h^2, \text{ and } \sup_{\sigma^* \in \mathcal{U}_n} \E(b_{nj}|\sigma^*, \bY)^2 < h^2$ for some constant $h>0$. Then with $P_1$ and $P_2$ defined by  
$$P_1 = \Big\| \frac{\sqrt{n}\lambda_n}{2} \bC^{-1}_{n(11)} \text{sign}(\btheta^0_{(1)})\Big\|_2, \quad P_2 = \| \bm{W}_{1n} \boldsymbol{\mu} \|_\infty,$$ 
we have,
\begin{align}
\label{A_n}
    \lefteqn{\sup_{\sigma^* \in \mathcal{U}_n} \Pi \big( A^c_n | \bY,\sigma^* \big) }\nonumber\\
    & = \sup_{\sigma^* \in \mathcal{U}_n} \Pi \big( \big\{ \big\| \bC^{-1}_{n(11)} \bZ_{n(1)} - \frac{\sqrt{n}\lambda_n}{2} \bC^{-1}_{n(11)} \text{sign}(\btheta^0_{(1)})\big\|_\infty \nonumber \\
    & \qquad \qquad> \sqrt{n} ( \|\btheta^0_{(1)}\|_\text{min}  ) \big\} \big| \bY, \sigma^* \big) \nonumber \\
    & \leq \sup_{\sigma^* \in \mathcal{U}_n} \Pi \big( \big\{ \big\| \bC^{-1}_{n(11)} \bZ_{n(1)} - \bm{W}_{1n} \boldsymbol\mu \big\|_\infty  \nonumber \\
    & \quad + \big\| \frac{\sqrt{n}\lambda_n}{2} \bC^{-1}_{n(11)} \text{sign}(\btheta^0_{(1)})\big\|_\infty \nonumber \\
    & \quad + \| \bm{W}_{1n} \boldsymbol{\mu} \|_\infty > \sqrt{n} ( \|\btheta^0_{(1)}\|_\text{min}  ) \big\} \big| \bY, \sigma^* \big) \nonumber \\
    & \leq \Pi \big( P_1 + P_2 > \frac{\sqrt{n}}{2}  \|\btheta^0_{(1)}\|_\text{min}  \big| \bY \big) \nonumber\\
    & \quad + \sup_{\sigma^* \in \mathcal{U}_n} \Pi \big( \bigcup_{j=1}^{s_0} \{ |a_{nj}|  > \frac{\sqrt{n}}{2} \|\btheta^0_{j} \|_\text{min} \} \big| \bY, \sigma^* \big) \nonumber \\
    & \leq \Pi \big( \frac{\sqrt{n} \lambda_n \sqrt{s_0}}{M_2} + \| \bm{W}_{1n} \boldsymbol{\mu} \|_\infty > \frac{\sqrt{n}}{2}  \|\btheta^0_{(1)}\|_\text{min} \big| \bY\big) \nonumber \\
    & \quad + \sum_{j=1}^{s_0} \sup_{\sigma^* \in \mathcal{U}_n} \Pi \big( |a_{nj}| > \frac{M_3}{2} n^{{b_2}/{2}} \big| \bY, \sigma^* \big)  
\end{align}
by the beta-min condition. Now, 
\begin{align*}
    &\Pi \big( \frac{\sqrt{n}\lambda_n \sqrt{s_0}}{ M_2} + \|\bm{W}_{1n} \boldsymbol \mu \|_{\infty} > \frac{\sqrt{n}}{2} \|\boldsymbol \theta^0_{(1)}\|_\text{min} \big)\\ 
    &\leq \Pi \big( \| \bm{W}_{1n}\|_F \|\bm{X} \big( \boldsymbol \hat{\theta}^\mathrm{R} - \boldsymbol{\theta}_0\big)\|_2 > \frac{\sqrt{n}}{2} \|\boldsymbol{\theta}_{0(1)}\|_\text{min}  + b_0 n^{\frac{b_4+b_1}{2}}\big)\\
    &\leq \frac{\|\bm{C}_{n(11)}^{-1} \bm{X}^{\mathrm{T}}_{n(1)}\|_F^2 \E \big( n^{-1}\|\bm{X}\hat{\boldsymbol \theta}^\mathrm{R} - \bm{X} \boldsymbol \theta_0\|_2^2\big)}{\big( {\sqrt{n}}\|\boldsymbol{\theta}_{0(1)}\|_\text{min}/2 + b_0 n^{{(b_4+b_1)}/{2}}\big)^2}
\end{align*}
for some constant $b_0>0$. Clearly, first term in \eqref{A_n} is free of $\sigma^*$ and goes to zero since ${\sqrt{n}}{2}\|\boldsymbol{\theta}_{0(1)}\|_\text{min}/2$ grows to infinity faster than $n^{b_2/2}$ and by the theory of ridge regression estimation in \cite{shao2012estimation}, we have $\E \big( \|\bm{X}\hat{\boldsymbol \theta}^\mathrm{R} - \bm{X} \boldsymbol \theta_0\|_2^2\big) = \mathcal{O}(n)$. The second term of \eqref{A_n} reduces to $s_0 \mathcal{O} \big(1 - \Phi \big((2h)^{-1} M_3 n^{b_2/2} \big) \big) = o(e^{-n^{b_3}})$ using the Gaussian tail bound $ 1 - \Phi(t) \leq t^{-1} e^{-{t^2/2}}$. Next, writing $\nu_* = \min\{ \nu_j: 1 \leq j \leq p - s_0\} $, we have, 
\begin{align*}
    \lefteqn{\sup_{\sigma^* \in \mathcal{U}_n} \Pi \big( B^c_n | \bY, \sigma^* \big)}\\
    & = \Pi \big( \big\{ \big| \bC_{n(21)}\bC^{-1}_{n(11)} \bZ_{n(1)} - \bZ_{n(2)} \big| > \frac{\sqrt{n}\lambda_n}{2} \boldsymbol\nu \big\} \big| \bY, \sigma^* \big)\\
    & \leq \sup_{\sigma^* \in \mathcal{U}_n} \Pi \big( \big\{ \big| \bC_{n(21)}\bC^{-1}_{n(11)} \bZ_{n(1)} - \bZ_{n(2)} - \bm{W}_{2n} \boldsymbol\mu \big|\\
    & \qquad \qquad + |\bm{W}_{2n} \boldsymbol\mu| > \frac{ \sqrt{n}\lambda_n}{2} \boldsymbol\nu \big\} \big| \bY, \sigma^* \big)\\
    & \leq \sup_{\sigma^* \in \mathcal{U}_n} \Pi \big( \overset{p - s_0}{\underset{j=1}{\cup}} \big\{|b_{nj}| > \frac{\sqrt{n}\lambda_n}{2} \nu_j\big\} \big| \bY, \sigma^* \big) \\
    & \qquad \qquad + \Pi \big( |\bm{W}_{2n} \boldsymbol\mu| > \frac{ \sqrt{n}\lambda_n}{2} \boldsymbol\nu \big| \bY \big)\\
    & \leq \sup_{\sigma^* \in \mathcal{U}_n} \Pi \big( \overset{p - s_0}{\underset{j=1}{\cup}} \big\{|b_{nj}| > \frac{\sqrt{n}\lambda_n}{2} \nu_*\big\} \big| \bY, \sigma^* \big) \\
    & \qquad \qquad + \Pi \big( \|\bm{W}_{2n}\|_F \|\bm{X}\big( \hat{\boldsymbol{\theta}}_R - \boldsymbol \theta_0 \big)\|_2 > \frac{ \sqrt{n}\lambda_n}{2} \boldsymbol\nu \big| \bY \big) \\
    & \leq \sum_{j = 1}^{p - s_0} \sup_{\sigma^* \in \mathcal{U}_n} \Pi \big( \big\{|b_{nj}| > \frac{\sqrt{n}\lambda_n}{2} \nu_*\big\} \big| \bY, \sigma^* \big) \\
    & \qquad \qquad + \frac{\|\bm{W}_{2n}\|_F^2 \|\bm{X}\big( \hat{\boldsymbol{\theta}}_R - \boldsymbol \theta_0 \big)\|^2_2}{\big(\frac{\lambda_n}{2 \sqrt{n}} \boldsymbol\nu \big)^2} \\
    & = (p - s_0) \mathcal{O}\big( 1 - \Phi \big( \frac{ \sqrt{n}\lambda_n \nu_*}{2 h} \big)\big) \\
    & \qquad + \frac{\|\bC_{n(21)} \bC_{n(11)}^{-1} \bX_{n(1)}^{\mathrm{T}} - \bX^{\mathrm{T}}_{n(2)}\|^2_F \E \big(\frac{1}{n} \| \bm{X}(\hat{\boldsymbol{\theta}}_R - \boldsymbol \theta_0)\|_2^2 \big)}{\big(\frac{\sqrt{n}\lambda_n}{2 } \boldsymbol\nu \big)^2}.
\end{align*}
Again, the second term goes to zero as the denominator grows like $n^{b_4}$, whereas the numerator is $\mathcal{O}(1)$, and the first term is $\mathcal{O}(p e^{-n \lambda_n^2 \nu_*^2/(2h)})= o(e^{n^{b_3}})$ by the Gaussian tail bound because $n\lambda_n^2 \asymp n^{b_4} \gg n^{b_3}$. Consequently, we have $\Pi \big( B_n^c | \bY \big) \leq o(e^{-n^{b_3}}).$ 
\end{proof}

\begin{proof}[Proof of \Cref{th:thm3}]
Using \eqref{debiased immersion}, decompose $ \sqrt{n}(\theta_j^{**} - \theta^0_{j})={T}_j + \Lambda_j$, where 
where 
$${T}_j = \sqrt{n} \big( \frac{{\bR^{(j)}}^{\mathrm{T}}\bX \btheta}{{\bR^{(j)}}^{\mathrm{T}}\bX^{(j)}} - \sum_{k = 1}^{p} P_{jk}\theta^0_{k} \big)$$
is the fluctuation term, and 
$$\Lambda_j = \sqrt{n}\sum_{k \neq j} P_{jk} (\theta^0_{k} - \theta^*_k)$$ 
is the bias term. From the relation $\bX \btheta | (\bY,\sigma^*) \sim \normal_n(\bX \hat{\boldsymbol{\theta}}^{\mathbb{R}}, {\sigma^*}^2\bm{H}(a_n) )$, and the posterior consistency of $\sigma^*$ at $\sigma_0$, it follows that the posterior distribution of ${T}_j$ given $\bY$ can be approximated arbitrarily closely in the total variation distance by  $\normal_p \big( m_j, \Sigma_{jj} \big)$ uniformly for $j=1,\ldots,p$, where $m_j$ and $\Sigma_{jj}$ are respectively defined in \eqref{DBLasso lim mean} and \eqref{DBLasso lim variance}. 
It remains to show that given the data, the bias term is asymptotically negligible, that is, for all $\epsilon >0$, $\max_j \prob(|\Lambda_j |> \epsilon| \bY )= o_P(1)$. Note that
\begin{align*}
    \Lambda_j = &  \sqrt{n} \sum_{k \neq j} \frac{{\bR^{(j)}}^{\mathrm{T}} \bX^{(k)}}{{\bR^{(j)}}^{\mathrm{T}} \bX^{(j)}}  (\theta^0_{k} - {\theta}^*_k) \\
    & \leq \sqrt{n} \max_{k \neq j} \Bigg| \frac{{\bR^{(j)}}^{\mathrm{T}} \bX^{(k)}}{{\bR^{(j)}}^{\mathrm{T}} \bX^{(j)}} \Bigg| \| \boldsymbol \theta^* - \boldsymbol \theta^{0} \|_1,
\end{align*}
which can be bounded by $$\sqrt{n}{\lambda^{\bX}_j/(2{|\bR^{(j)}}^{\mathrm{T}} \bX^{(j)}/n|)} \mathcal{O}_p \big( s_0 \sqrt{({\log p})/{n}} \big)$$ using the KKT condition for the LASSO of $\bX^{(j)}$  on $ \bX^{(-j)}$ given by  $|{\bX^{(k)}}^{\mathrm{T}} \bR^{(j)} / n| \leq \lambda^{\bX}_j/2$ and using \Cref{consis_thm}. Finally, if $\lambda^{\bX}_j = \mathcal{O}(\sqrt{({\log p})/n})$, $|{\bR^{(j)}}^{\mathrm{T}} \bX^{(j)}/n|$ can be shown to be bounded away from zero \citep{zhang2014confidence}. Consequently, $\Lambda_j = \mathcal{O}_p((s_0 {\log p})/ \sqrt{n})$, and choosing $s_0 = o(\sqrt{n}/{\log p})$ we can prove $\Delta_j$ negligible. 
\end{proof}

\begin{proof}[Proof of Corollary~\ref{coverage}]
It suffices to show that the mean and variance of the approximating normal distribution of $\theta_j^{**}$ given by $\theta^0_j+m_j/\sqrt{n}$ and $\Sigma_{jj}/n$ asymptotically agree with $\hat{\theta}^\mathrm{DB}_j$ and $\sigma_0^2 \|\bR^{(j)}\|^2/|\bX^\mathrm{T}_{(j)}\bR^{(j)}|^2$ respectively.
 
 Consider the singular value decomposition of the design matrix $\bX$ as $\bX = \bm{U}\bm{D}\bm{V}^{\mathrm{T}}$, where $\bm{U} \in \R^{n \times n}$ and $\bm{V} \in \R^{n \times p}$ satisfy $\bm{U}^{\mathrm{T}}\bm{U} = \bm{I}_n$ and $\bm{V}^{\mathrm{T}} \bm{V} = \bm{I}_p$, and $\bm{D}=\mathrm{diag}(d_1, d_2, \dots, d_n)$, where $d_1,\dots,d_n$ are the singular values of $\bX$. Then, we have $\mathrm{tr}\big(\bm{H}(a_n) \big) = \sum_{j=1}^{n} {d_j^2}/({d_j^2 + a_n})$ and so, 
\begin{align*}
     \Big|\frac{{\bR^{(j)}}^{\mathrm{T}} \bm{H}(a_n) \bR^{(j)}}{{\bR^{(j)}}^{\mathrm{T}} \bR^{(j)}} - 1 \Big|  &= \Big| \frac{{\bR^{(j)}}^{\mathrm{T}} \big(\bm{I} - \bm{H}(a_n) \big) \bR^{(j)}}{{\bR^{(j)}}^{\mathrm{T}} \bR^{(j)}}\Big|\\
     &\leq  \mathrm{tr}\big(\bm{I}_n - \bm{H}(a_n) \big)\\
     &= n \big( 1 - n^{-1} \sum_{j=1}^{n} {d_j^2}/{(d_j^2 + a_n)} \big)\to 0
\end{align*} 
by \Cref{lemma3}. Thus, the ratio of the variances
converges to 1. Next, we show that the difference between the means of the two Gaussian distributions, normalized by the standard deviation, vanishes as $n$ grows. By \Cref{th:thm3}, the induced posterior distribution of $\sqrt{n}  \big(\theta^{**}_j - \hat{\theta}_j^{\mathrm{DB}}\big)$ given $\bm{Y}$ is approximately $\normal_p \big( m_j - \sqrt{n}(\hat{\theta}_j^{\mathrm{DB}}-\theta_j^0), \Sigma_{jj} \big)$. Then, from \Cref{lemma5}, we can write 
$$n^{-1/2}\frac{ |{\bX^{(j)}}^{\mathrm{T}} \bR^{(j)}|}{\|\bR^{(j)}\|} \big|\hat{\theta}_j^{\mathrm{DB}} - (\theta^0_{j}+m_j/\sqrt{n}) \big| \leq A_1 + A_2 = o(1),$$ 
where $A_1 = n^{-1/2} \|\bY - \bX \hat{\bm \theta}^\mathrm{R} \|_1=o_P(1)$ by \Cref{lemma4} and the $A_2 = n^{-1/2} \frac{\lambda^{\bX_j}/2}{\|\bR^{(j)}\|/n} \|\btheta^0 - \hat{\bm \theta}^L\|_1 = o_P(1)$. The latter follows as the KKT condition for the LASSO of $\bX^{(j)}$ on $\bX^{(-j)}$ is given by $|{\bX^{(k)}}^{\mathrm{T}} \bR^{(j)} / n| \leq \lambda^{\bX}_j/2$. Finally, $\|\bR^{(j)}\|/\sqrt{n}$ is bounded away from zero if  and $s_0 \ll \sqrt{n}/(\log p)$, noting that $\|\btheta^0 - \hat{\bm \theta}^L\|_1 = \mathcal{O}_P(s_0 \lambda_n)$ from  (see Theorem 6.1 in \cite{buhlmann2011statistics}).
\end{proof}

\begin{proof}[Proof of Theorem~\ref{thm:coverage ellipsoid}]

By \Cref{th:thm2}, with high probability, the post-variable selection posterior is supported on $\R^{S_0}\times \{ \bm{0}_{S_0^c}\}$, and 
\begin{align*}
\btheta^{\mathrm{PS}}_{S_0}|(\bY,\hat S=S_0,\sigma)\sim \normal_{s_0}(\hat{\btheta}^{\mathrm{R,PS}}_{S_0}, \sigma^2 (\bX_{S_0}^\mathrm{T} \bX_{S_0} + a_n \bm{I}_{s_0})^{-1}).
\end{align*}
Since the post-selection setup reduces to a fixed dimensional normal regression, the marginal posterior distribution of $\sigma$ is consistent. Hence, with $\sigma$ integrated out, the posterior distribution of $\btheta^{\mathrm{PS}}_{S_0}$ given $(\bY,\hat S=S_0)$ is 
approximated in the total variation distance by $\normal_{s_0}(\hat{\btheta}^{\mathrm{R,PS}}_{S_0}, \sigma_0^2 (\bX_{S_0}^\mathrm{T} \bX_{S_0})^{-1})$, using  Assumption~\ref{sd_assum}. 
Consequently, by the definition of $r_{\alpha}$, we have as $n\to\infty$, 
    \begin{align}
    \label{r_alpha limit}
        r_\alpha \to \sigma_0^2 \chi^2_{s_0,\alpha}\mbox{ in } \prob_{\btheta_0}\mbox{-probability},
        \end{align}
    where $\chi^2_{s_0,\alpha}$ is the $(1-\alpha)$-quantile of the chi-square distribution with $s_0$ degrees of freedom. 

By \Cref{ridge CLT}, 
  $\sqrt{n}(\hat{\btheta}^\mathrm{R,PS}_{S_0} - \btheta^0_{S_0}) \rightsquigarrow \normal_{s_0}(\bm{0}_{s_0}, \sigma_0^2 \bC_{11}^{-1})$, which, together with \Cref{sd_assum}, implies that $$(\hat{\btheta}^\mathrm{R,PS}_{S_0} - \btheta^0_{S_0})^\mathrm{T}(\bX_{(1)}^\mathrm{T}\bX_{(1)}+ a_n\bm{I}_{S_0})(\hat{\btheta}^\mathrm{R,PS}_{S_0} - \btheta^0_{S_0}) \rightsquigarrow \sigma_0^2\chi^2_{s_0}.$$ 
    
    Define the event $$E_n = \{\sigma_0^2 \chi^2_{s_0,\alpha}(1-\delta_n) \leq r_{\alpha} \leq \sigma_0^2\chi^2_{s_0,\alpha}(1+\delta_n)\},$$ where  $\delta_n \to 0$ sufficiently slowly. Then $\prob(E_n^c) \to 0$ as $n \to \infty$ by \eqref{r_alpha limit}. On $E_n$, we have 
    $$\underline{D}_n \times \{\bm{0}_{S_0^c}\}\subset \{ \btheta^0_{S_0} \in D_{n,1-\alpha}^{\mathrm{PS}} \} \subset \overline{D}_n\times \{\bm{0}_{S_0^c}\},$$     
    where 
    \begin{align*} 
    \underline{D}_n &=\{(\hat{\btheta}^\mathrm{R}_{S_0} - \btheta^0_{S_0})^\mathrm{T}(\bX_{(1)}^\mathrm{T}\bX_{(1)}+a_n\bm{I}_{S_0})(\hat{\btheta}^\mathrm{R}_{S_0} - \btheta^0_{S_0}) \\ & \qquad \qquad \qquad \leq (1-\delta_n) \sigma_0^2\chi^2_{s_0,\alpha}\}\\ 
    \overline{D}_n &=\{(\hat{\btheta}^\mathrm{R}_{S_0} - \btheta^0_{S_0})^\mathrm{T}(\bX_{(1)}^\mathrm{T}\bX_{(1)}+a_n\bm{I}_{S_0})(\hat{\btheta}^\mathrm{R}_{S_0} - \btheta^0_{S_0}) \\ & \qquad \qquad \qquad \leq (1+\delta_n) \sigma_0^2\chi^2_{s_0,\alpha}\}.
    \end{align*}
    Thus the coverage probability $\prob_{\btheta^0} (\btheta^0 \in D_{n, 1-\alpha}^{\mathrm{PS}})$  is upper bounded by
    \begin{align*}
         \prob_{\btheta^0}(\{\btheta^0 \in D_{n,1-\alpha}^{\mathrm{PS}}\}\cap E_n) + \prob_{\btheta^0}(E_n^c)
        \leq \prob_{\btheta^0}(\overline{D}_n) + \prob_{\btheta^0}(E_n^c),
    \end{align*}
     which converges to  $1-\alpha$. 
    Similarly, the coverage probability is lower bound by 
    \begin{align*}
        \prob_{\btheta^0}(\{\btheta^0_{S_0} \in D_{n,1-\alpha}^{\mathrm{PS}}\}\cap E_n) - \prob_{\btheta^0}(E_n^c)
        \geq \prob_{\btheta^0}(\underline{D}_n) - \prob_{\btheta^0}(E_n^c), 
    \end{align*}
    which also converges to  $1-\alpha$. 
\end{proof}

\bibliography{reference}

\end{document}